\documentclass[acmtog,authorversion]{acmart} %

\AtBeginDocument{%
  }

\setcopyright{rightsretained}%
\copyrightyear{2024}
\acmYear{2024}
\acmDOI{10.1145/3659577}

\acmJournal{TOG}
\acmVolume{43}
\acmNumber{3}
\acmArticle{33}
\acmMonth{6}

\usepackage[export]{adjustbox}%
\usepackage{enumitem}
\usepackage{booktabs} %
\usepackage{tabularx}
\usepackage{xargs}
\usepackage{mathtools}
\usepackage{multirow}
\usepackage{amsmath}
\usepackage{amsfonts}
\usepackage{bm}
\usepackage{microtype}
\usepackage{siunitx}
\usepackage[percent]{overpic}
\usepackage{cancel}
\usepackage{wrapfig}
\usepackage{placeins}
\usepackage[normalem]{ulem}
\usepackage{grffile}  %
\usepackage{xspace}

\usepackage{listings}

\definecolor{clr-background}{gray}{0.95}
\definecolor{clr-text}{RGB}{0,0,0}
\definecolor{clr-string}{RGB}{163,21,21}
\definecolor{clr-namespace}{RGB}{0,0,0}
\definecolor{clr-preprocessor}{RGB}{128,128,128}
\definecolor{clr-keyword}{RGB}{0,0,255}
\definecolor{clr-type}{RGB}{43,145,175}
\definecolor{clr-variable}{RGB}{0,0,0}
\definecolor{clr-constant}{RGB}{111,0,138} %
\definecolor{clr-comment}{RGB}{0,128,0}

\lstdefinestyle{VS}{
	language=C++,
	backgroundcolor=\color{clr-background},
	basicstyle=\color{clr-text}\linespread{1.1}\ttfamily\footnotesize,
	stringstyle=\color{clr-string},
	identifierstyle=\color{clr-variable}, %
	commentstyle=\color{clr-comment},
	directivestyle=\color{clr-preprocessor}, %
	keywordstyle=\color{clr-type},
	keywordstyle={[2]\color{clr-constant}}, %
	showspaces=false,
	showstringspaces=false,
	showlines=true,
	xleftmargin=0.0cm,
	frame=t,
	framesep=0.15cm,
	framerule=0pt,
	tabsize=4
}

\def\equationautorefname~#1\null{%
  Equation~(#1)\null
}

\iftrue
\newcommandx{\customComment}[3]{\textcolor{#2}{#1: #3}}
\else
\newcommandx{\customComment}[3]{}
\fi

\definecolor{amber}{rgb}{1.0, 0.49, 0.0}
\definecolor{darkgreen}{rgb}{0.0, 0.5, 0.0}
\newcommandx{\petrik}[1]{\customComment{Petrik}{blue}{#1}}
\newcommandx{\jan}[1]{\customComment{Jan}{magenta}{#1}}
\newcommandx{\fabrice}[1]{\customComment{Fabrice}{amber}{#1}}
\newcommandx{\alex}[1]{\customComment{Alex}{orange}{#1}}
\newcommandx{\benedikt}[1]{\customComment{Benedikt}{red}{#1}}
\newcommandx{\tizian}[1]{\customComment{Tizian}{darkgreen}{#1}}
\newcommandx{\andrea}[1]{\customComment{Andrea}{purple}{#1}}
\newcommandx{\aaron}[1]{\customComment{Aaron}{blue}{#1}}

\newcommandx{\revision}[1]{#1}

\gdef\useCroppedImages{0}
\gdef\cropInsets{0}

\usepackage[export]{adjustbox}
\usepackage{calc}
\usepackage{tabularx}
\usepackage{tikz}
\usepackage{xstring}

\newlength{\beautyHeight}
\newlength{\beautyPixWidth}
\newlength{\beautyPixHeight}
\newlength{\insetvsep}
\gdef\useInsetA{0}
\gdef\useInsetB{0}
\gdef\useInsetC{0}

\newcommand{\setInset}[6]{%
    \expandafter\gdef\csname useInset#1\endcsname{1}%
    \expandafter\gdef\csname inset#1Color\endcsname{#2}%
    \expandafter\gdef\csname crop#1X\endcsname{#3}%
    \expandafter\gdef\csname crop#1Y\endcsname{#4}%
    \expandafter\gdef\csname crop#1W\endcsname{#5}%
    \expandafter\gdef\csname crop#1H\endcsname{#6}%
}

\newcommand{\unsetInset}[1]{%
    \expandafter\gdef\csname useInset#1\endcsname{0}%
}

\newcommand{\addBeautyCrop}[8]{%
    \pdfpxdimen=\dimexpr 1 in/72\relax
    \def\beauty{%
        \let\cropR\relax%
        \let\cropB\relax%
        \newlength\cropR%
        \newlength\cropB%
        \setlength\cropR{{#3 px}-{#5 px}-{#7 px}}%
        \setlength\cropB{{#4 px}-{#6 px}-{#8 px}}%
        \sbox0{\includegraphics[width=#2\textwidth,trim={#5px {\cropB} {\cropR} #6px},clip]{#1}}%
        \begin{tikzpicture}
            \node[anchor=north west,inner sep=0] at (0,0) {\usebox0};
            \begin{scope}[x=\wd0/#7, y=\ht0/#8]
            \if\useInsetA1{
                \draw[\insetAColor,thick] (\cropAX-#5,-\cropAY+#6) rectangle + (\cropAW,-\cropAH);
            }\fi
            \if\useInsetB1{
                \draw[\insetBColor,thick] (\cropBX-#5,-\cropBY+#6) rectangle + (\cropBW,-\cropBH);
            }\fi
            \if\useInsetC1{
                \draw[\insetCColor,thick] (\cropCX-#5,-\cropCY+#6) rectangle + (\cropCW,-\cropCH);
            }\fi
            \end{scope}
        \end{tikzpicture}
    }%
    \setlength\beautyHeight{\heightof{\beauty}}%
    \setlength\beautyPixWidth{#3px}%
    \setlength\beautyPixHeight{#4px}%
    \global\beautyHeight=\beautyHeight%
    \global\beautyPixWidth=\beautyPixWidth%
    \global\beautyPixHeight=\beautyPixHeight%
    \begin{adjustbox}{valign=t}
        \beauty{}
    \end{adjustbox}
}

\newcommand{\trimInset}[6]{%
    \let\cropR\relax%
    \let\cropB\relax%
    \newlength\cropR%
    \newlength\cropB%
    \setlength\cropR{{\beautyPixWidth}-{#3 px}-{#5 px}}%
    \setlength\cropB{{\beautyPixHeight}-{#4 px}-{#6 px}}%
    \color{#2}%
    \fbox{\includegraphics[width=1\linewidth,trim={{#3 px} {\cropB} {\cropR} {#4 px}},clip]{#1}}%
}

\newcommand{\addInset}[2]{%
    \color{#2}%
    \fbox{\includegraphics[width=1\linewidth]{#1}}%
}

\newcommand{\auxtimes}{x}
\newcommand{\auxplus}{+}
\newcommand{\auxspace}{ }

\newcommand{\addInsets}[1]{%
    \begin{adjustbox}{valign=t}
        \StrSubstitute{#1}{.}{-}[\baseFileName]
        \begin{adjustbox}{totalheight=1\beautyHeight,tabular={c}}
            \if\useInsetA1%
                \def\cropfile{\baseFileName-\cropAW\auxtimes\cropAH\auxplus\cropAX\auxplus\cropAY-crop}
                \if\cropInsets1
                    \immediate\write18{convert #1 -crop \cropAW\auxtimes\cropAH\auxplus\cropAX\auxplus\cropAY\auxspace -filter point -resize 800\% \cropfile.jpg}
                \fi
                \if\useCroppedImages1
                    \addInset{\cropfile.jpg}{\insetAColor}
                \else
                    \trimInset{#1}{\insetAColor}{\cropAX}{\cropAY}{\cropAW}{\cropAH}%
                \fi%
            \fi%
            \if\useInsetB1%
                \if\useInsetA1\\[\insetvsep]\fi%
                \def\cropfile{\baseFileName-\cropBW\auxtimes\cropBH\auxplus\cropBX\auxplus\cropBY-crop}
                \if\cropInsets1
                    \immediate\write18{convert #1 -crop \cropBW\auxtimes\cropBH\auxplus\cropBX\auxplus\cropBY\auxspace -filter point -resize 800\% \cropfile.jpg}
                \fi
                \if\useCroppedImages1
                    \addInset{\cropfile.jpg}{\insetBColor}
                \else
                    \trimInset{#1}{\insetBColor}{\cropBX}{\cropBY}{\cropBW}{\cropBH}%
                \fi%
            \fi%
            \if\useInsetC1%
                \if\useInsetB1\\[\insetvsep]\fi%
                \def\cropfile{\baseFileName-\cropCW\auxtimes\cropCH\auxplus\cropCX\auxplus\cropCY-crop}
                \if\cropInsets1
                    \immediate\write18{convert #1 -crop \cropCW\auxtimes\cropCH\auxplus\cropCX\auxplus\cropCY\auxspace -filter point -resize 800\% \cropfile.jpg}
                \fi
                \if\useCroppedImages1
                    \addInset{\cropfile.jpg}{\insetCColor}
                \else
                    \trimInset{#1}{\insetCColor}{\cropCX}{\cropCY}{\cropCW}{\cropCH}%
                \fi%
            \fi%
        \end{adjustbox}
    \end{adjustbox}
}

\newcommand{\x}{\mathbf{x}}
\newcommand{\w}{\bm{\omega}}
\newcommand{\wi}{{\w_\mathrm{i}}}
\newcommand{\wo}{{\w_\mathrm{o}}}
\newcommand{\wh}{{\w_\mathrm{h}}}
\newcommand{\normal}{\mathbf{n}}
\newcommand{\tang}{\mathbf{t}}
\newcommand{\bitang}{\mathbf{b}}
\newcommand{\latents}{\mathbf{z}}
\newcommand{\albedo}{\alpha}
\newcommand{\attribs}{\mathbf{k}}

\newcommand{\brdf}{f}
\newcommand{\pdf}{p}

\newcommand{\warp}{W}
\newcommand{\Li}{{L_\mathrm{i}}}
\newcommand{\Lo}{{L_\mathrm{o}}}

\newcommand{\decoder}{g}

\newcommand{\decoderParams}{\theta}

\newcommand{\inkwell}{\textsc{Inkwell}}
\newcommand{\stage}{\textsc{Stage}}
\newcommand{\cheesegraterBlade}{\textsc{Cheese slicer} blade}
\newcommand{\cheesegraterHandle}{\textsc{Cheese slicer} handle}
\newcommand{\cheesegrater}{\textsc{Cheese slicer}}
\newcommand{\teapot}{\textsc{Teapot}}
\newcommand{\cakebox}{\textsc{Cake box}}

\newcommand{\FLIP}{\protect\reflectbox{F}LIP\xspace}

\begin{document}

\title{Real-Time Neural Appearance Models}

\author{Tizian Zeltner}\authornote{Equal contribution. Order determined by a rock-paper-scissors tournament.}
\email{tzeltner@nvidia.com}
\affiliation{
 \institution{NVIDIA}
 \country{Switzerland}
}
\author{Fabrice Rousselle}\authornotemark[1]
\affiliation{
 \institution{NVIDIA}
 \country{Switzerland}
}
\author{Andrea Weidlich}\authornotemark[1]
\affiliation{
 \institution{NVIDIA}
 \country{Canada}
}
\author{Petrik Clarberg}\authornotemark[1]
\affiliation{
 \institution{NVIDIA}
 \country{Sweden}
}
\author{Jan Novák}\authornotemark[1]
\affiliation{
 \institution{NVIDIA}
 \country{Czech Republic}
}
\author{Benedikt Bitterli}\authornotemark[1]
\affiliation{
 \institution{NVIDIA}
 \country{USA}
}
\author{Alex Evans}
\affiliation{
 \institution{NVIDIA}
 \country{United Kingdom}
}
\author{Tomáš Davidovič}
\affiliation{
 \institution{NVIDIA}
 \country{Czech Republic}
}
\author{Simon Kallweit}
\affiliation{
 \institution{NVIDIA}
 \country{Switzerland}
}
\author{Aaron Lefohn}
\affiliation{
 \institution{NVIDIA}
 \country{USA}
}

\begin{abstract}

We present a complete system for real-time rendering of scenes with complex appearance previously reserved for offline use. This is achieved with a combination of algorithmic and system level innovations.

Our appearance model utilizes learned hierarchical textures that are interpreted using neural decoders, which produce reflectance values and importance-sampled directions.
To best utilize the modeling capacity of the decoders, we equip the decoders with two graphics priors.
The first prior---transformation of directions into learned shading frames---facilitates accurate reconstruction of mesoscale effects. The second prior---a microfacet sampling distribution---allows the neural decoder to perform importance sampling efficiently.
The resulting appearance model supports anisotropic sampling and level-of-detail rendering, and allows baking deeply layered material graphs into a compact unified neural representation.

By exposing hardware accelerated tensor operations to ray tracing shaders, we show that it is possible 
to inline and execute the neural decoders efficiently inside a real-time path tracer.
We analyze scalability with increasing number of neural materials and propose to improve performance using code optimized for coherent and divergent execution. 
Our neural material shaders can be over an order of magnitude faster than 
non-neural layered materials. This opens up the door for using film-quality visuals in real-time applications 
such as games and live previews.

\end{abstract}

\begin{CCSXML}
<ccs2012>
<concept>
<concept_id>10010147.10010371.10010372.10010376</concept_id>
<concept_desc>Computing methodologies~Reflectance modeling</concept_desc>
<concept_significance>500</concept_significance>
</concept>
</ccs2012>
\end{CCSXML}

\ccsdesc[500]{Computing methodologies~Reflectance modeling}

\keywords{appearance models, neural networks, real-time rendering}

\begin{teaserfigure}
    \center
    \includegraphics[width=0.70\linewidth,trim=150 143 40 12,clip,valign=bh]{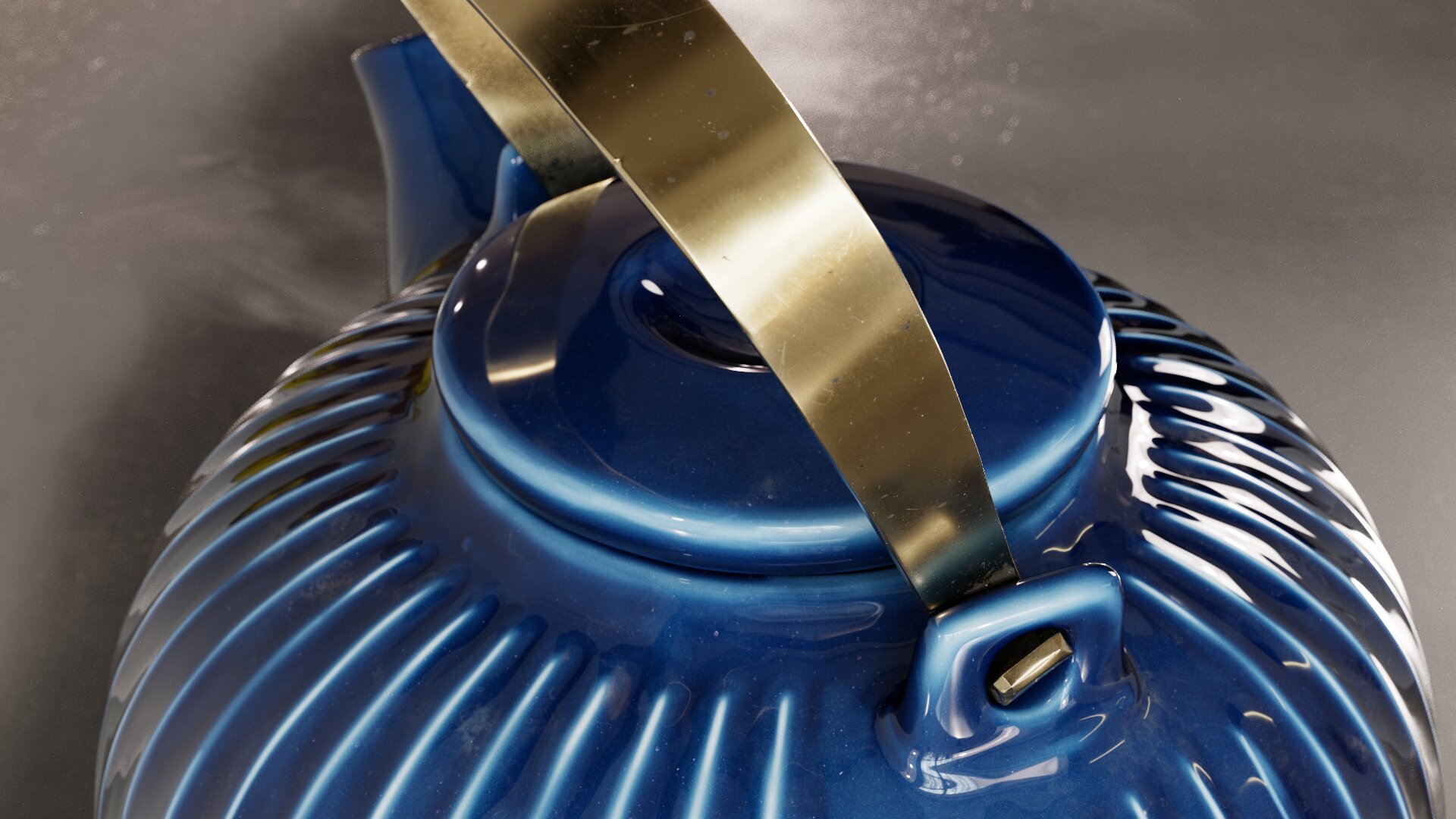}
    \begin{minipage}[b]{.29\textwidth}
  		  \includegraphics[width=\linewidth,trim=0 0 0 0]{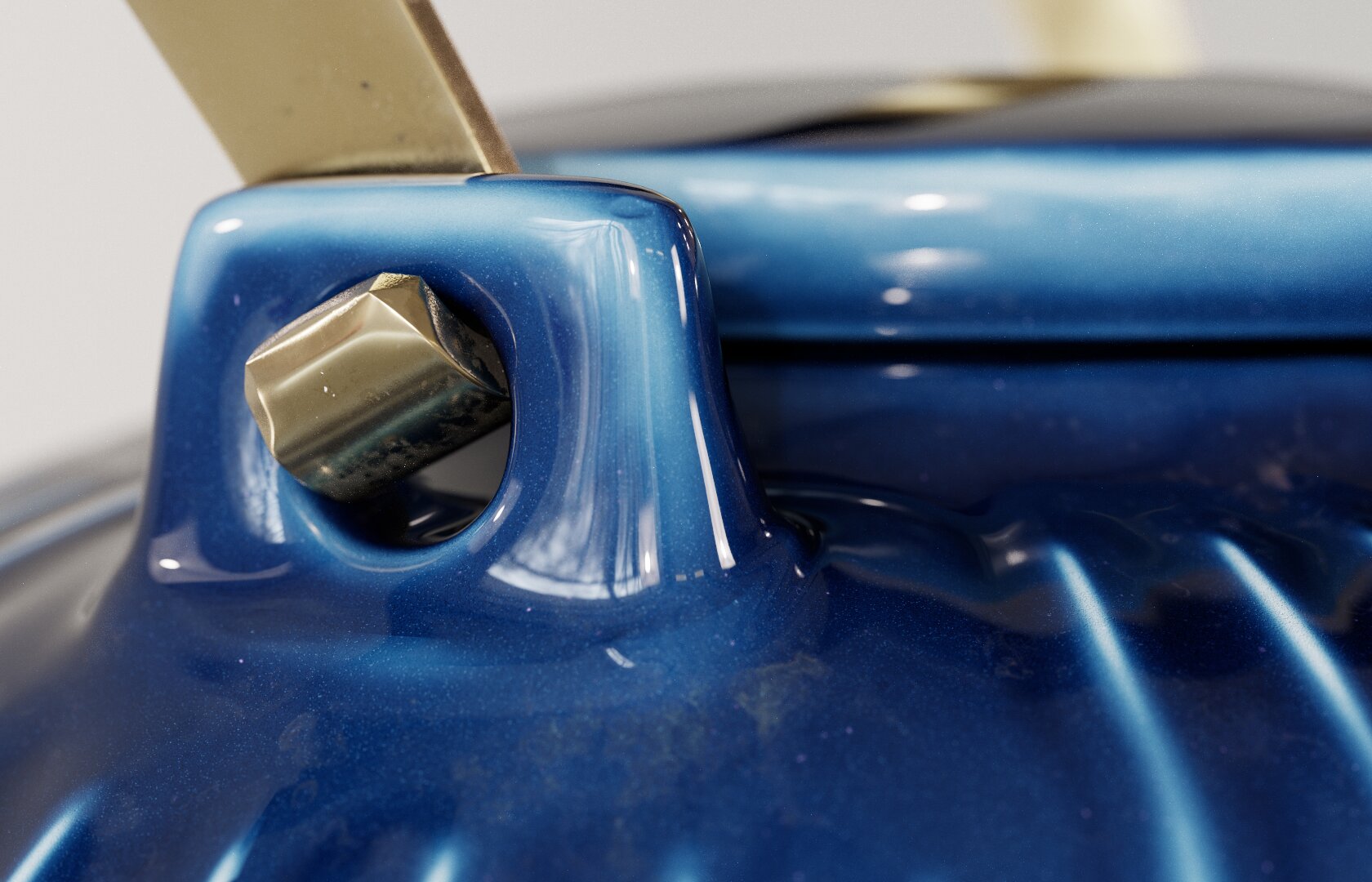}
  		  \includegraphics[width=\linewidth]{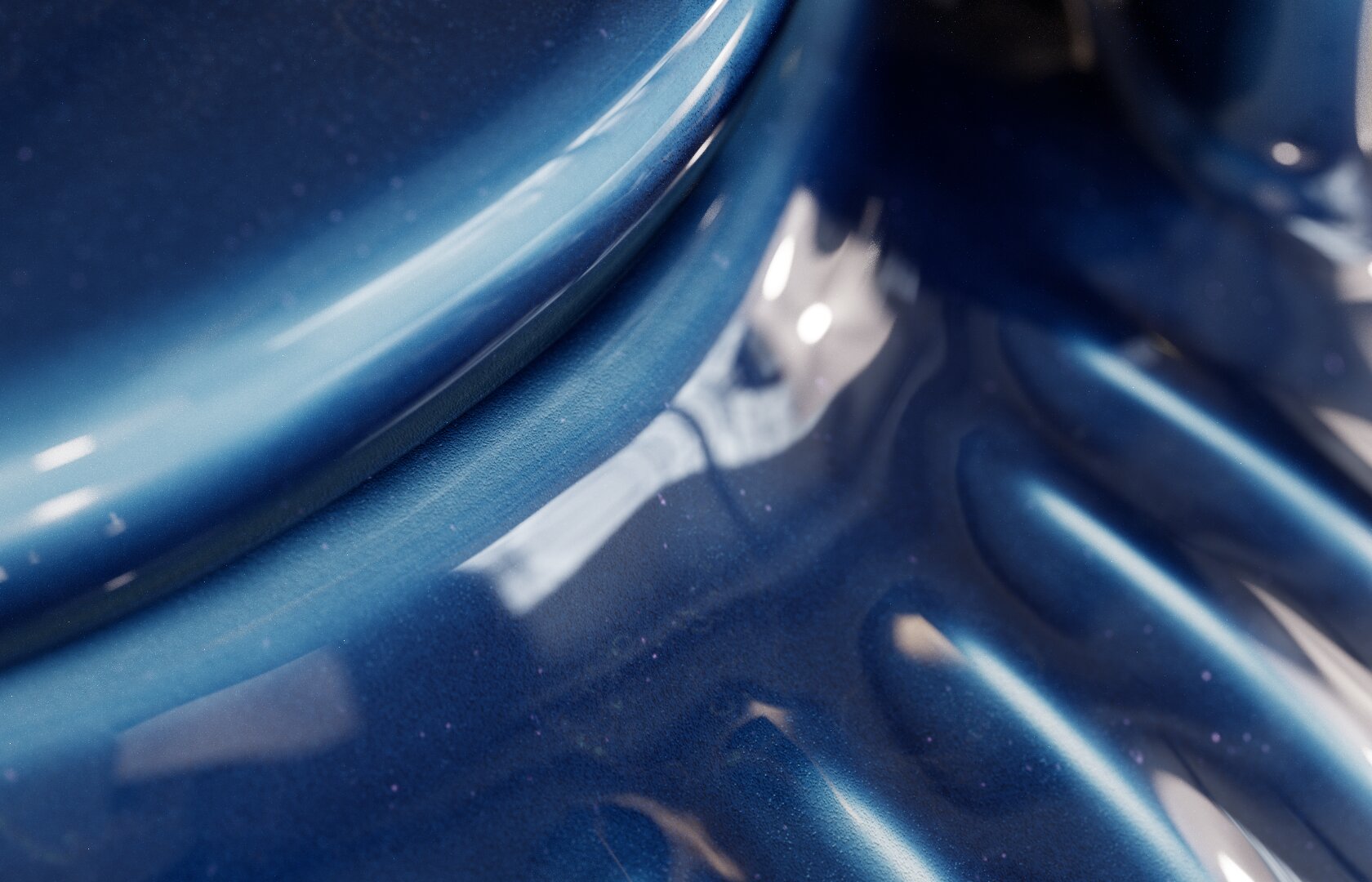}
    \end{minipage}
    \caption{
      Close-up renderings of a \teapot{} asset with our neural BRDF. Our model learns the intricate details and complex multi-layered material behavior of the ceramic, fingerprints, smudges, and dust which are responsible for the realism of the object while being faster to evaluate than traditional non-neural models of similar complexity. The system we present allows us to include such high-fidelity objects in real-time renderers in a scalable way.}\label{fig:teaser}
    \Description{Teaser image}
\end{teaserfigure}

\maketitle

\begin{figure*}[t]
    \setlength{\tabcolsep}{2pt}
    \footnotesize
    \begin{tabular}{ccccc}
        \includegraphics[width=0.19\linewidth,trim=0 0 0 0,clip,valign=b]{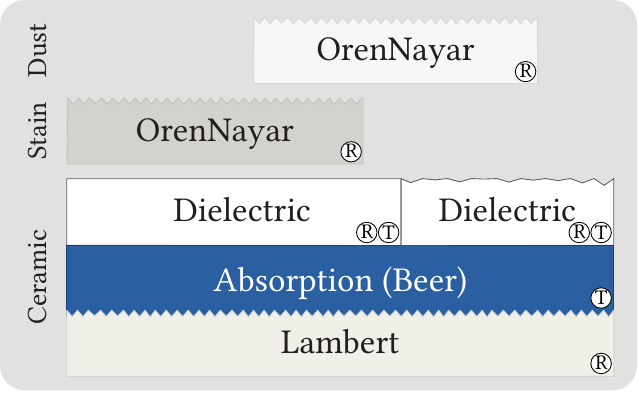} &
        \includegraphics[width=0.19\linewidth,trim=0 0 0 0,clip,valign=b]{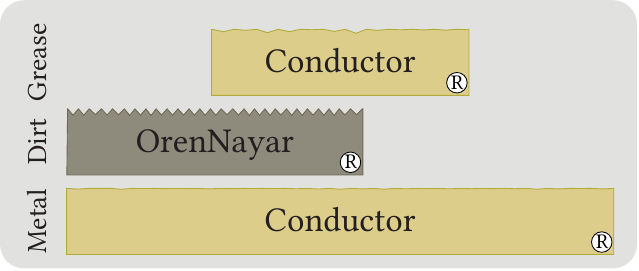} &
        \includegraphics[width=0.19\linewidth,trim=0 0 0 0,clip,valign=b]{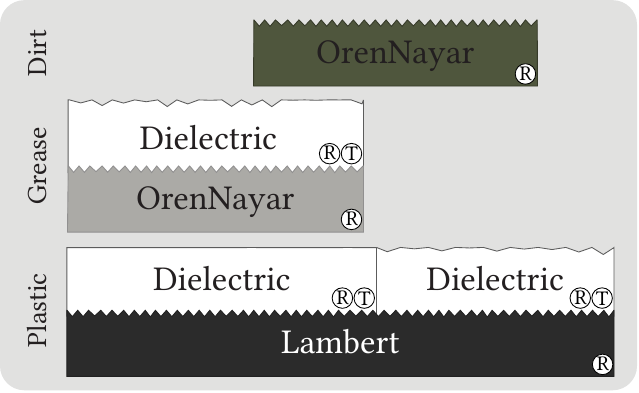} &
        \includegraphics[width=0.19\linewidth,trim=0 0 0 0,clip,valign=b]{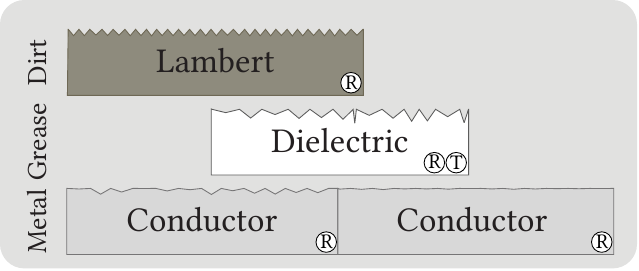} &
        \includegraphics[width=0.19\linewidth,trim=0 0 0 0,clip,valign=b]{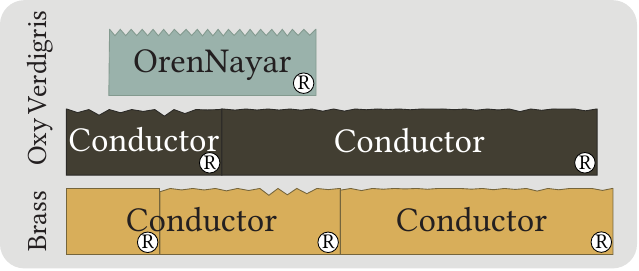}\\
        \includegraphics[width=0.19\linewidth,trim=235 57 0 0,clip,valign=b]{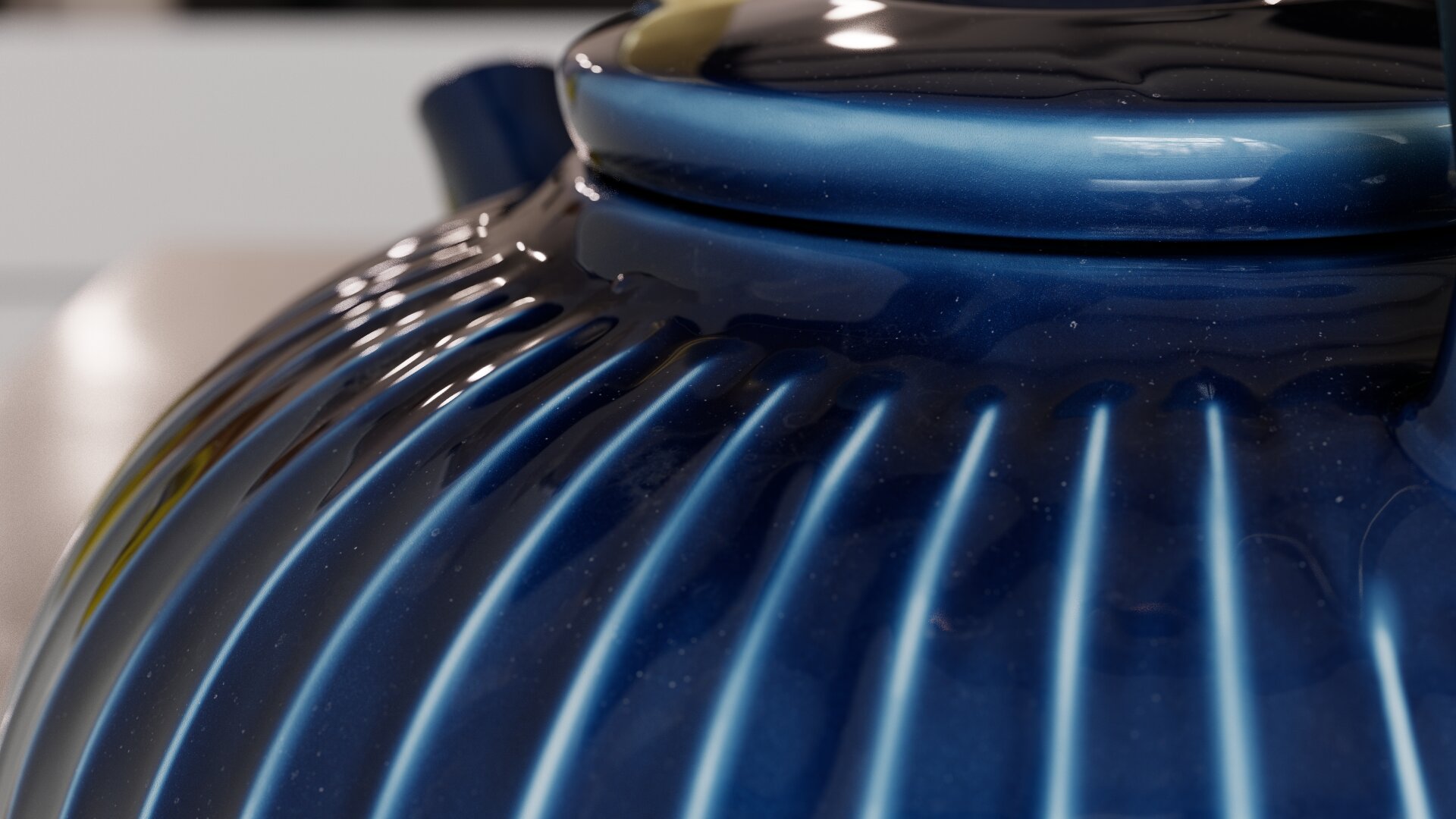}&
        \includegraphics[width=0.19\linewidth,trim=700 300 100 100,clip,valign=b]{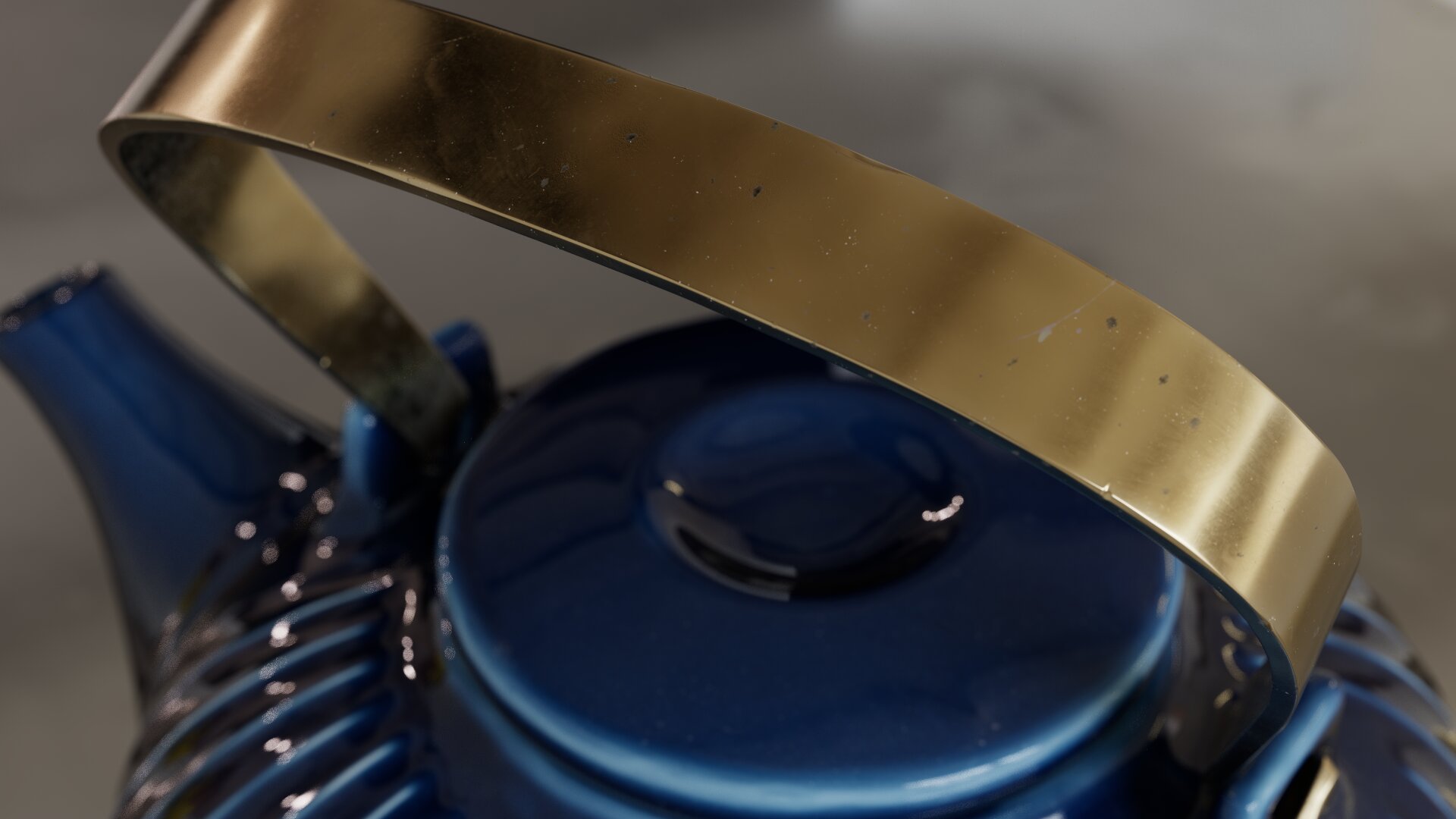}&
        \includegraphics[width=0.19\linewidth,trim=0 0 300 98, clip,valign=b]{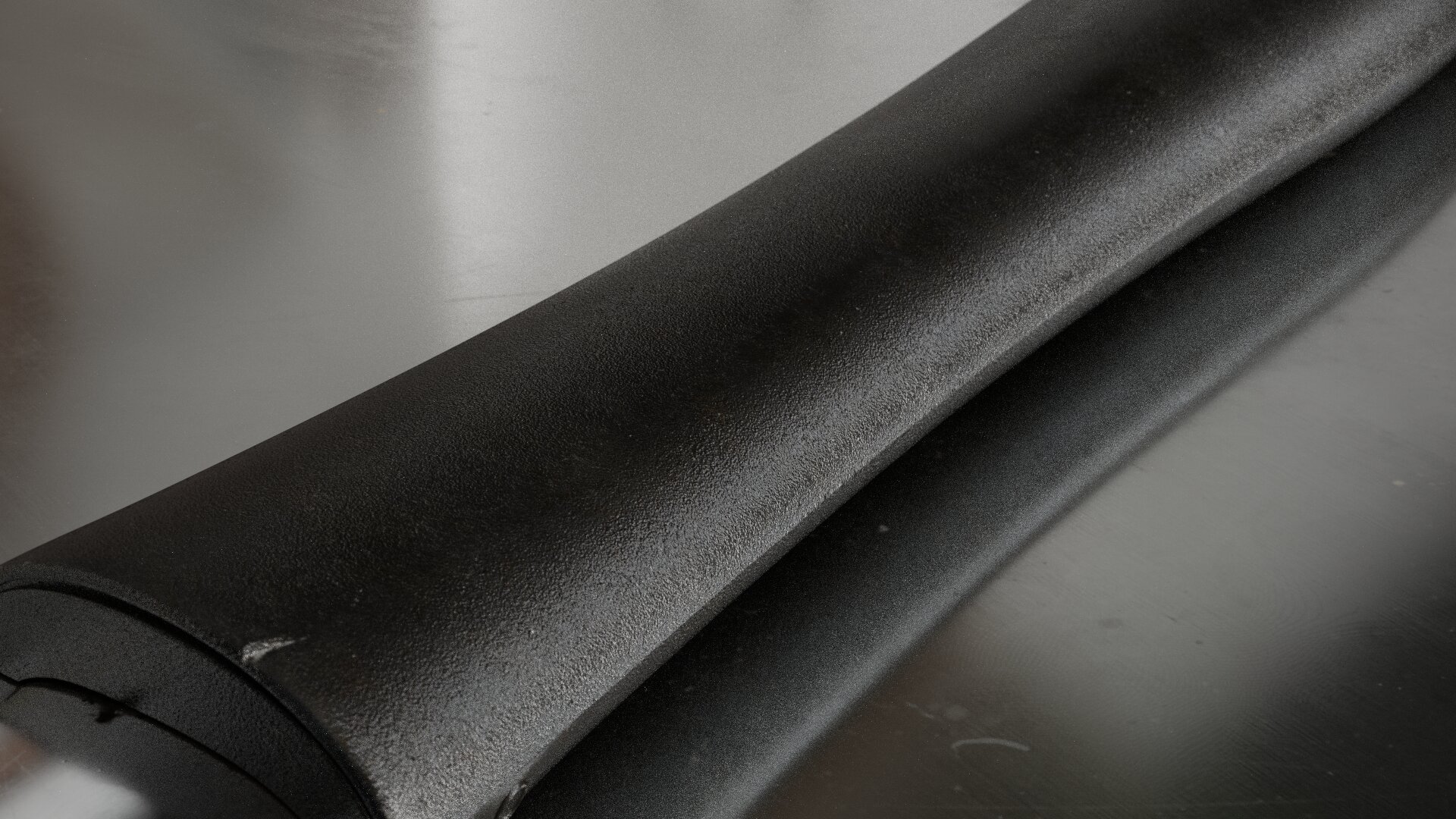}&
        \includegraphics[width=0.19\linewidth,trim=0 0 138 0,clip,valign=b]{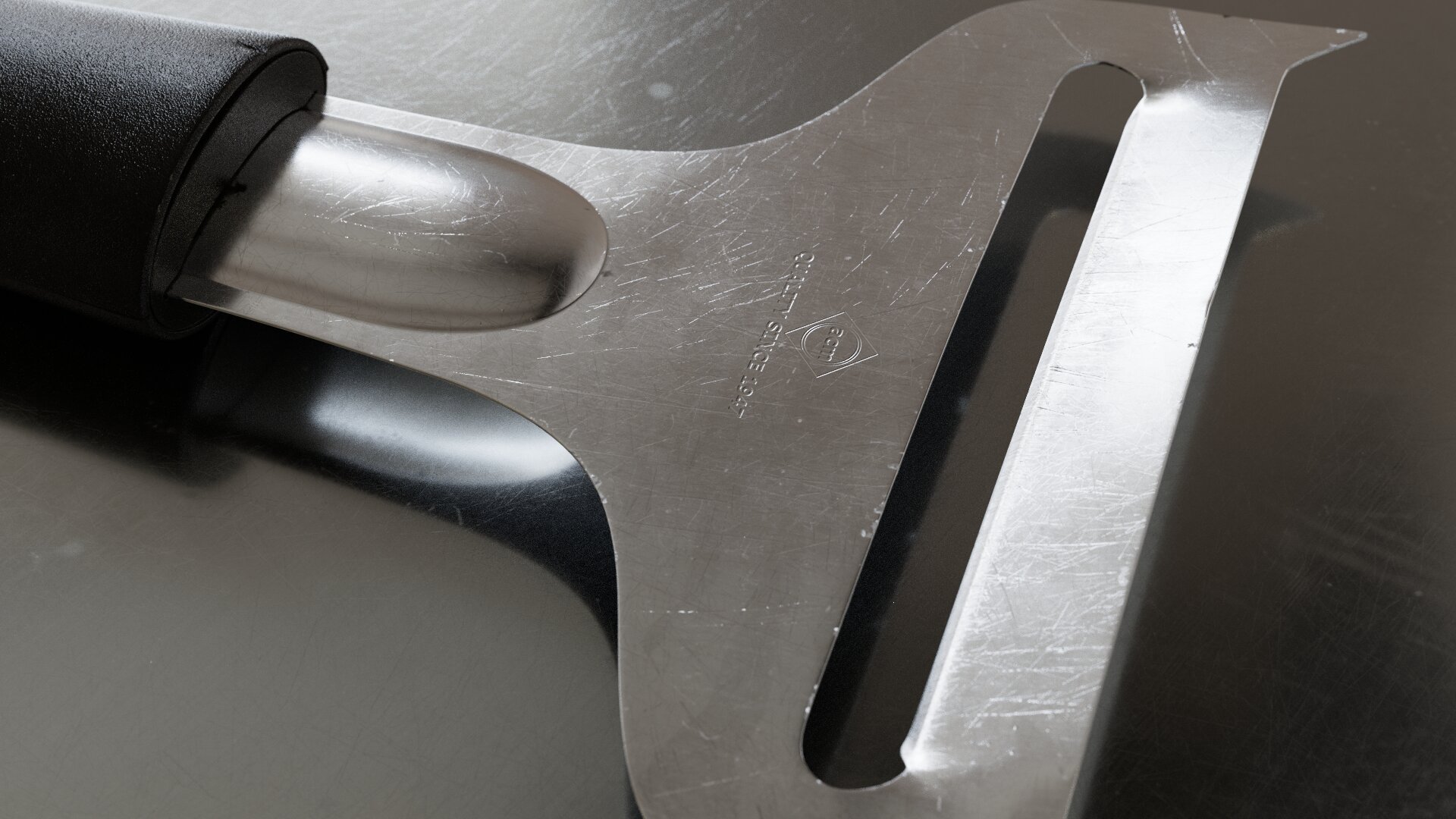}&
        \includegraphics[width=0.19\linewidth,trim=500 300 300 100,clip,valign=b]{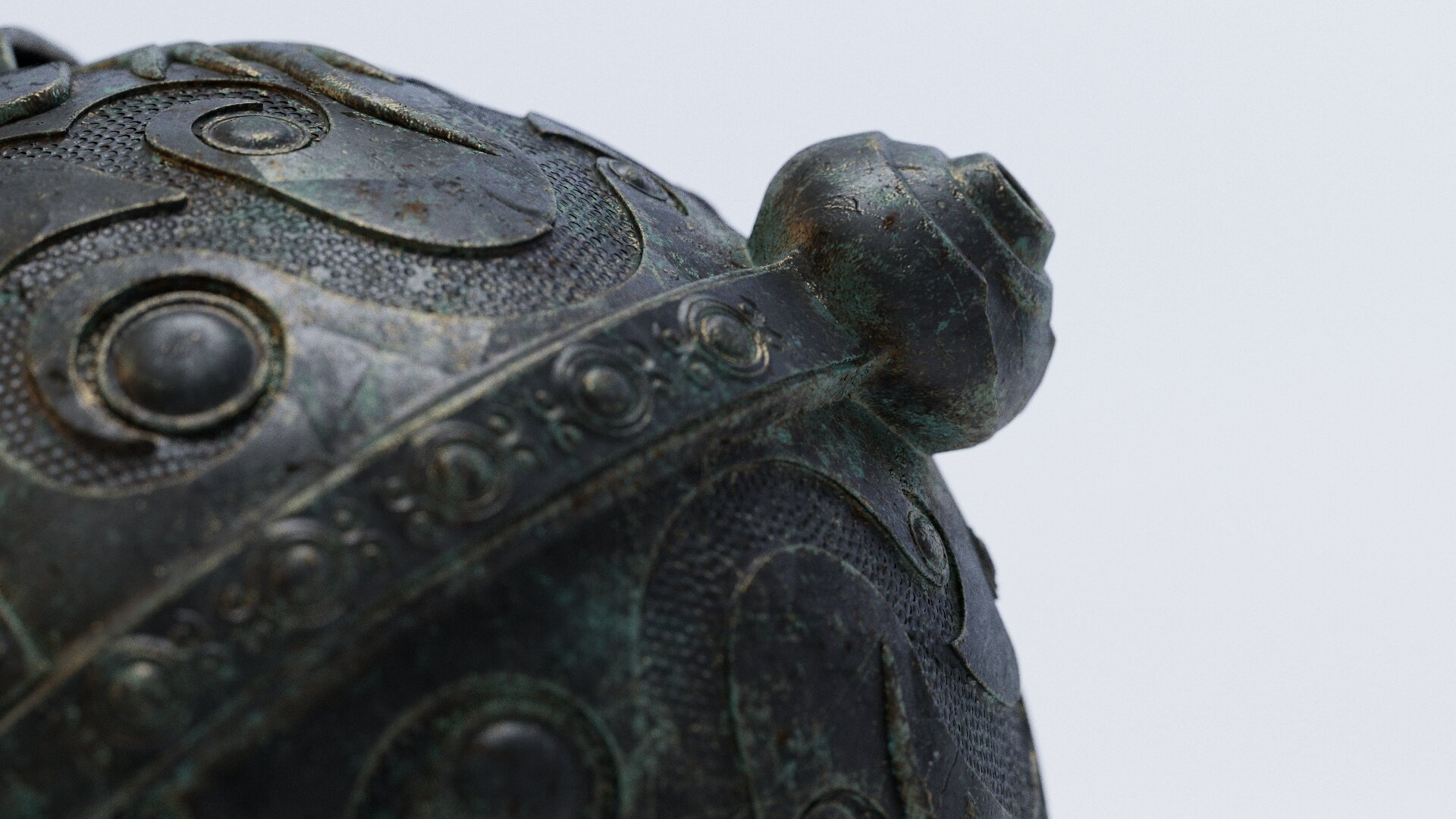}\\[4pt]
        Ceramic body & Metal handle & Plastic handle & Metal blade & Metal body\\
        \cmidrule(l){1-2}\cmidrule(l){3-4}\cmidrule(l){5-5}
        \multicolumn{2}{c}{\teapot}&
        \multicolumn{2}{c}{\cheesegrater}&
        \inkwell
    \end{tabular}
    \vspace{-2mm}
    \caption{
          We show rendered images of five reference materials created with a layering approach similar to \cite{Jakob2019} that we approximate with neural models for representing the BRDF and importance sampling. All objects are challenging for real-time renderers due to their complex reflection behavior and high resolution textures (see \autoref{tab:material-stats}). The corresponding shading graphs are provided in the supplementary material.
    }\label{fig:material-overview}
\end{figure*}

\section{Introduction}

Recent progress in rendering algorithms, light transport methods, and ray tracing hardware have pushed the limits of image quality that can be achieved in real time.
However, progress in real-time material models has noticeably lagged behind. While deeply layered materials and sophisticated shading graphs are commonplace in off-line rendering, such approaches are often far too costly to be used in real-time applications.
Aside from computational cost, sophisticated materials pose additional challenges for importance sampling and filtering: highly detailed materials will alias severely under minification, and the complex multi-lobe reflectance of layered materials causes high variance if not sampled properly.

Recent work in neural appearance modelling~\cite{Kuznetsov2022,Sztrajman2021,Zheng2021} has shown that multi-layer perceptrons (MLPs) can be an effective tool for appearance modelling, importance sampling, and filtering.
Nevertheless, these models do not support film-quality appearance and a scalable solution for high-fidelity visuals in real time has yet to be demonstrated.

\begin{figure*}[t]
    \setlength{\tabcolsep}{0.003\textwidth}%
\renewcommand{\arraystretch}{1}%
\footnotesize%
\begin{tabular}{cccc}
Numerically optimized analytical BRDF & Manually optimized analytical BRDF & Our neural BRDF & Reference \\
8-channel texture + diffuse \& specular lobes & 8-channel texture + diffuse \& specular lobes & 8-channel latent texture + MLP ($3\times 64$) & 11-ch. texture + sh. graph \\
\begin{overpic}[width=0.14157142857142857\textwidth,trim=168 0 220 20,clip]{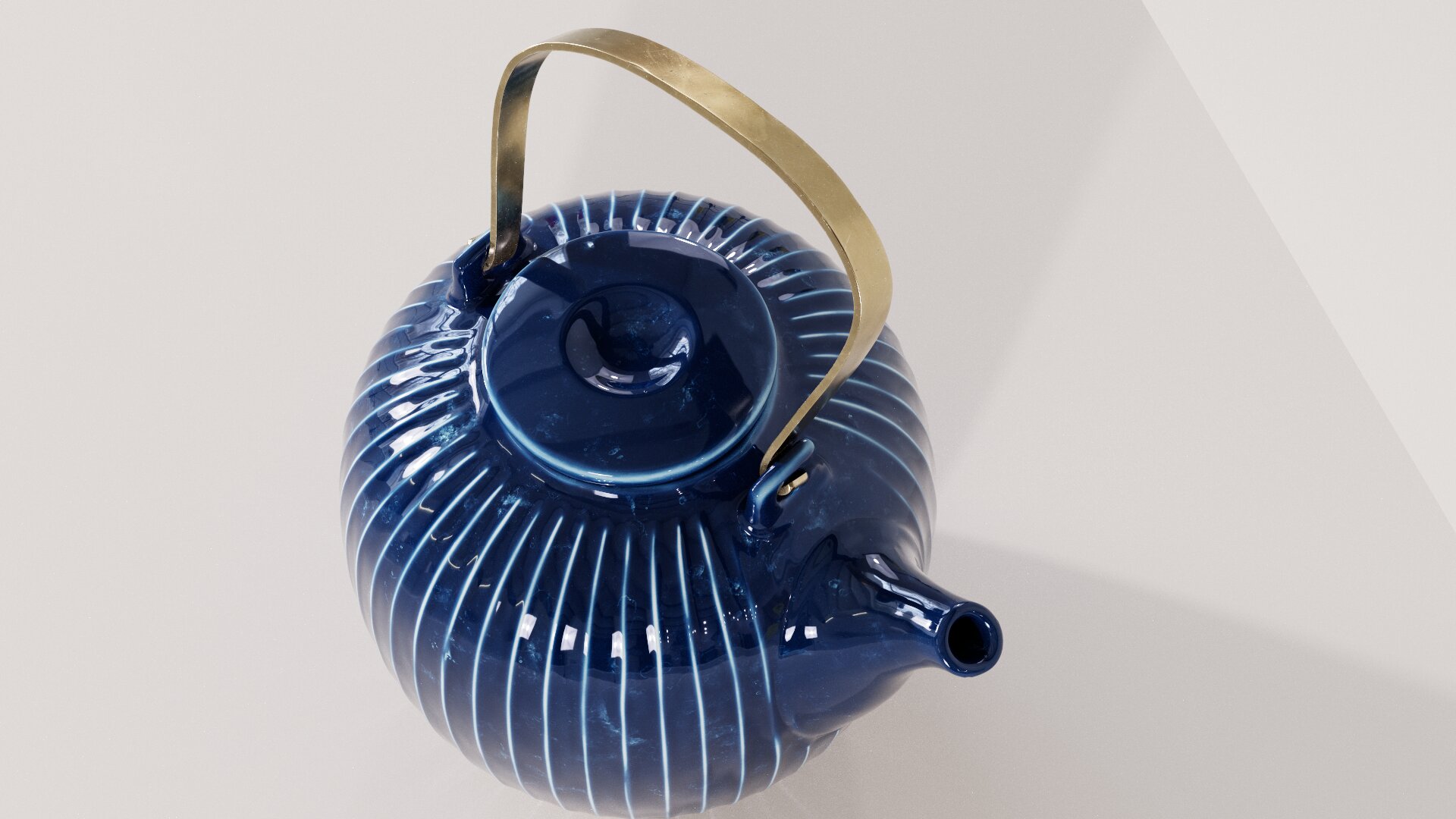}
\put(-9,0){\rotatebox{90}{\hspace{11mm}View 1}}
\end{overpic}%
\begin{overpic}[width=0.14157142857142857\textwidth,trim=168 0 220 20,clip]{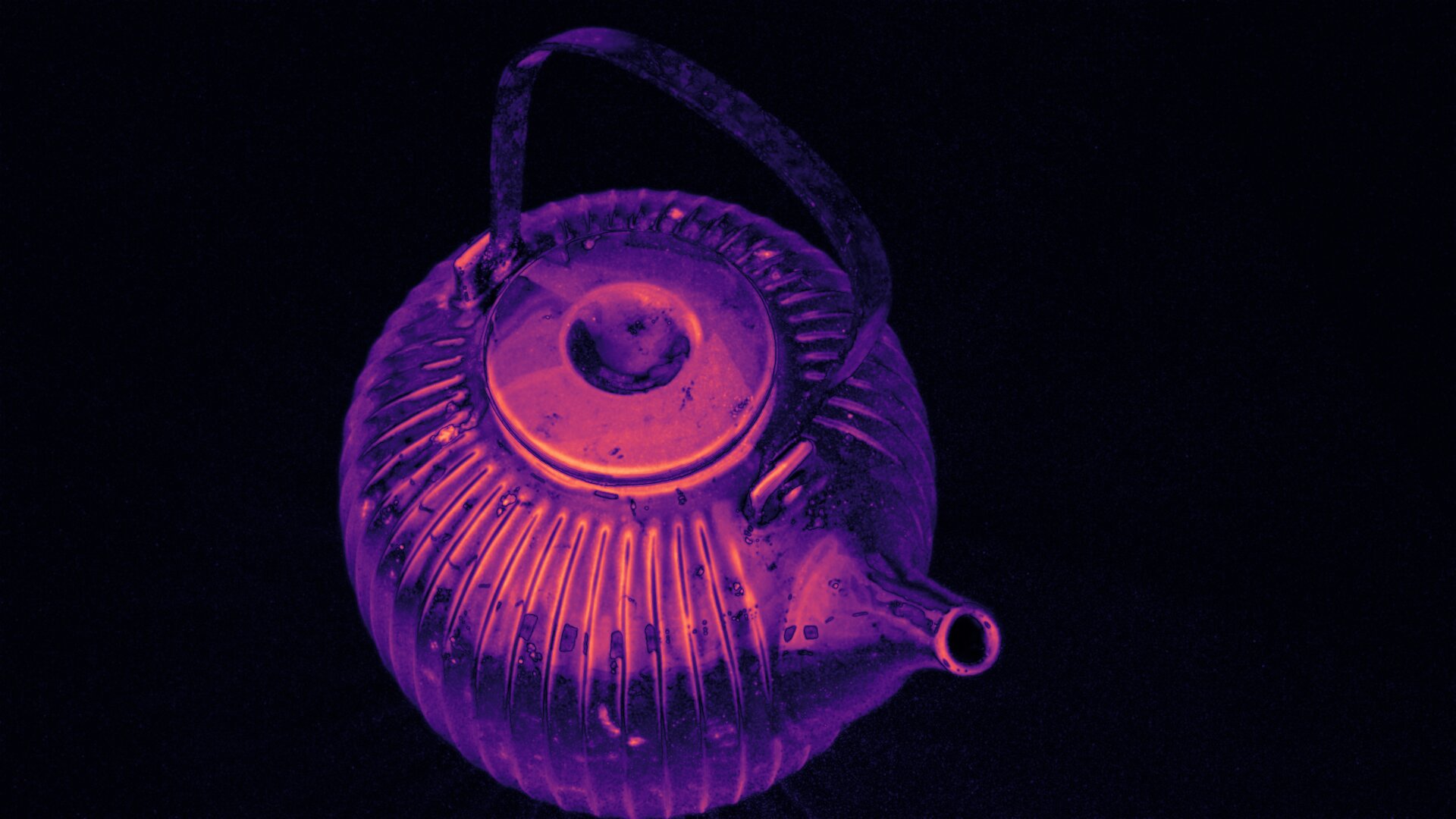}
\end{overpic}%
&\begin{overpic}[width=0.14157142857142857\textwidth,trim=168 0 220 20,clip]{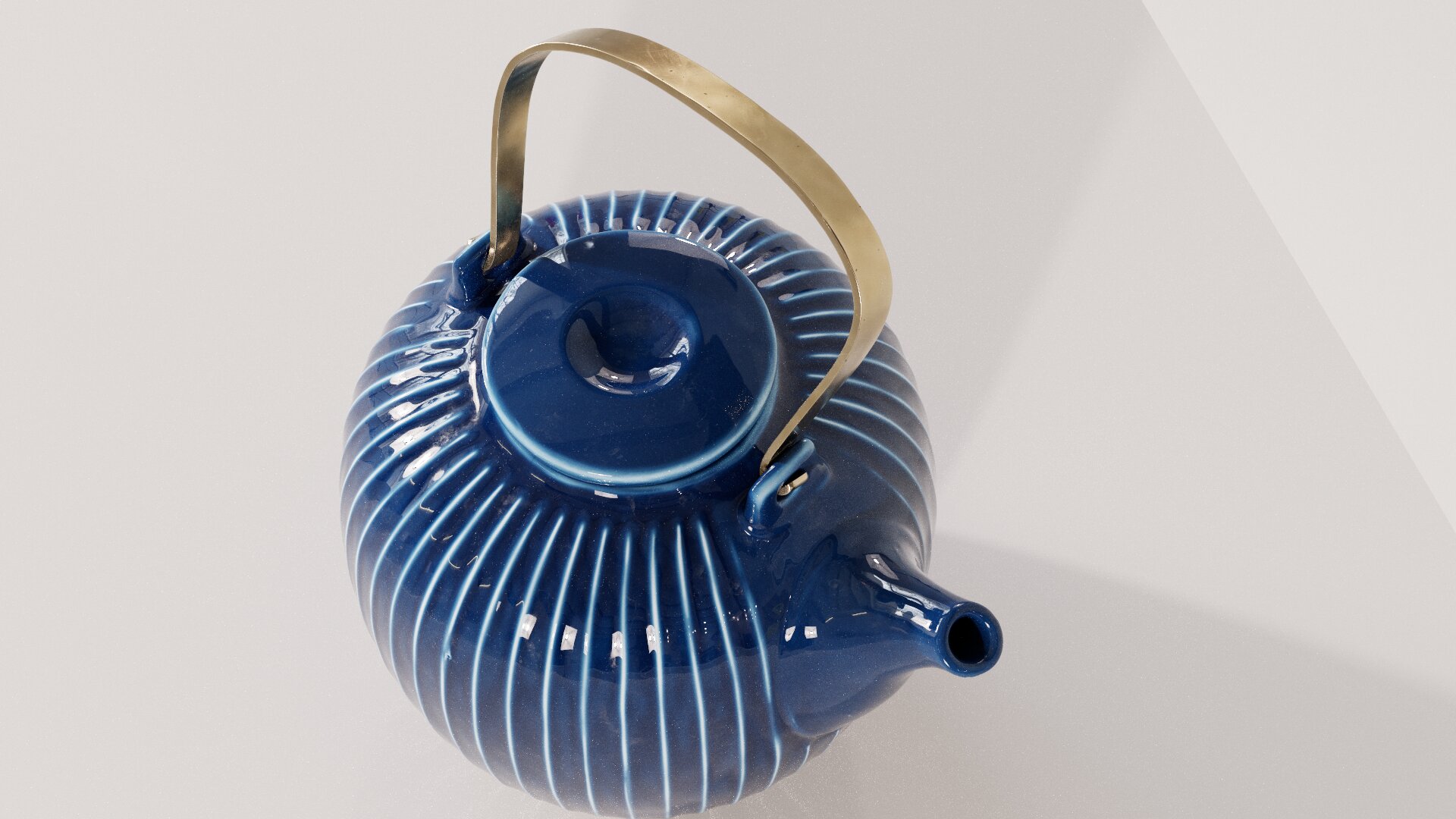}
\end{overpic}%
\begin{overpic}[width=0.14157142857142857\textwidth,trim=168 0 220 20,clip]{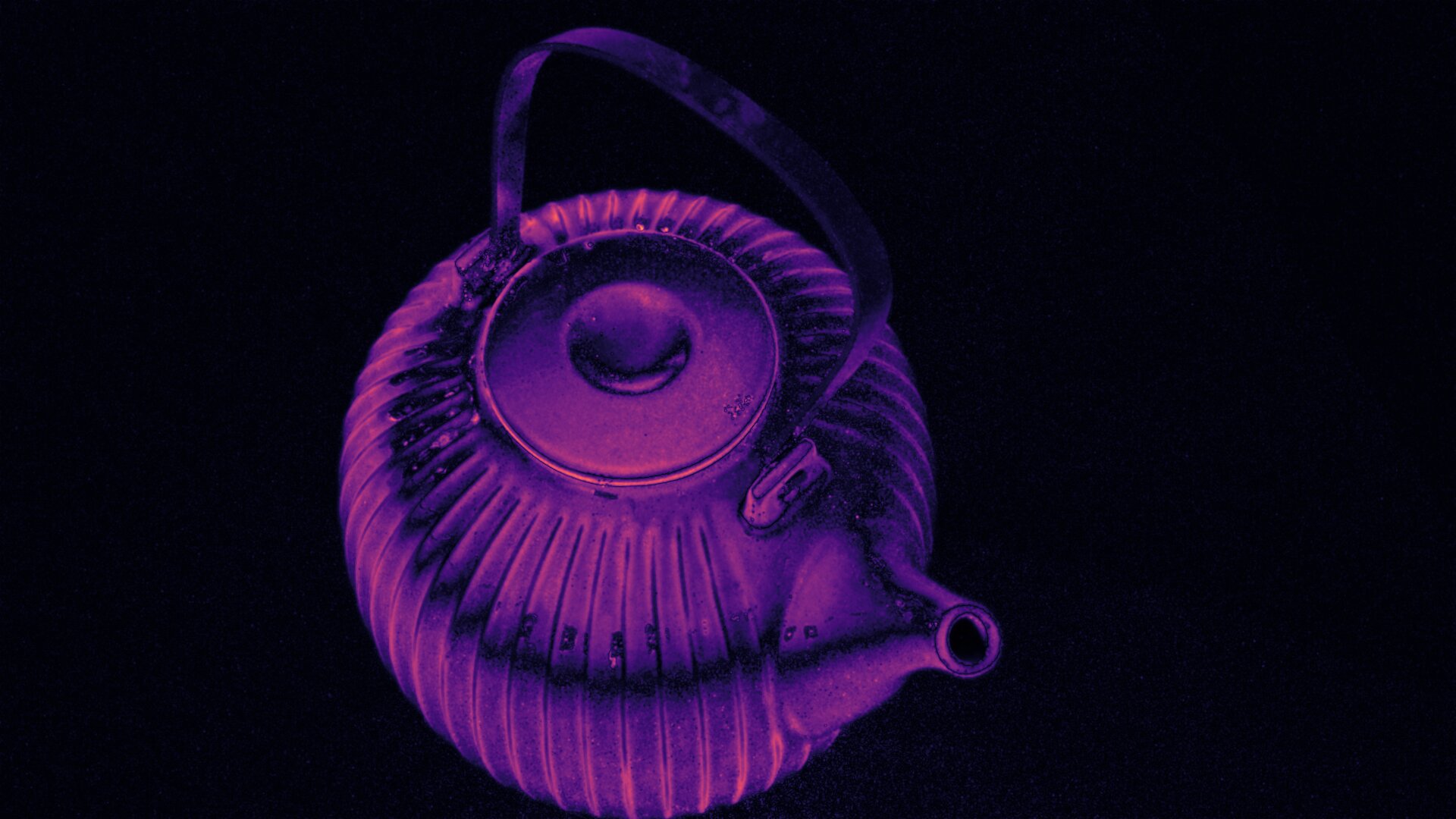}
\end{overpic}%
&\begin{overpic}[width=0.14157142857142857\textwidth,trim=168 0 220 20,clip]{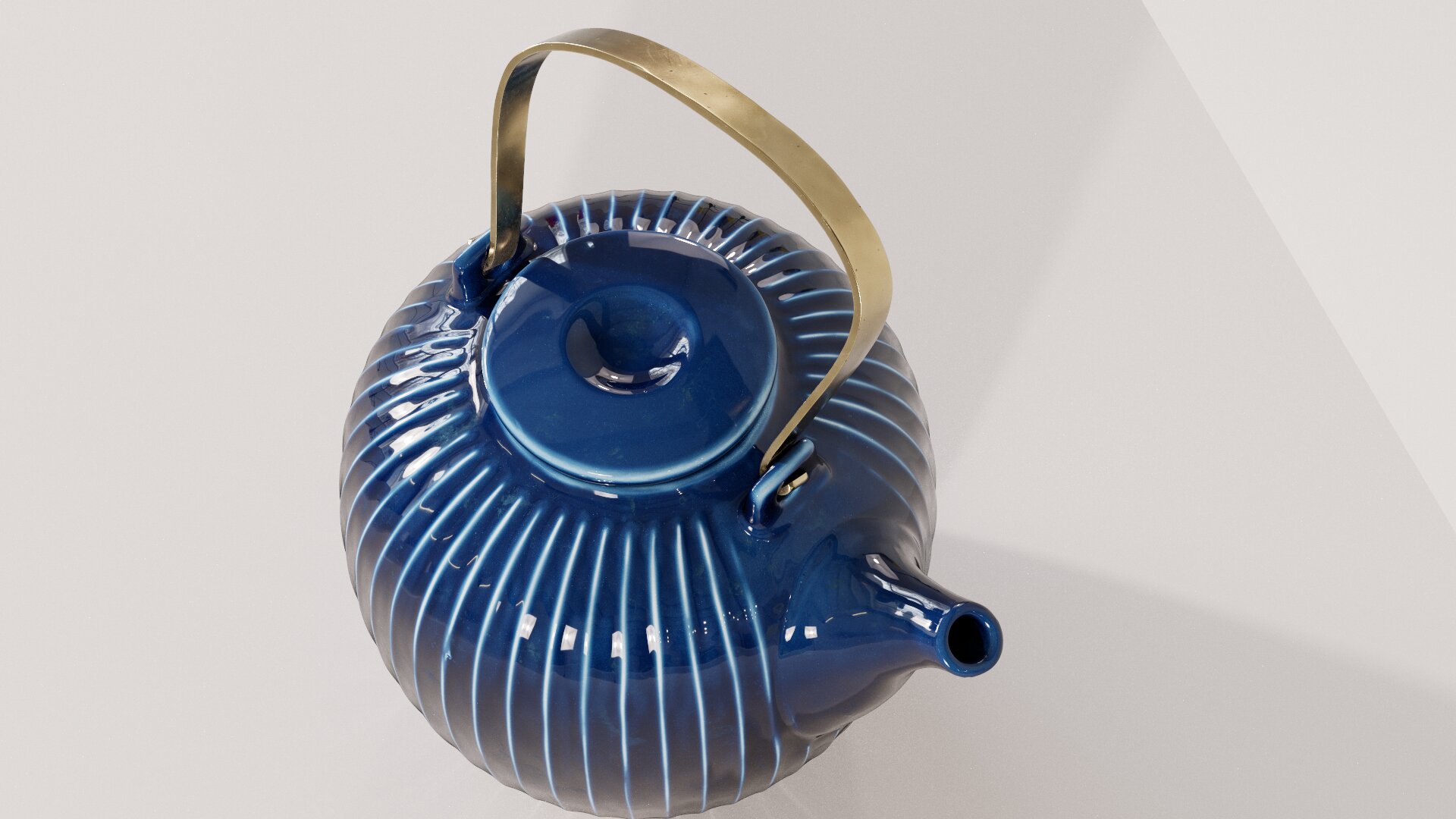}
\end{overpic}%
\begin{overpic}[width=0.14157142857142857\textwidth,trim=168 0 220 20,clip]{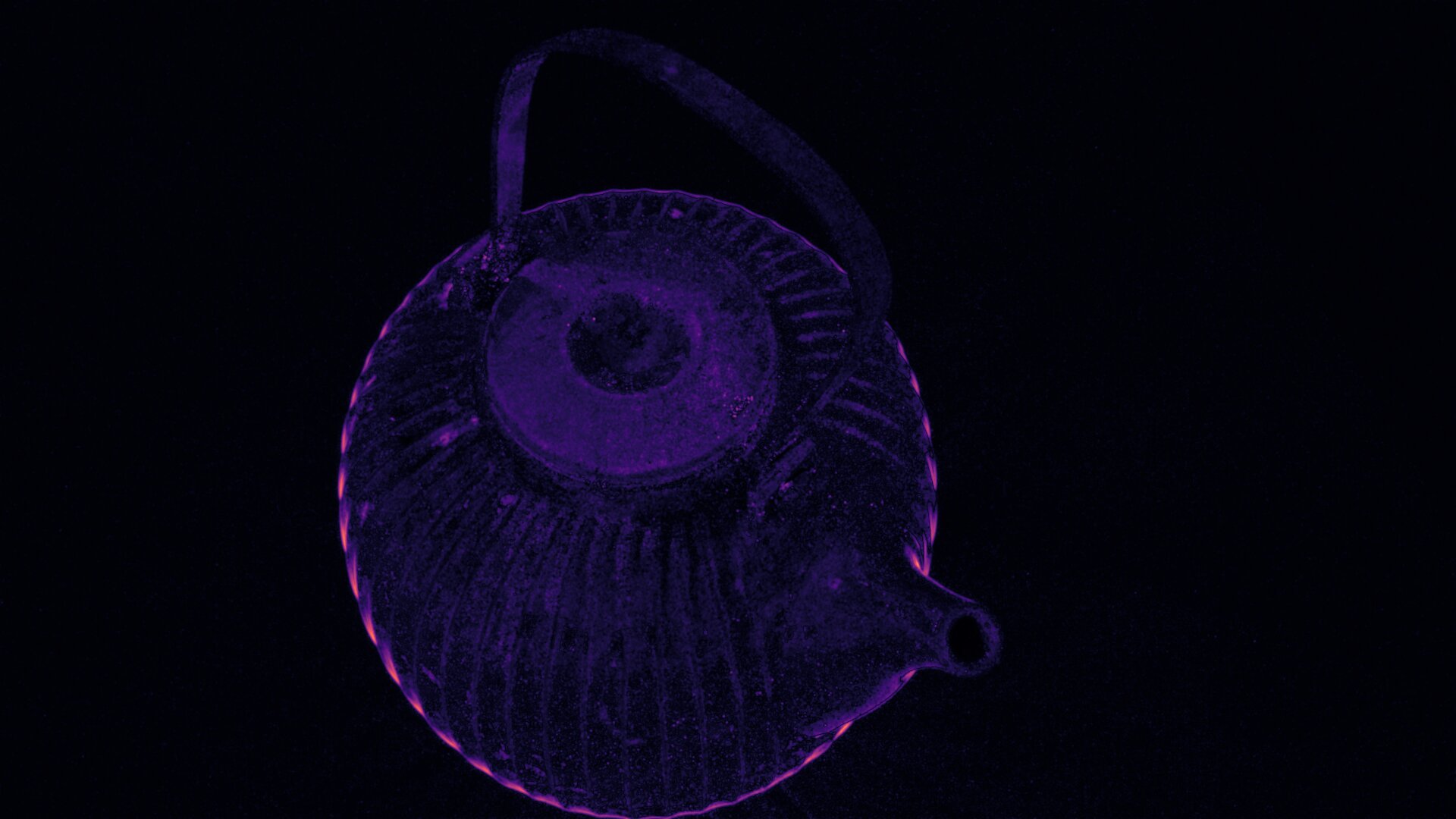}
\end{overpic}%
&\begin{overpic}[width=0.14157142857142857\textwidth,trim=168 0 220 20,clip]{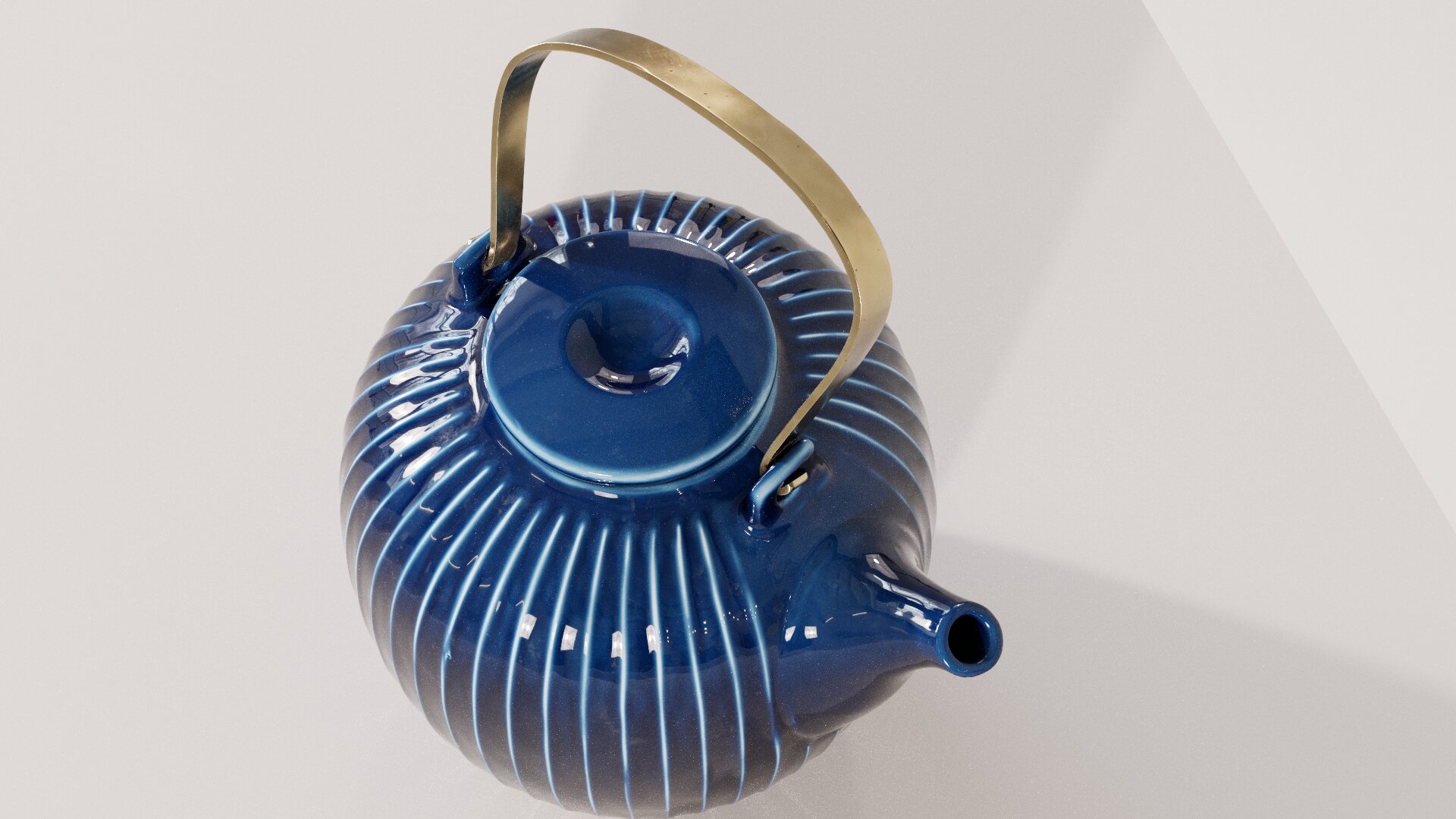}
\end{overpic}%
\\
\begin{overpic}[width=0.14157142857142857\textwidth,trim=30 20 358 0,clip]{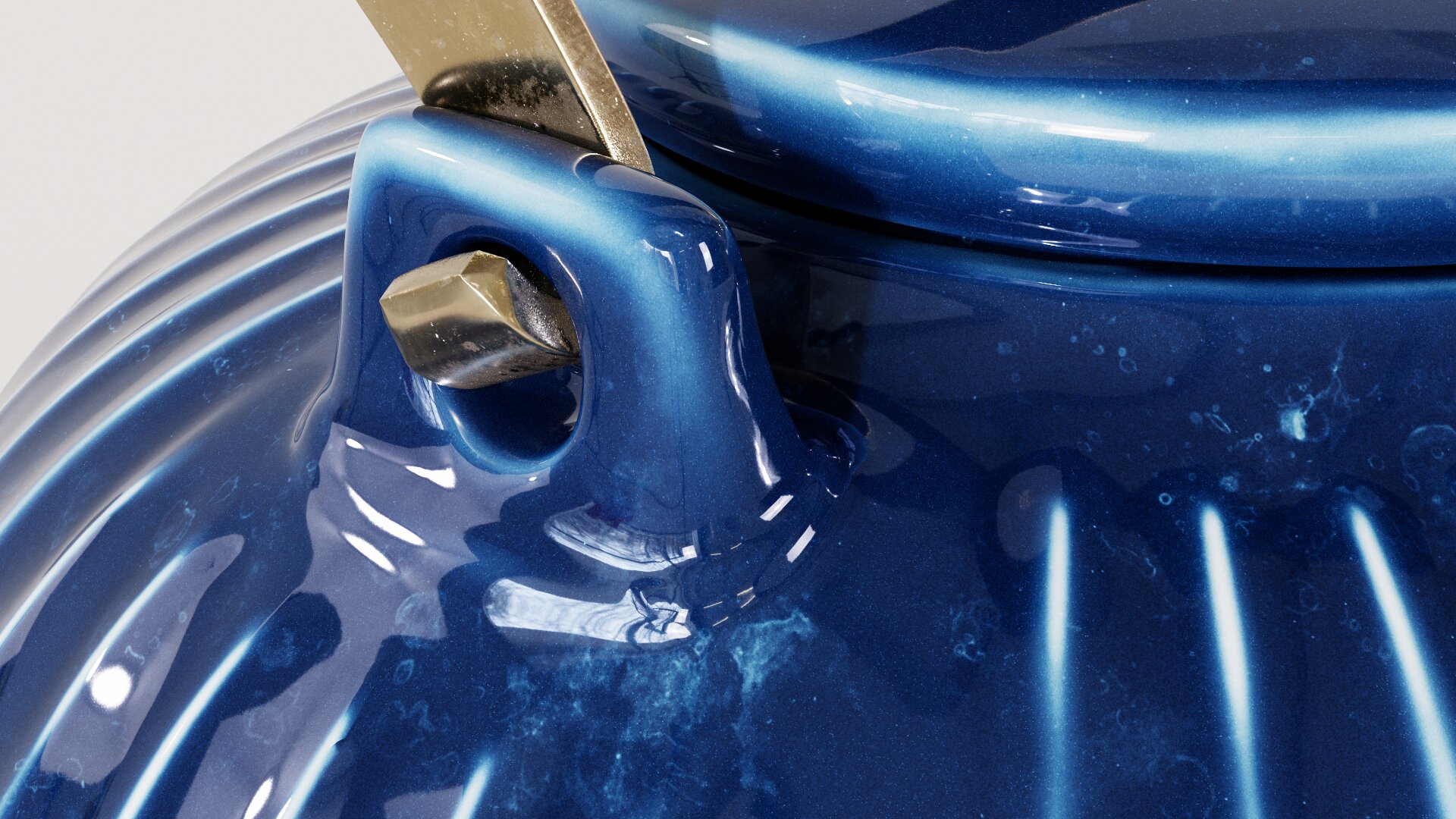}
\put(-9,0){\rotatebox{90}{\hspace{11mm}View 2}}
\end{overpic}%
\begin{overpic}[width=0.14157142857142857\textwidth,trim=30 20 358 0,clip]{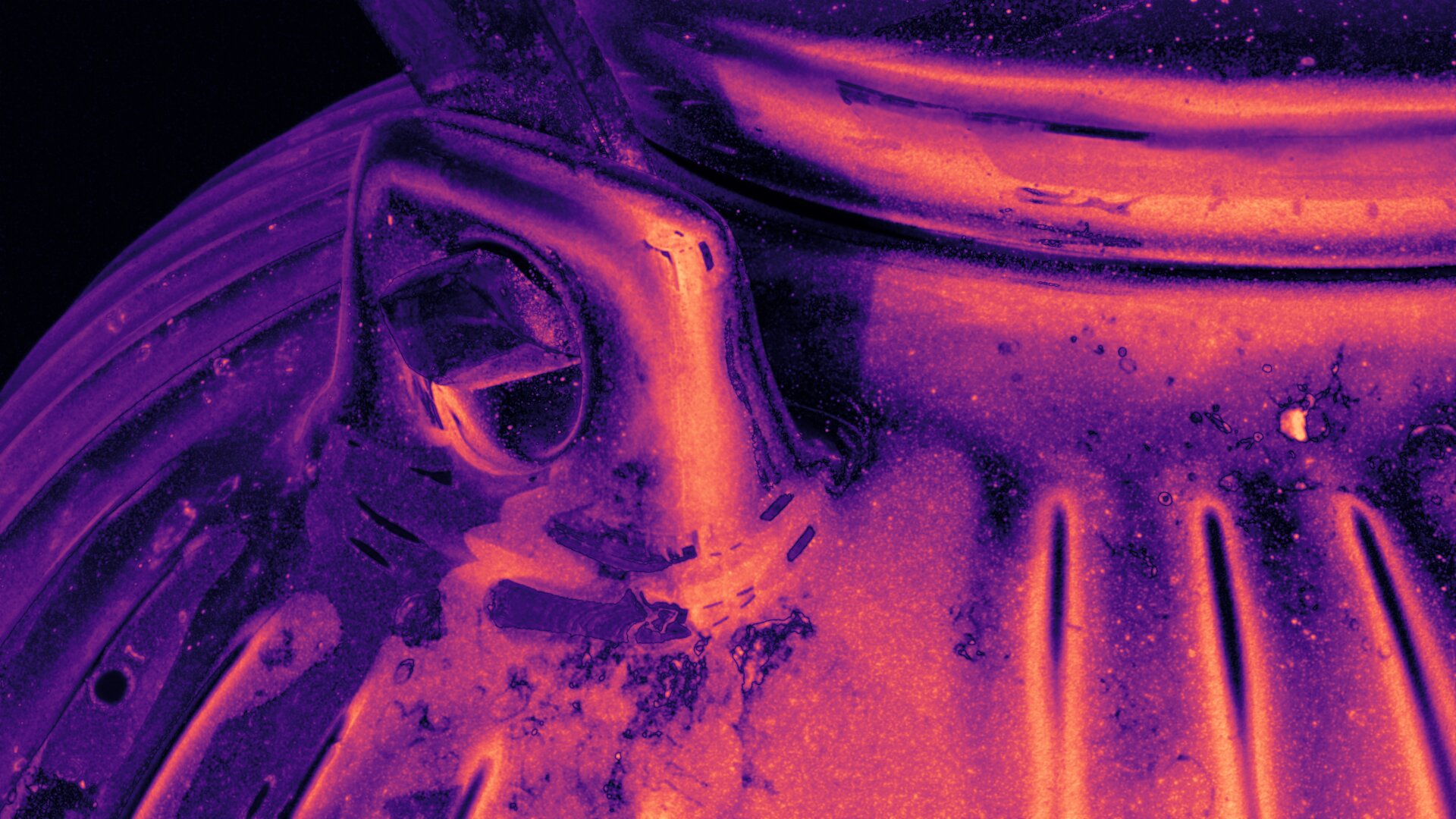}
\end{overpic}%
&\begin{overpic}[width=0.14157142857142857\textwidth,trim=30 20 358 0,clip]{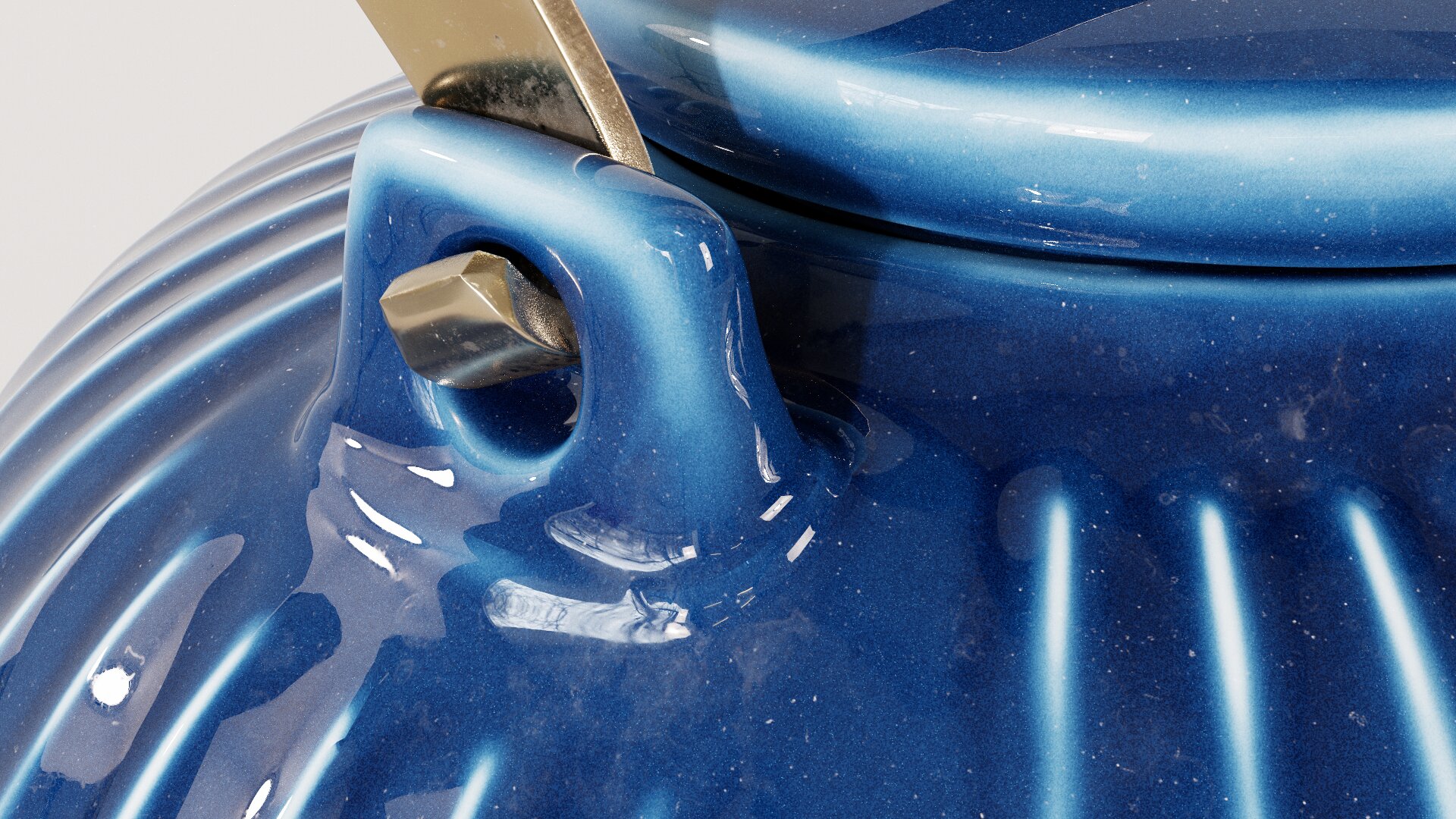}
\end{overpic}%
\begin{overpic}[width=0.14157142857142857\textwidth,trim=30 20 358 0,clip]{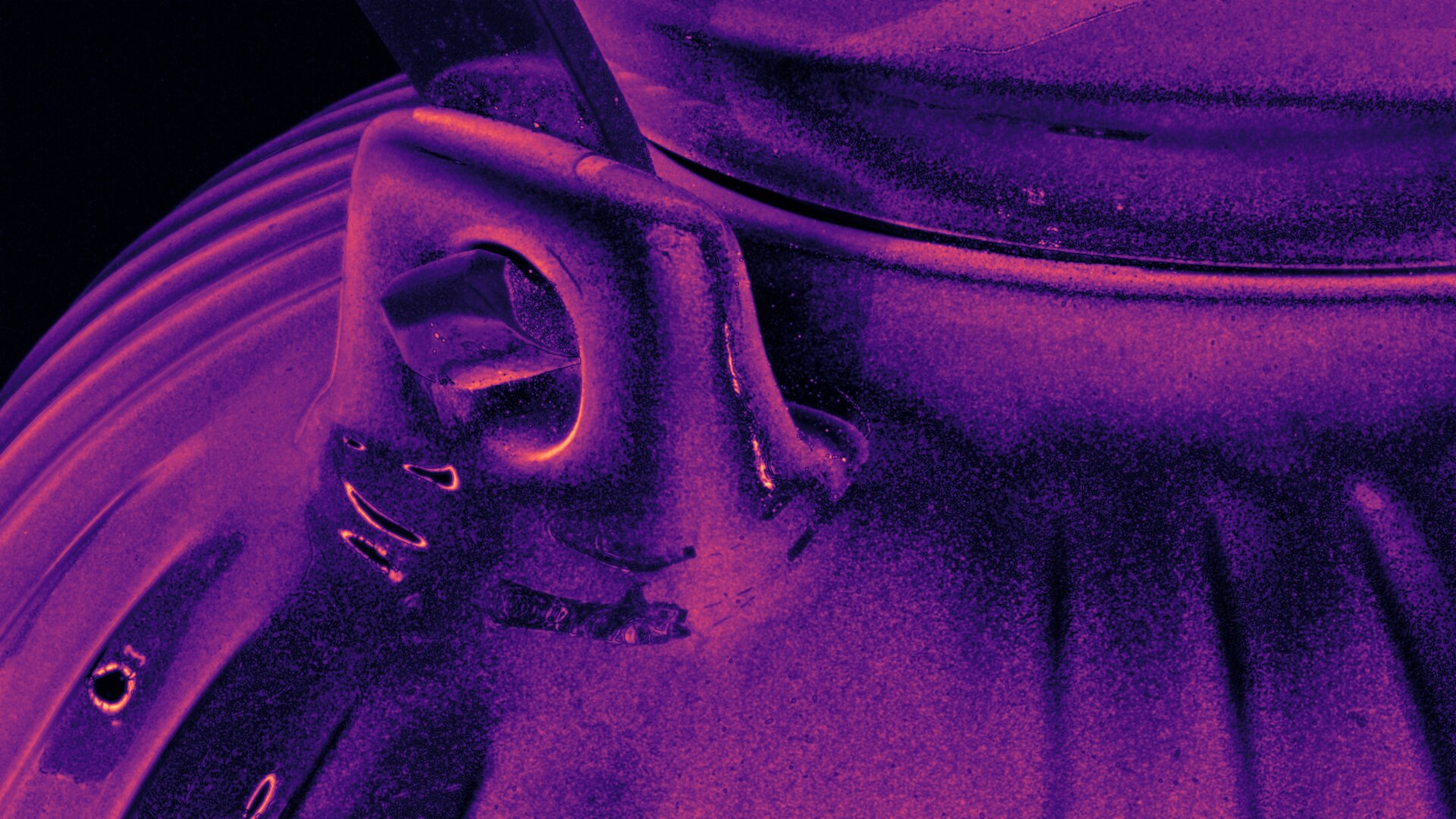}
\end{overpic}%
&\begin{overpic}[width=0.14157142857142857\textwidth,trim=30 20 358 0,clip]{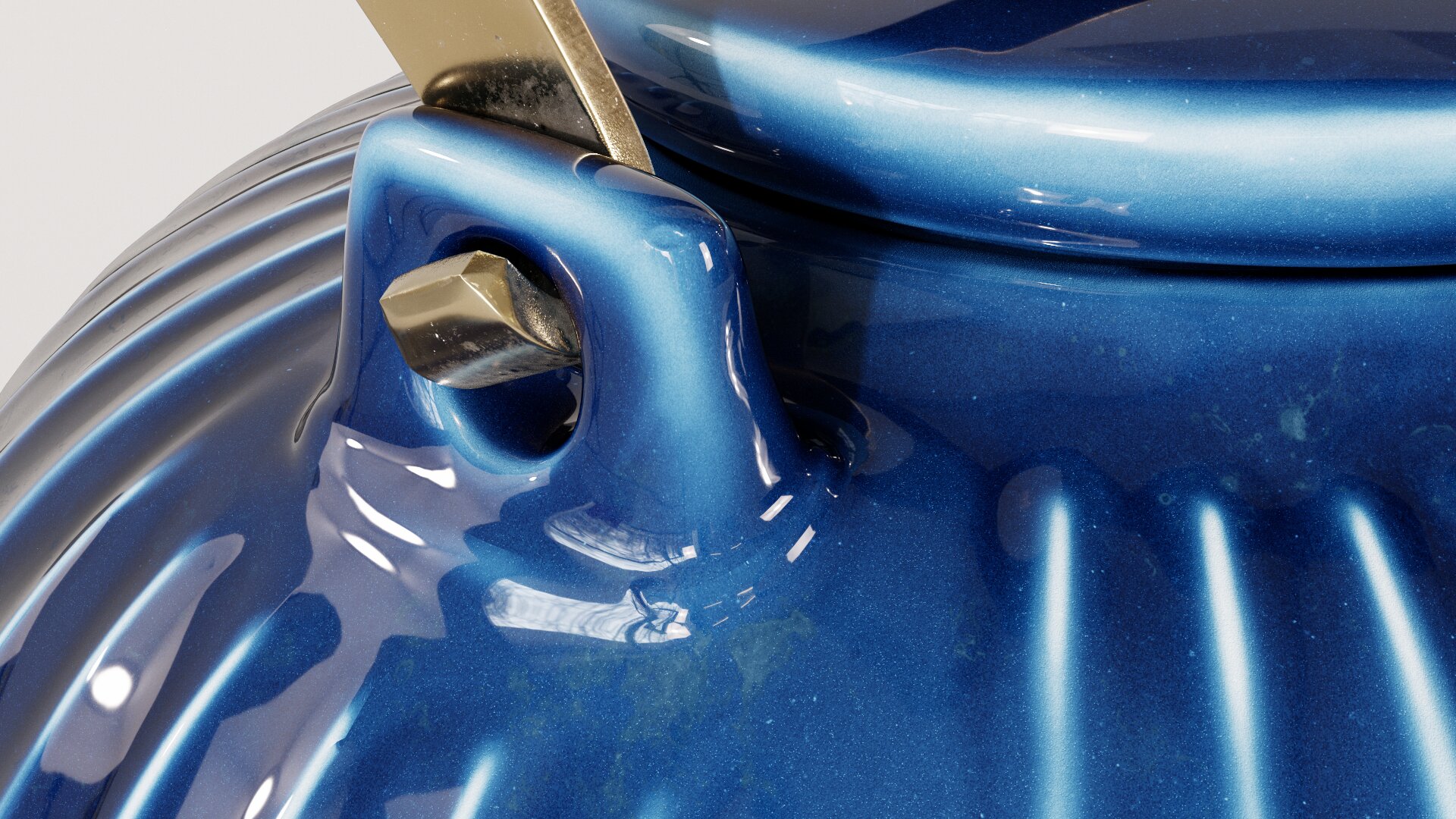}
\end{overpic}%
\begin{overpic}[width=0.14157142857142857\textwidth,trim=30 20 358 0,clip]{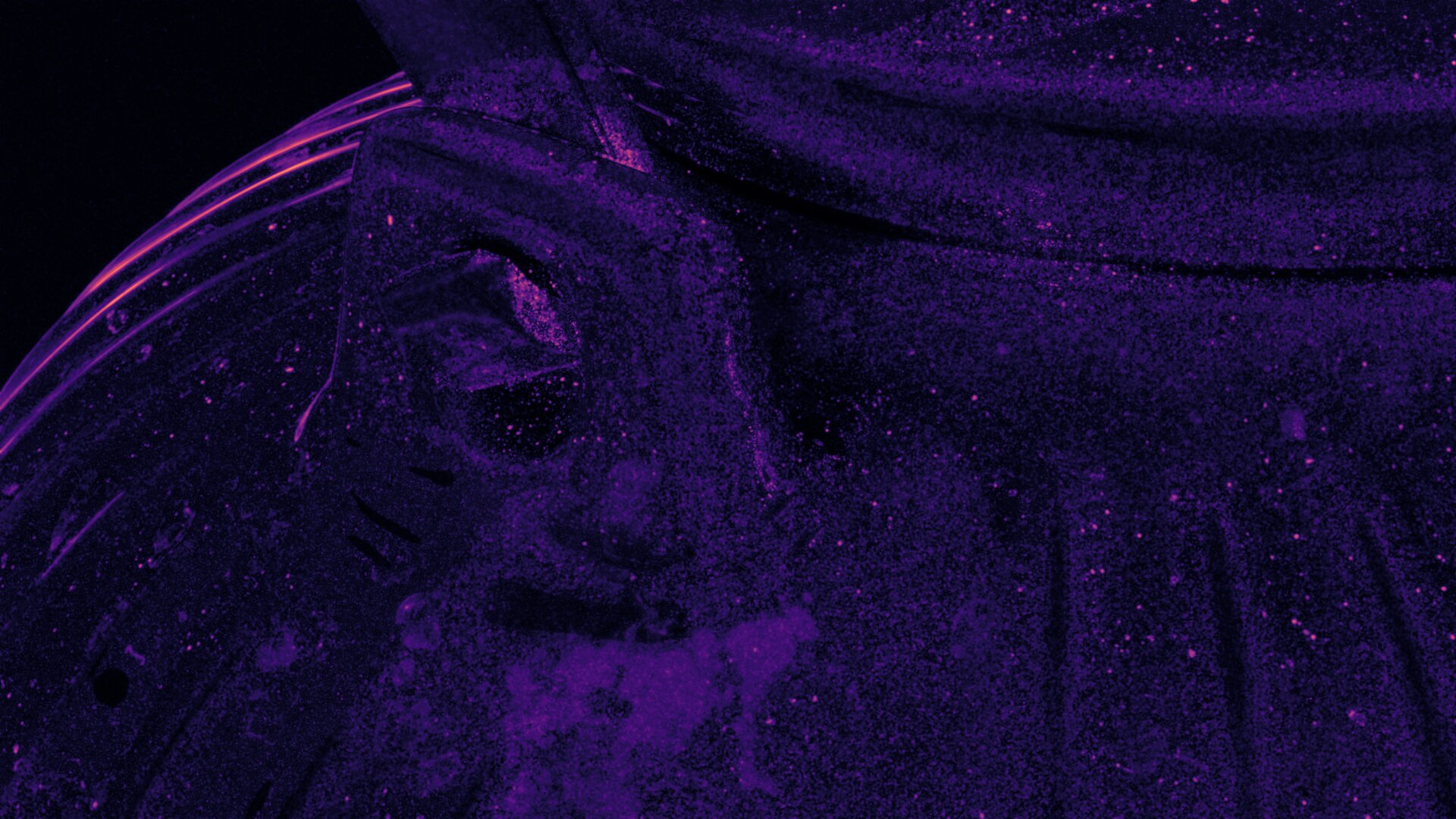}
\end{overpic}%
&\begin{overpic}[width=0.14157142857142857\textwidth,trim=30 20 358 0,clip]{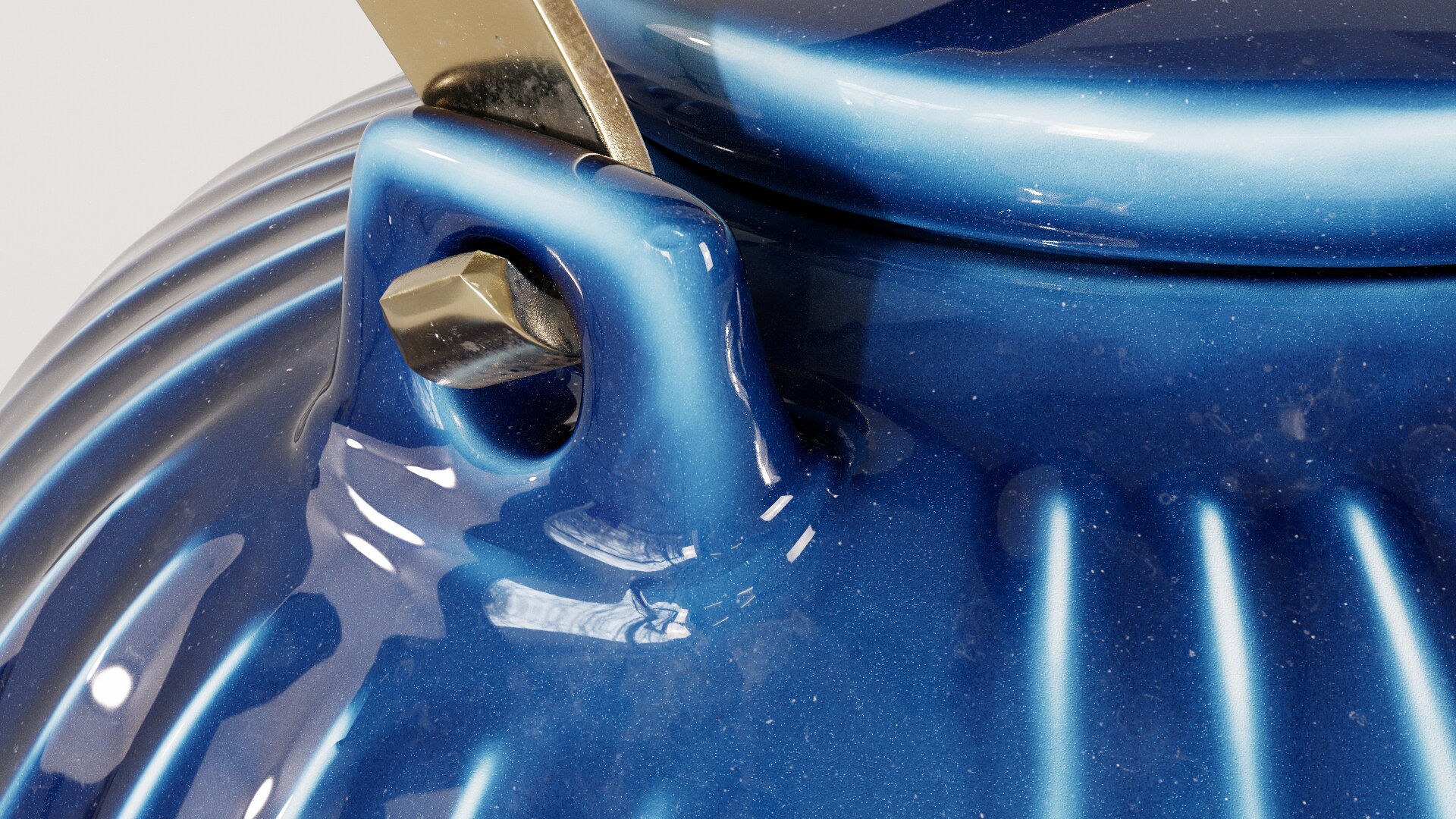}
\end{overpic}%
\\
\end{tabular}

    \caption{
        \revision{First two columns: approximations of the multi-layer \teapot{} materials from \autoref{fig:material-overview} using a simple analytical BRDF, parameterized by only 8 spatially-varying input channels: base color (3), specular roughness (1), specular normal map (2), specularity (1), and metallness (1). Third column: our neural BRDF parameterized by an 8-channel latent texture. \FLIP visualizations emphasize the perceptual differences against the reference (last column, \autoref{fig:material-overview}, \autoref{tab:material-stats}). The parameters for the analytic BRDF are either numerically optimized or tuned manually. In both cases, we see a much larger approximation error as it lacks the expressive power to capture the complexity of the reference, e.g. the view-dependent blue color of the ceramic glazing.}
    }\label{fig:8d-approximation}
\end{figure*}

In this paper, we set our goal accordingly: to render film-quality materials, such as those used in the VFX industry exemplified in \autoref{fig:material-overview} with statistics in \autoref{tab:material-stats}, in real time. These materials prioritize realism and visual fidelity, relying on very high-resolution textures. Layering of reflectance components, rather than an uber-shader, is used to generate material appearance yielding arbitrary BRDF combinations with tens of parameters. \revision{Approximating such materials with simple analytical models is inaccurate (see \autoref{fig:8d-approximation}) and porting to real-time applications is therefore challenging.}

In order to render film-quality appearance in real time we i)~carefully cherry-pick components from prior works, ii)~introduce algorithmic innovations, and iii)~develop a scalable solution for inlining neural networks in the innermost rendering loop, both for classical rasterization and path tracing.
We choose to forgo editability in favor of performance, effectively ``baking'' the reference material into a neural texture interpreted by neural networks. Our model can thus be viewed as an optimized representation for fast rendering, which is baked (via optimization) after editing has taken place.

Our model consists of an \emph{encoder} and two \emph{decoders}, with the neural (latent) texture in between. 
The encoder maps BRDF parameters to a latent space, thereby converting a set of traditional textures (per-layer albedo, normal map, etc.) into a single multi-channel latent texture. 
Using the encoder is key to support materials with high-resolution textures.
The latent texture is decoded using two networks: an evaluation network that infers the BRDF value for a given pair of directions, and a sampling network that maps random numbers to sampled (outgoing) directions.

Our main algorithmic contributions can be characterized as embedding fixed-function elements---graphics priors---in the two neural decoders.
First, we insert a standard rotation operation between trainable components of the BRDF decoder to handle normal mapped surfaces. Second, we utilize a network-driven microfacet distribution for importance sampling.
These priors are necessary to efficiently utilize the (limited) expressive power of small networks.

\begin{table}[t]
    \centering
    \small
    \caption{
        \revision{Statistics of our reference materials from \autoref{fig:material-overview}. The shading graph with shading nodes~(a) is programmatically converted to a number of BRDF layers~(b) controlled by parameters~(c), which are varied spatially using RGB textures (with the total number of used channels in parenthesis)~(d); the total number of RGB megatexels is reported in column~(e).}
    }\label{tab:material-stats}
    \begin{tabular}{lrrrrr}
        \toprule
        & \hspace{-2mm}Nodes & Layers & \revision{Parameters} & Textures & MTexels \\
        & (a)                & (b)    & (c)                   & (d)      & (e)     \\
        \midrule
        \teapot{} ceramic & 37 & 5 & \revision{121} & \revision{5 (11)} & 1174 \\
        \teapot{} handle  & 41 & 2 & \revision{91} & 11 \revision{(19)} & 152 \\
        \textsc{Slicer} handle & 20 & 5 & \revision{43} & 3 \revision{(7)} & 201 \\
        \textsc{Slicer} blade & 54 & 3 & \revision{114} & 16 \revision{(40)} & 324 \\
        \inkwell & 49 & 5 & \revision{143} & 4 \revision{(11)} & 201 \\
        \bottomrule
    \end{tabular}
\end{table}

On the system level, we present an efficient method for inlining fully fused neural networks
in rendering code. To the best of our knowledge, this is the first complete and scalable system for running
neural material shaders inside real-time shading languages.
A key contribution is an execution model that utilizes tensor operations whenever possible and efficiently handles divergent code paths.
This allows fast inferencing in any shader stage including ray tracing and fragment shaders, which is important for adoption in game engines and interactive applications.

Our neural model has a fixed evaluation cost, independent of the material complexity, allowing us to render complex materials in a real-time path tracer.
To that end, we authored highly detailed assets with layered materials (\autoref{fig:material-overview}) that provide visual detail down to a 10 cm viewing distance.
We can reproduce the visual fidelity of such complex assets, with shading being up to 10$\times$ faster than the original, moderately optimized shading models, while also providing additional sampling and filtering facilities (\autoref{fig:teaser}).

Achieving the desired visual fidelity at real-time rates required innovations both in the neural model and at the system level:
\begin{itemize}
	\setlength{\itemindent}{-3mm}
	\item a complete and scalable system for film-quality neural materials,
	\item tractable training for gigatexel-sized assets using an encoder,
	\item decoders with priors for normal mapping and sampling, and
	\item efficient execution of neural networks in real-time shaders.
\end{itemize}
We believe the joint evolution of models and systems to be crucial to bringing neural shaders to real-time, and we built our system to serve as a solid foundation in this regard.

\section{Related Work}

In this section, we review previous work related to neural material representation, filtering, and sampling,
and refer to \citet{Pharr2016} for a detailed overview of classical material models.

\subsection{Neural appearance modeling}
\label{sec:neural_appearance_modeling}

We focus on representing existing materials neurally and rendering them in real time on classical geometry.
We therefore do not utilize ray marched neural fields~\cite{Mildenhall2020,Baatz2022,Mueller2022}, although these could present a viable alternative in the future.
Our goals generally align with prior work on neural BRDFs~\cite{Zheng2021,Sztrajman2021,Fan2022,Rainer2019,Rainer2020,Kuznetsov2019,Kuznetsov2021}. Common to these methods is a conditioning of a neural network on a pair of directions, and optionally a trained latent code. Latent codes are typically stored in a texture~\cite{Thies2019} and sampled using classical UV mapping to support spatially varying BRDFs.

However, we differ from prior work on a number of key axes:

\paragraph{Obtaining latent textures.} 
\citet{Kuznetsov2019} in their NeuMIP work employ \emph{direct optimization}, updating a randomly-initialized latent texture via backpropagation---a simple but costly solution for large textures with millions of texels.
In contrast, \citet{Rainer2019} rely on an auto-encoder architecture to \emph{encode} a set of reflectance measurements into latent codes.
We pursue a hybrid approach: we first train an encoder and, partway through training, we use it to create a hierarchical latent texture, which we then \emph{finetune} through direct optimization.
This approach combines the speed of the encoder-decoder architecture with the flexibility of direct optimization.
Contrary to \citet{Rainer2019}, we do not encode the reflectance measurements, but the set of corresponding material parameters (albedo, roughness, normal, etc.).

\paragraph{Encodings and priors.} Both \citet{Zheng2021} and \citet{Sztrajman2021} reparametrize input directions into a half-angle coordinate system~\cite{Rusinkiewicz1998}.
While this specific encoding did not provide much benefit in our case, we leverage the principle and incorporate a novel graphics prior---rotation to learned shading frames---to better handle normal-mapped, layered materials.

\begin{figure*}[t]
    \begin{overpic}[width=0.98\linewidth]{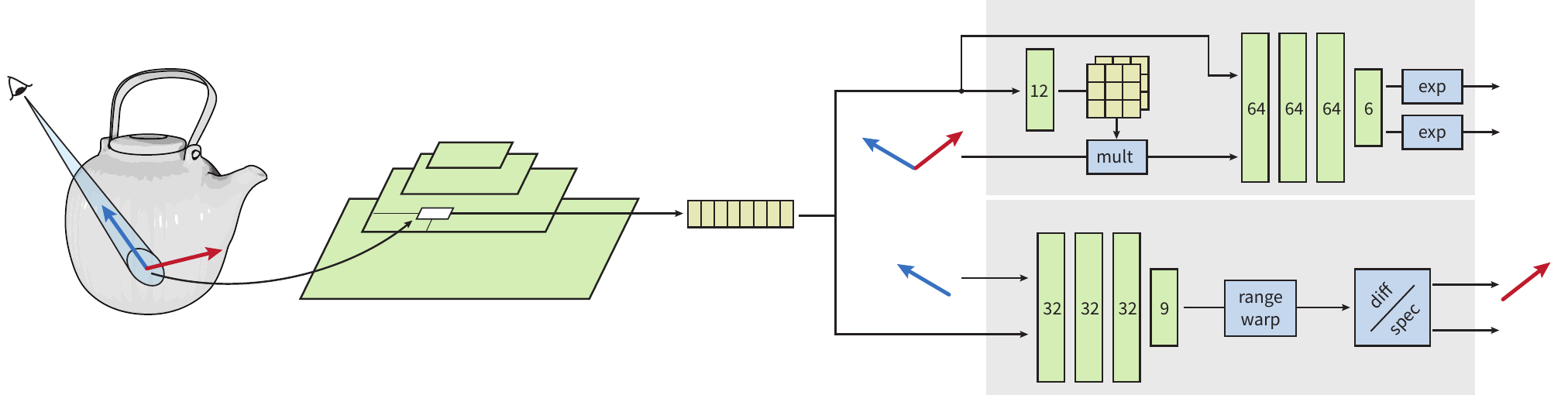}
        \footnotesize
        \put(7.5,22){Geometry}
        \put(23,8){$(u,v,l)$}
        \put(26,17){Latent texture $\latents$}
        \put(43,13.5){Latent code $\latents(\x)$}
        \put(52,21){\begin{minipage}{2cm}\centering BRDF\\evaluation\end{minipage}}
        \put(52,1.5){\begin{minipage}{2cm}\centering Importance\\sampling\end{minipage}}
        \put(73.7,18.2){\begin{minipage}{2cm}Shading\\frames\end{minipage}}
        \put(55.2,14){$\wi$}
        \put(59.8,14){$\wo$}
        \put(57,6){$\wi$}
        \put(97.8,6){$\wo$}
        \put(65,23.7){Frame extraction}
        \put(79,23.7){Decoding MLP}
        \put(66.5,11){Decoding MLP}
        \put(83,9.5){\begin{minipage}{2cm}\centering Analytic\\sampler\end{minipage}}
        \put(96,19.2){BRDF $\decoder$}
        \put(96,16.2){Albedo $\albedo$}
        \put(96,3.7){PDF $\pdf$}
    \end{overpic}
    \vspace{-2mm}
    \caption{
        We use our neural BRDFs in a renderer as follows: for each ray that hits a surface with a neural BRDF, we perform standard $(u,v)$ and MIP level $l$ computation, and query the latent texture of the neural material.
        Then we input the latent code $\latents(\x)$ into one or two neural decoders, depending on the needs of the rendering algorithm.
        The BRDF decoder (top box) first extracts two shading frames from $\latents(\x)$, transforms directions $\wi$ and $\wo$ into each of them, and passes the transformed directions and $\latents(\x)$ to an MLP that predicts the BRDF value (and optionally the directional albedo).
        The importance sampler (bottom box) extracts parameters of an analytical, two-lobe distribution, which is then sampled for an outgoing direction $\wo$, and/or evaluated for PDF $\pdf(\x,\wi,\wo)$.
    }\label{fig:method-illustration}
\end{figure*}

\paragraph{Rendering novel BRDFs.} \citet{Fan2022} are able to render novel BRDFs not part of the training set through layering of latents.
However, this requires large neural networks unsuitable for real-time.
We focus on small networks that render only materials they were trained on and do not pursue generalization.
We support layered materials by capturing the joint effect of \emph{all} layers at once, dispensing with the explicit layering of the original material, and avoiding any layering of neural components.

\subsection{Neural material filtering}

Aliasing due to shading is commonly addressed with mipmapping, but requires special care for non-diffuse materials as their appearance can change significantly with linear filtering.
Methods such as LEAN~\cite{Olano2010}, LEADR~\cite{Dupuy2013} and MIPNet~\cite{Gauthier2022} use statistical methods or neural downsampling to more closely match the prefiltered ground truth.
While these approaches tune the parameters of traditional BRDFs, we instead train neural models and hierarchical textures to represent the filtered appearance directly, similarly to \citet{Kuznetsov2021} and \citet{Bako2022}, albeit with a different interpolation scheme (see \autoref{sec:latent-texture}).
However, we still leverage LEAN~\cite{Olano2010} as a graphics prior to filter the inputs of our encoder.

\subsection{Neural material importance sampling}

Prior work on the importance sampling of neural materials can classified as: i) utilizing an analytical proxy distribution, ii) leveraging normalizing flows, and iii) warping samples with a network directly. See \citet{Xu2023NeuSample} for an overview of neural materials samplers.

We utilize the first approach, in which a network parameterizes an analytical distribution.
In contrast to \citet{Sztrajman2021} and \citet{Fan2022}, who use the Phong-Blinn model or an isotropic Gaussian, we leverage a standard microfacet model~\cite{Trowbridge1975,Walter2007}.
The microfacet model better handles anisotropy that is prevalent in (filtered) realistic materials.

Normalizing flows for importance sampling~\cite{Dinh2017,Mueller2019} were first utilized for neural BRDFs by \citet{Zheng2021}.
With sufficiently large networks, these can accurately match intricate distributions but we found it challenging to match the quality of the analytical proxy at comparable runtime performance.

The third approach, using the network directly to warp samples, has been recently explored by \citet{Bai2023} who aid training of the network with 2D optimal transport.
This method has the drawback that the learned density only approximately matches the true Jacobian determinant of their warp. This leads to potentially unbounded bias, and we exclude this option to maintain compatibility with physically based renderers.

\section{Overview}

Our goal is to reproduce the appearance of real materials that stems from the interaction of light with matter. It can be described using the spatially varying bidirectional reflectance distribution function (SVBRDF) $\brdf(\x, \wi, \wo)$ that quantifies the amount of scattered differential radiance $\mathrm{d}\Lo(\x,\wo)$ due to incident radiance $\Li(\x,\wi)$:
\begin{align}
    \brdf(\x,\wi,\wo) &= \frac{\mathrm{d}\Lo(\x,\wo)}{\Li(\x,\wi) \cos\theta_i \mathrm{d}\wi}\,,
\end{align}
where $\x$ is a surface point, and $\wi$, $\wo$ are incident and outgoing directions, respectively. 
The SVBRDF can be integrated over the upper hemisphere $H^2$ to produce directional albedo $\albedo(\x,\wo)$:
\begin{align}
    \quad \albedo(\x,\wo) &= \int_{H^2} \brdf(\x,\wi,\wo) \cos\theta_i \mathrm{d} \wi\,.
    \label{eq:albedo}
\end{align}
Our model represents both of these quantities; see \autoref{fig:method-illustration}.

We design our model to serve as an optimized representation of existing (reference) SVBRDFs.
That is, given a target material $\brdf(\x, \wi, \wo)$, we provide a function $\decoder \approx \brdf$ that closely approximates the reference material and can be evaluated in real time. To be useful, our system must satisfy a number of properties:

\paragraph{Visual fidelity.} Our main goal is to faithfully reproduce a broad range of challenging materials, including multi-layer materials with low-roughness dielectric coatings, conductors with glints, stains, and anisotropy. We wish to go beyond fitting to spatially uniform measured material datasets \cite{Matusik2003,Dupuy2018}, and want to explicitly address materials with high resolution textures (4k and above) with detailed normal maps. %

\paragraph{Level of detail.} Unfiltered high-resolution materials tend to alias under minification and properly filtered reflectance can change significantly within a pixel footprint. 
We seek to support filtered lookups to enable level-of-detail rendering at low sample counts.

\paragraph{Importance sampling.} In addition to representing the BRDF, we need an effective importance sampling strategy to permit  deployment in Monte Carlo estimators, such as path tracing. This includes the traditionally challenging problem of importance sampling filtered versions of the material.

\paragraph{Performance.} Our neural representation is geared towards real-time applications, where material evaluation may only use a small fraction of the total frame time.
We require compatibility with path tracing, where materials are evaluated at random locations over many bounces. This precludes large networks and models relying on convolutions.

\paragraph{Practicality} While the optimization of our neural material happens in an offline process, training times have to remain reasonable even for high material resolutions (4k and beyond) for the system to remain practical. Days of training time are not acceptable.
\\

Our main focus is on developing a system that fits the aforementioned criteria. 
Like prior works on neural materials, we forgo explicit constraints on energy conservation and reciprocity relying on the MLP learning these from data.
We also set aside certain special cases, such as BRDFs with delta components, and (rough) refraction, although preliminary experiments show that our model can handle the latter.

In Sections \ref{sec:neural-material-intro} and \ref{sec:training}, we describe the architecture of our neural model and its training procedure, following with a comparative analysis of individual components in \autoref{sec:analysis}.
Since real-time performance is one of our main goals, we dedicate \autoref{sec:system} to the task of efficiently evaluating the neural model from inside ray tracing shaders.
We conclude by demonstrating the quality and runtime performance on a number of challenging scenes in \autoref{sec:results}.

\section{Neural BRDF Decoder}
\label{sec:neural-material-intro}

In this section, we describe the architecture of our appearance model illustrated in \autoref{fig:method-illustration}.
The model consists of two main components: a \emph{latent texture} and two \emph{neural decoders}. All these components are jointly optimized to represent a specific material or a set of materials; details of the optimization procedure (e.g., encoding of the latent texture) follow in the next section.

The latent texture represents spatial variations of the material with a compact, eight-dimensional code denoted $\latents$.
Given a query location $\x$ and the corresponding latent code $\latents(\x)$, the BRDF value is inferred by a neural decoder $\decoder$ with trainable parameters $\decoderParams$:
\begin{align}
    \brdf(\x, \wi, \wo) \approx \decoder\left(\latents(\x), T\cdot\wi, T\cdot\w ; \decoderParams\right)\,,
    \label{eq:neural-brdf}
\end{align}
where $T$ represents a transformation of incident and outgoing directions to a number of learned shading frames. Next, we discuss the properties of the latent texture $\latents$ and then describe the procedure of extracting $T$.

\subsection{Latent texture}
\label{sec:latent-texture}

Similarly to prior works ~\cite{Thies2019,Kuznetsov2021},
we store latent codes in a UV-mapped, hierarchical texture, where each texel characterizes the appearance of the object at a given spatial location and scale. To maintain the fidelity of the original material, we set the resolution of the finest level to the texture resolution of the original material, and we leverage its UV-parametrization to preserve the original texel density.

\begin{figure}[t]
    \vspace{-1mm}
    \newlength{\myimg}\setlength{\myimg}{0.072\textwidth}%
    \setlength{\tabcolsep}{1pt}
    \renewcommand{\arraystretch}{1}
    \begin{center}
        \begin{tabular}{ccc} 
            \footnotesize Unfiltered & \footnotesize Ours & \footnotesize Ground truth \\
            \includegraphics[width=\myimg]{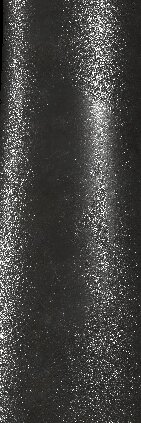}&%
            \includegraphics[width=\myimg]{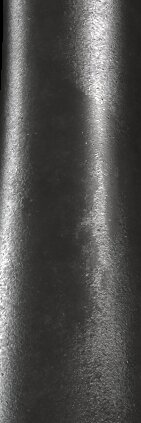}&%
            \includegraphics[width=\myimg]{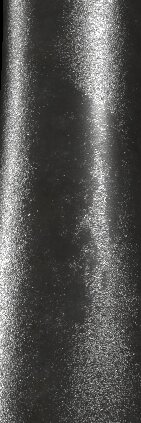} \\
            \multicolumn{3}{c}{\small (a) \cheesegrater, close}
        \end{tabular}
        \begin{tabular}{ccc} 
            \footnotesize Unfiltered & \footnotesize Ours & \footnotesize Ground truth \\
            \includegraphics[width=\myimg]{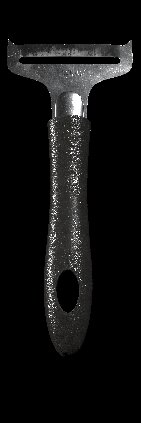}&%
            \includegraphics[width=\myimg]{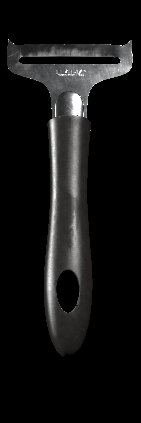}&%
            \includegraphics[width=\myimg]{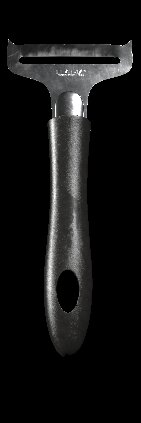} \\
            \multicolumn{3}{c}{\small (b) \cheesegrater, far}
        \end{tabular}
    \end{center}
    \vspace{-2.5mm}
    \caption{
        Highly detailed materials will alias significantly when rendered without supersampling (left columns, unfiltered). Supersampling averages high frequency glints and produces a filtered material, but at impractical sample cost for real-time (right columns, ground truth at 512 SPP). Our neural material can render filtered materials without aliasing at any distance, without supersampling (middle columns, ours).
    }\label{fig:lod-aliased}
    \vspace{-3.5mm}
\end{figure}

Highly detailed materials may cause severe aliasing under minification (\autoref{fig:lod-aliased}, left columns in (a) and (b)). By default, our neural decoder would reproduce such aliasing. To avoid this, the hierarchical latent texture stores the latent codes in a texture pyramid~\cite{Thies2019, Kuznetsov2021}.
Each level of the pyramid contains latent codes that characterize the original material filtered with a specific filter radius. 
The decoder is trained to infer the properly filtered BRDF value for all levels of the pyramid (\autoref{fig:lod-aliased}, middle columns in (a) and (b)).

During rendering, we first determine the pixel footprint at the intersection point, and project it into UV space~\cite{Moller2021}. We then determine the appropriate level of the texture pyramid to sample based on the area of the footprint.

The level index may be fractional and lie between two levels of the pyramid. We probabilistically select one of them using Russian roulette, and fetch the latent code via bilinear interpolation within the level. This introduces a small, but bounded amount of variance. We found this to yield higher quality than the more commonly used method of trilinearly interpolating the latent codes. This is likely because the latter strategy induces the additional constraint that the latent interpolation produce plausible BRDF values across levels, even though they may store very different content.

\begin{figure*}[t]
    \newcommand{\circled}[1]{{\textcircled{\footnotesize #1}}}
    \begin{overpic}[width=1\linewidth]{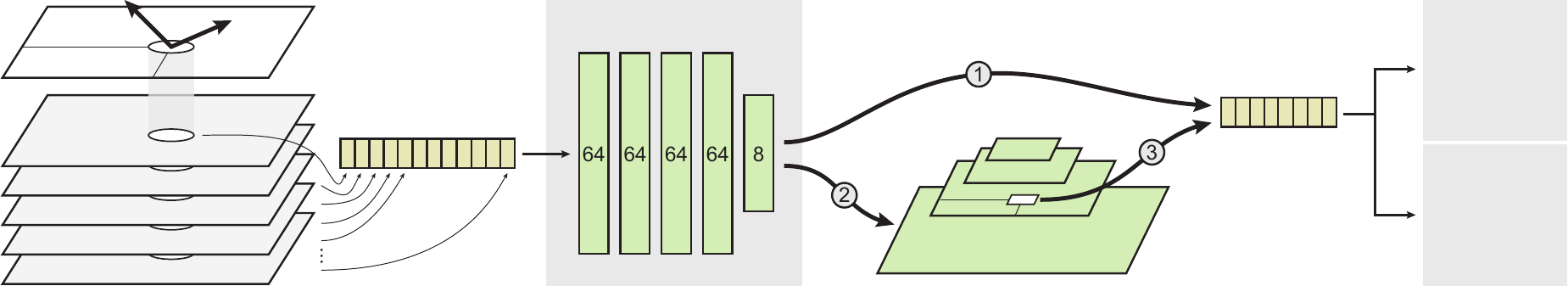}
        \footnotesize
        \put(2,13.8){$(u,v)$ space}
        \put(2,8){Albedo}
        \put(2,6.1){Normal}
        \put(2,4.3){Tangent}
        \put(2,2.4){Roughness}
        \put(2,0.6){...}
        \put(39.6,16){Encoder}
        \put(21,10.5){Surface parameters $\attribs(\x)$}
        \put(61,10.5){Latent texture $\latents$}
        \put(77.5,13){Latent code $\latents(\x)$}
        \put(90,13.5){\begin{minipage}{2cm}\centering BRDF\\evaluation\end{minipage}}
        \put(90,4){\begin{minipage}{2cm}\centering Importance\\sampling\end{minipage}}
    \end{overpic}
    \caption{
        We optimize our model by uniformly sampling the UV domain of the reference material. We start by fetching surface parameters (e.g., albedo) encoding them using an MLP to a latent code, and interpreting it as a BRDF value using the decoder (path marked with \circled{1}). Once the encoder is sufficiently trained, we construct the latent texture \circled{2} by processing all texels, and then drop the encoder. We continue ``finetuning'' the latent texture by sampling the UV space and MIP levels of the texture and optimizing the texels directly \circled{3}. We sample exponentially distributed filter footprints to optimize all levels of the latent texture, and train the decoder with prefiltered versions of the input material.
    }\label{fig:optimization-illustration}
\end{figure*}

\subsection{Transformation to learned shading frames}
\label{sec:rotation-op}

Our focus on real-time applications severely constrains the size of the decoder network. This makes it all the more important to incorporate graphics priors into the architecture to handle realistic materials, such as those exemplified in \autoref{fig:material-overview}.
These layered materials produce intricate SVBRDFs, where reflection lobes shift in direction as we move over the surface. Such effects are readily modeled in classical materials via textured transformations, e.g., using normal maps, but are hard to achieve for a standard MLP.

A material may feature as many normal maps as scattering layers.
We aim to compress the stack of layers, but still provide the model with enough room to represent multiple normal maps.
We therefore incorporate a transformation module into the network, which transforms incident and outgoing directions into a number of learned shading frames (\emph{mult} operation in \autoref{fig:method-illustration}).
Specifically, we use a single trainable layer to extract a fixed number $N$ of normals $(\normal_1 \dots \normal_N)$ and tangent vectors $(\tang_1 \dots \tang_N)$ from the latent code. Then we construct a basis 
$(\tang_i, \bitang_i, \normal_i)$ for each $i$-th pair of normalized normals and tangents, and construct a combined transformation matrix $T$:
\begin{align}
    T = 
    \begingroup %
\left(
    \setlength\arraycolsep{2pt}
  \begin{array}{ccccccc}
    t_{1,x} & b_{1,x} & n_{1,x} & \dots & t_{N,x} & b_{N,x} & n_{N,x}\\
    t_{1,y} & b_{1,y} & n_{1,y} & \dots & t_{N,y} & b_{N,y} & n_{N,y}\\
    t_{1,z} & b_{1,z} & n_{1,z} & \dots & t_{N,z} & b_{N,z} & n_{N,z}
  \end{array}
\right)^\intercal.
\endgroup
\label{eq:frame-rotation}
\end{align}
The transformation layer then computes the product $T \cdot \w_i$ and $T \cdot \w_o$, resulting in $N$ new incident and outgoing vectors, one pair for each of the learned shading frames. The vectors are then fed to the decoder.
The transformation allows the model to rotate the input directions into multiple, spatially varying shading frames in a single operation, improving the representational power of the network. We analyze the benefits in \autoref{sec:analysis}.

\paragraph{Discussion.} 
It may not be immediately obvious why a vanilla MLP struggles with rotating directions.
This is because, even though MLPs are built from matrix operations, they can only perform multiplicative transformations of the inputs with the (fixed) \emph{network weights}. 
They cannot readily multiply the input dimensions with \emph{each other}. 
In our case, a decoder with a vanilla MLP cannot easily multiply $\wi, \wo$ with the latent code, which stores spatial variations of the material. The decoder is forced to approximate the multiplicative transform using its trainable layers, depleting its modeling capacity.
Our approach is conceptually similar to (self-)attention models that augment neural networks with multiplicative transforms between activations \cite{Rebain2022AttentionBeatsConcat, Vaswani2017}.

\subsection{Importance sampling}
\label{sec:importance-sampling}

We focus on samplers suitable for representation by a network: an invertible transform $\warp$ from random variates $\mathbf{u}\in[0,1)^2$ into outgoing directions $\wo = \warp(\mathbf{u}; \x, \wi)$, and its associated probability density function (PDF) $\pdf(\wo; \x, \wi)$. Low variance results are achieved whenever the shape of $\pdf$ closely matches $\brdf$.

Optimizing an MLP to perform the sample transform $\warp$ does not guarantee invertibility of $\warp$ and tractable PDF evaluations. Importance sampling thus requires a different approach than BRDF evaluation.
We draw inspiration from prior work and utilize a neural network to drive an existing analytic proxy distribution that is invertible in closed form. Like \citet{Sztrajman2021} and \citet{Fan2022}, we use a linear blend between a cosine-weighted hemispherical density and a specular reflection component, but we differ in the choice of the specular component.

Instead of the isotropic models proposed earlier (e.g., Blinn-Phong model \citep{Sztrajman2021} or a 2D Gaussian
\citep{Fan2022}) we use the more general, state-of-the-art microfacet model based on a Trowbridge-Reitz (GGX) NDF~\cite{Trowbridge1975,Walter2007} including elliptical anisotropy and non-centered mean surface slopes~\cite{Dupuy2015}. This is well-suited both to the strongly normal-mapped materials represented in our target materials, as well as filtered BRDFs that naturally produce anisotropic distributions; we demonstrate the advantage in \autoref{sec:analysis} and provide additional details of the sampler in \autoref{app:sampling}.

We train an additional \emph{importance sampling decoder} MLP that infers parameters of the analytic model from the same latent code as used for the BRDF evaluation. This is conceptually similar to \citet{Sztrajman2021}, though we additionally feed $\wi$ into the decoder to capture Fresnel-like effects where, e.g., the diffuse-specular mixing weights vary as a function of the incident angle.

\section{Training}
\label{sec:training}

We now discuss the training procedure for our decoder and latent texture (see \autoref{fig:optimization-illustration}), and how our training data is generated.

One major challenge in training detailed materials is the sheer number of parameters to be optimized. Although the number of network weights is small, the resolution of the latent texture matches that of the source material and can be considerable: the ceramic body of the \teapot{} (\autoref{fig:material-overview}) is defined using 14 4k~$\times$~4k texture tiles totaling 235 million texels, or 2.5 billion latent parameters. Optimizing these parameters independently using backpropagation is impractical. %
Instead, we make use of an \emph{encoder} in the first training phase to bootstrap latent codes, which we describe next.

\subsection{Encoder}
\label{sec:encoder}

The encoder is a simple MLP that takes the parameters $\attribs(\x)$ of the original material (albedo, roughness, normal maps, etc. for all material layers) at a given query location $\x$ as input, and outputs the corresponding latent vector $\latents(\x)$. 
To bootstrap the filtering, we prefilter the material parameters $\attribs(\x)$ (using LEAN~\cite{Olano2010}) for coarse MIP levels of the hierarchy.

In the first training phase, the model is trained end-to-end by forwarding the latent code from the encoder directly to the decoder, bypassing the latent texture.

After the decoder converges, we switch to the finetuning phase. The latent texture is initialized by evaluating the encoder for all texels, after which the encoder is dropped. The contents of the latent texture are then trained directly using backpropagation through the decoder. Because the encoder only participates in training, it has no impact on the evaluation cost during rendering.

The encoder also improves the structure of the latent space: it guarantees that similar material parameters are mapped to similar points in the latent space. This leads to better results under interpolation, and makes the job of the decoder easier. In contrast, direct optimization is prone to leaving portion of the random initialization noise in the latent texture, as analyzed in \autoref{sec:latent-texture-optimization}. %

The encoder can be optimized to encode multiple materials, or even the full appearance space spanned by the reference BRDF (by sampling its parameters uniformly).
Since our latent textures have a large memory footprint, in practice we train each one individually along with its own encoder, unless stated otherwise.

\subsection{Data generation and optimization}
\label{sec:data-generation}

We generate training data by uniformly sampling the UV space of the target (multi-layered) material.
For each sample, we generate random directions $\wi$ and $\wo$ by uniformly sampling their half and difference vectors~\citep{Rusinkiewicz1998,Sztrajman2021}, and evaluate the reference BRDF value.
Each sample additionally contains: normal, tangent, albedo, roughness, and layer weight, exported for each of the layers. 
Depending on the layer count a single sample may require over a hundred floating point numbers.
We generate the samples on the GPU online during training.

\paragraph{Filtering.} We discretely sample a pyramid level for each training sample from an exponential distribution, favoring finer levels. We average multiple sample points drawn from a Gaussian with appropriate footprint for the level, and choose the number of samples proportional to the filter area. This sampling process is fast enough that it does not significantly impact training time.

\paragraph{Mollification} Materials with very narrow peaks (e.g.\ the smooth glaze of the \teapot{}) lead to large training errors early in training and are challenging to learn for the network. To solve this, we initially blur the material directionally by averaging multiple samples from a small cone centered on $\wo$. The angle of the cone decreases during training, so that the network initially learns broad features of the material before converging to the reference.

\paragraph{Optimization.}
We train the BRDF decoder and the importance sampler simultaneously to establish a shared latent space.
The BRDF prediction is optimized using the $L_1$ loss in log space~\cite{Zheng2021}.
The PDF of samples $\wo$ drawn from the learned sampler is scored using the KL divergence against the current state of the learned BRDF. We found that training stability is improved when the latent code is detached from the KL loss computation. This way, the sampler MLP learns how to interpret the latents without interfering with the main BRDF evaluation decoder.

Albedo predictions, if enabled, are optimized using the $L_2$ loss against one-sample MC estimates of \autoref{eq:albedo}.

We optimize our models using 300k iterations, processing two batches of 65k training samples in each iteration; one for optimizing the BRDF decoder and one for the sampler. This amounts to nearly 40 billion (online-generated) material samples in total, with training times lasting around 4--5 hours per material on a single NVIDIA GeForce RTX 4090. Further details of the training procedure are provided in the supplemental document.

\begin{figure}[t]
    \vspace{-0.5mm}
    \setlength{\tabcolsep}{0.002\textwidth}%
    \renewcommand{\arraystretch}{1}
    \begin{center}
        \footnotesize
        \begin{tabular}{ccc}
            MIP 0 (4k~$\times$~4k) & MIP 3 (512~$\times$~512) & MIP 5 (128~$\times$~128)\\
            \includegraphics[width=0.32\columnwidth]{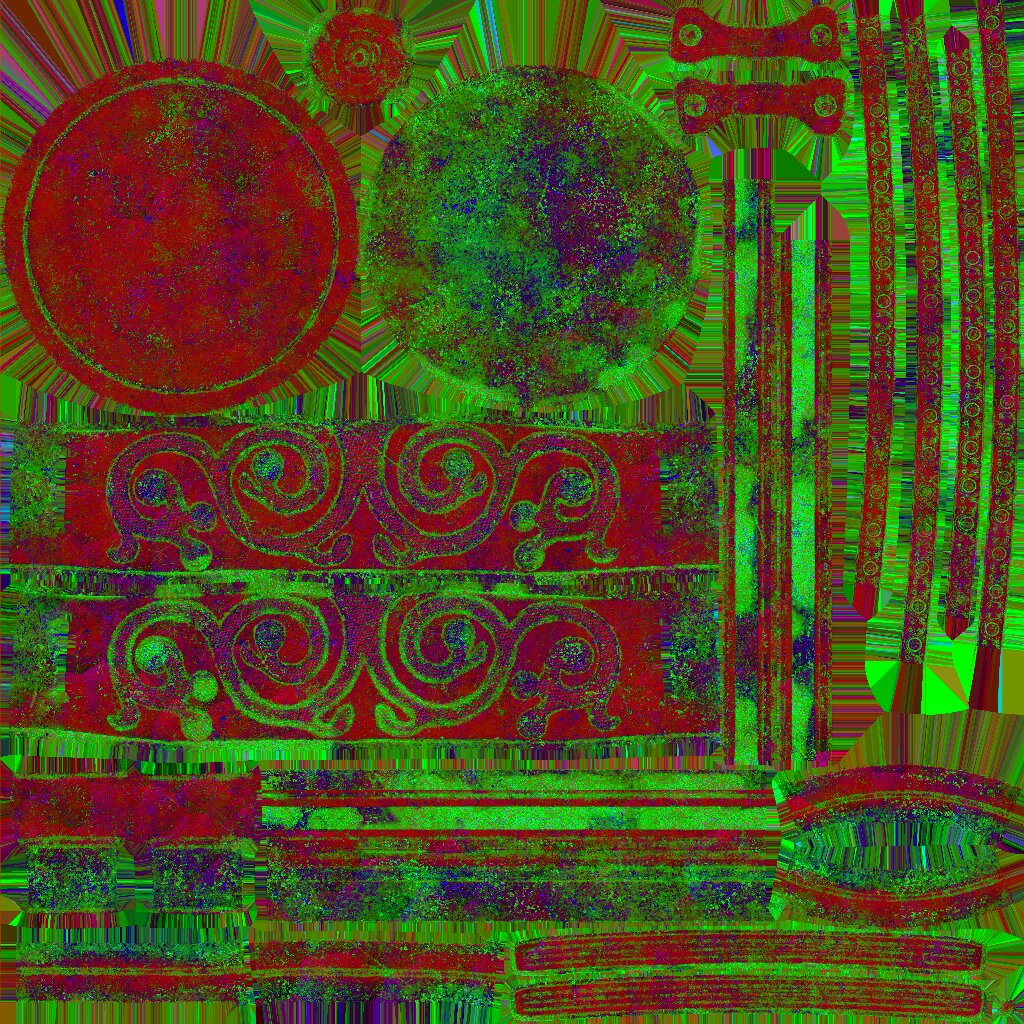}&
            \includegraphics[width=0.32\columnwidth]{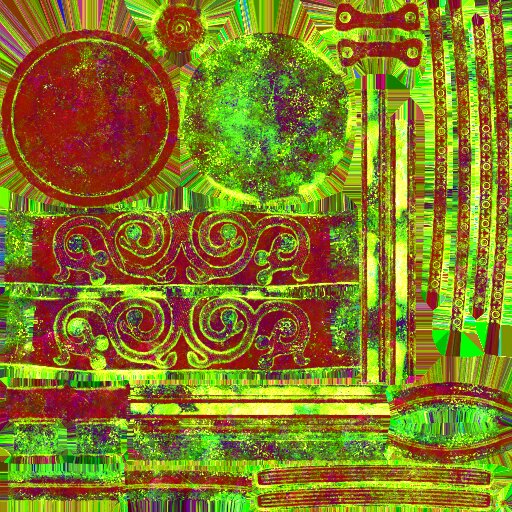}&
            \includegraphics[width=0.32\columnwidth]{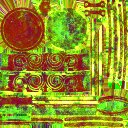}\\[3mm]
        \end{tabular}
        \begin{tabular}{cc}
            Latent texture distribution & Network weight distribution \\
            \includegraphics[width=0.49\columnwidth]{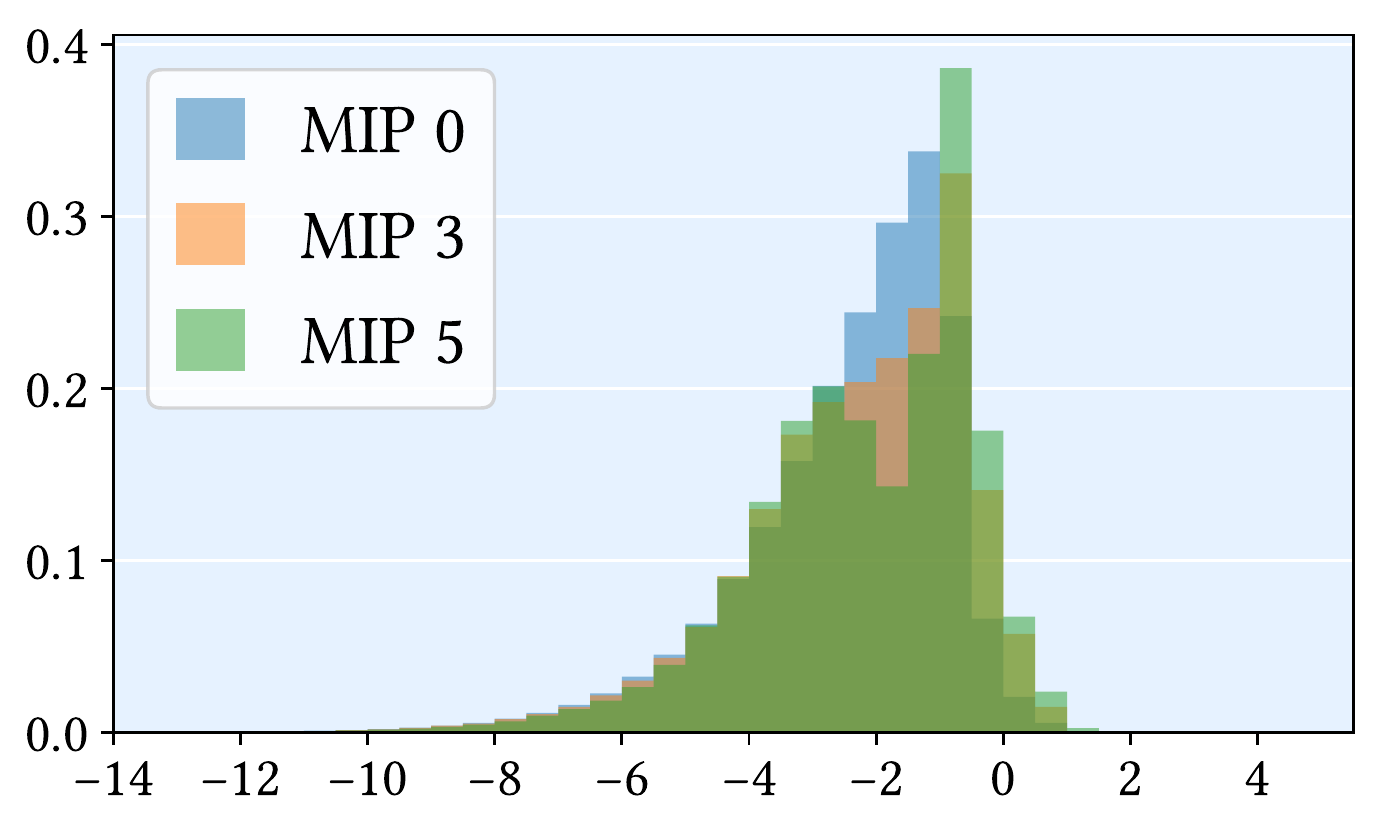}&
            \includegraphics[width=0.49\columnwidth]{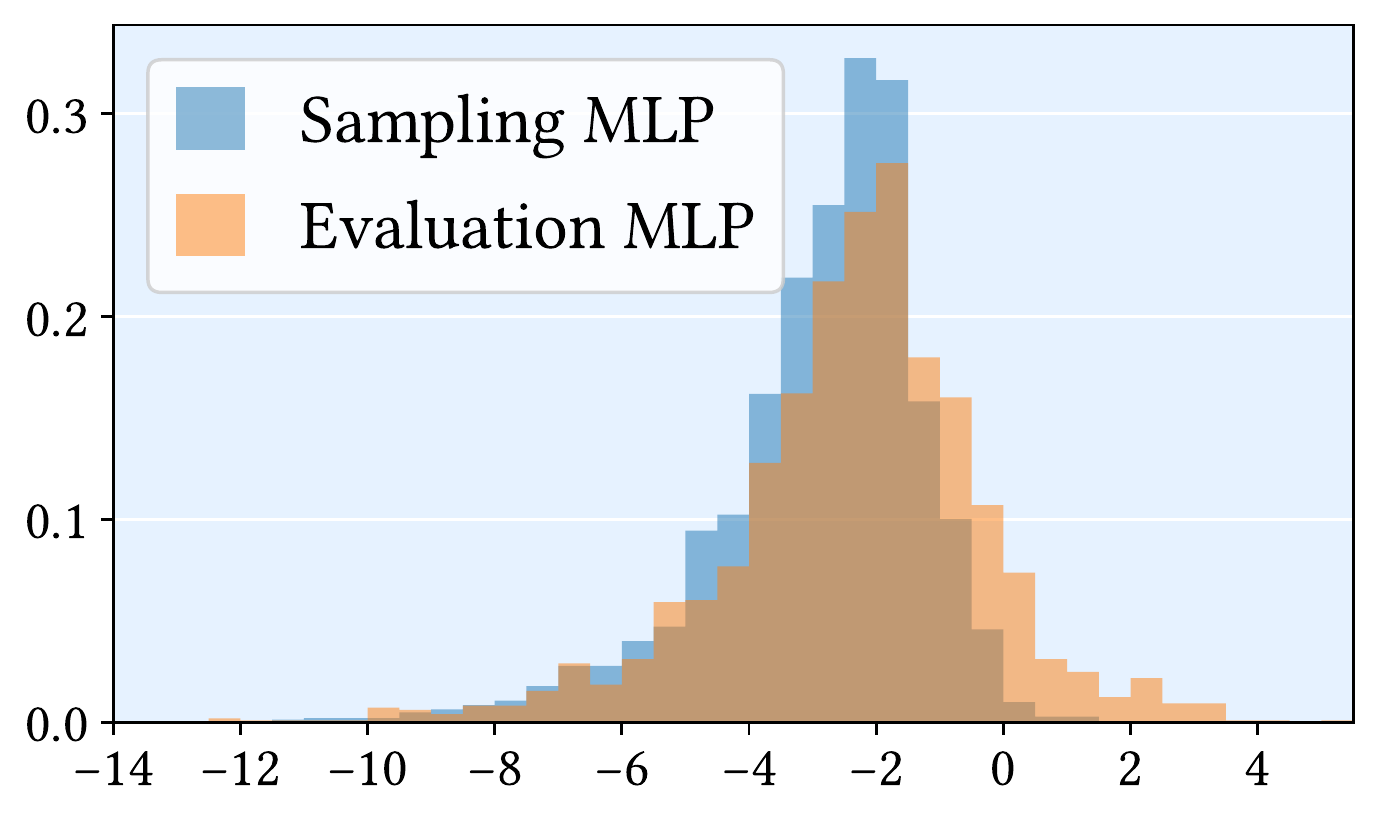}\\
            $\log_2(\text{magnitude})$ & $\log_2(\text{magnitude})$
        \end{tabular}
    \end{center}
    \vspace{-2mm}
    \caption{
        Top row: Optimized latent textures (3 channels shown as RGB) for the neural \inkwell{} material at three levels of the MIP hierarchy. Bottom row: The corresponding distribution of latent (left) and network parameter magnitudes (right). All parameters lie comfortably within the $(2^{-14},2^{16})$ numerical range
        of FP16 normal numbers (excluding denorms), making quantization easy. The other materials show very similar distributions.
    }\label{fig:param_visualization}
\end{figure}

\paragraph{Precision}
We train master parameters for the BRDF decoder and sampler in 32-bit floating-point (FP32) precision.
It is possible to make careful use of mixed precision training to further improve training performance
without losing accuracy, but due to the small sizes of our MLPs we did not explore this option.
For efficient inferencing, we use post-training quantization to convert the parameters to half precision (FP16) at load time.
\autoref{fig:param_visualization} shows a representative example of the distribution of parameters
for the evaluation and sampling models. In all our example configurations, 
the numerical range of network parameters lie within the normalized range of FP16.
In future work, we plan to explore quantization aware training to further reduce
runtime precision to INT8 or lower.

\section{Model analysis and ablation}
\label{sec:analysis}

Now that we have introduced our appearance model and its training procedure, we will analyze the main technical novelties: i)~the transformation into learned shading frames, ii)~the anisotropic importance sampler, and iii)~the use of the encoder. We also demonstrate the filtering capabilities and the option of inferring albedo.

A number of neural appearance models have been published in the past, addressing various aspects of appearance modeling, e.g., geometric level of detail~\cite{Kuznetsov2021,Kuznetsov2022}, interpretability of the latent space~\cite{Zheng2021}, or layering of neural components~\cite{Fan2022}.
These are complementary to our system and could be incorporated in the future.
In this work, we focus on accommodating film-quality visuals and efficient execution on modern GPUs (presented in \autoref{sec:system}).

\begin{figure*}[t]
    \setlength{\tabcolsep}{0.002\textwidth}%
\renewcommand{\arraystretch}{1}%
\footnotesize%
\begin{tabular}{cccc}
Vanilla MLP decoder with latent texture & With latent texture encoder & With transformed $\wi$, $\wo$---full model & \\
(basic variant) & (improved training) & (improved training and decoding) & Reference\\[1pt]
\begin{overpic}[width=0.24575\textwidth,trim=270 200 330 138,clip]{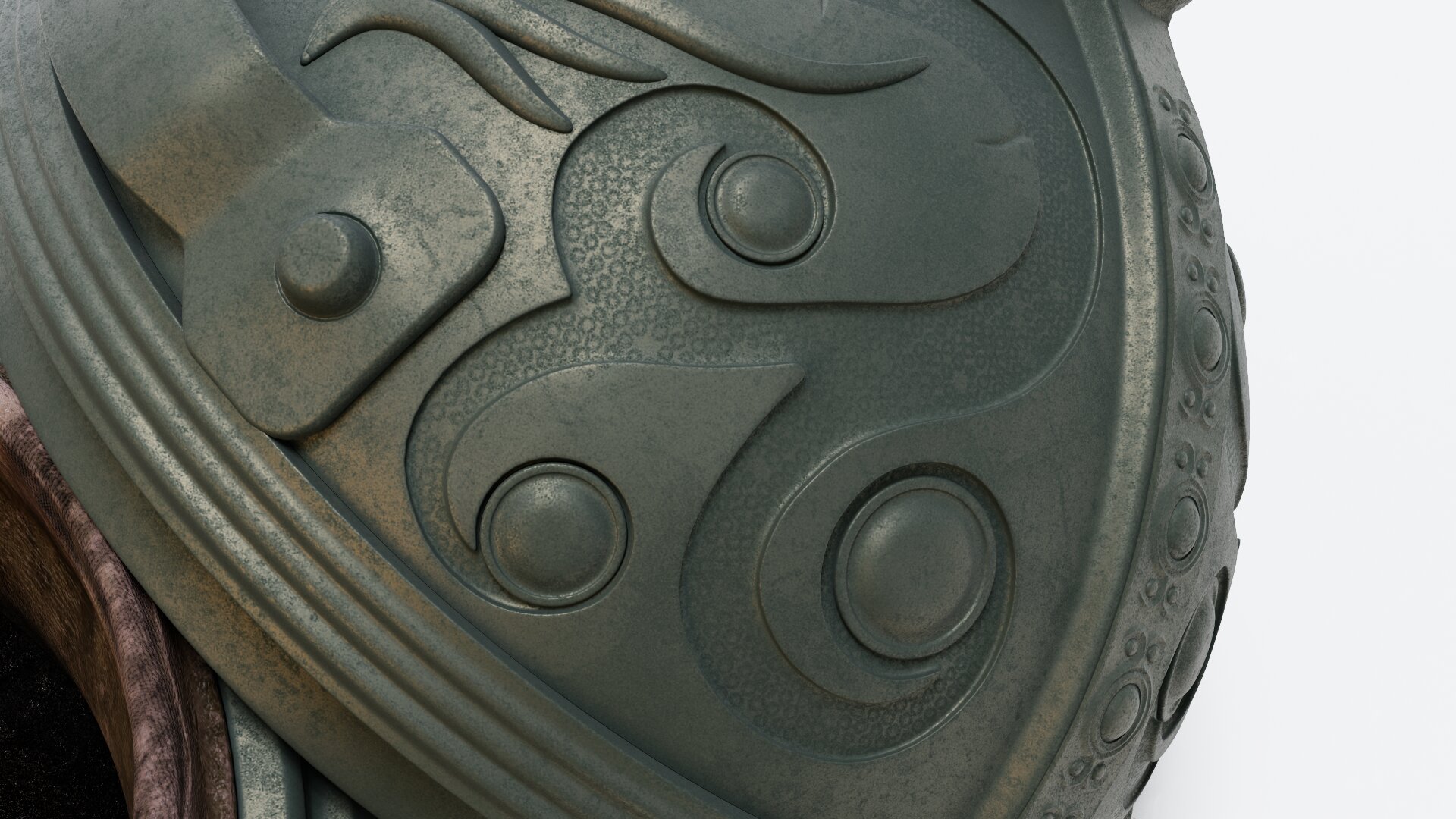}
\put(0,0){\includegraphics[width=0.07\linewidth,trim=270 200 330 138,clip]{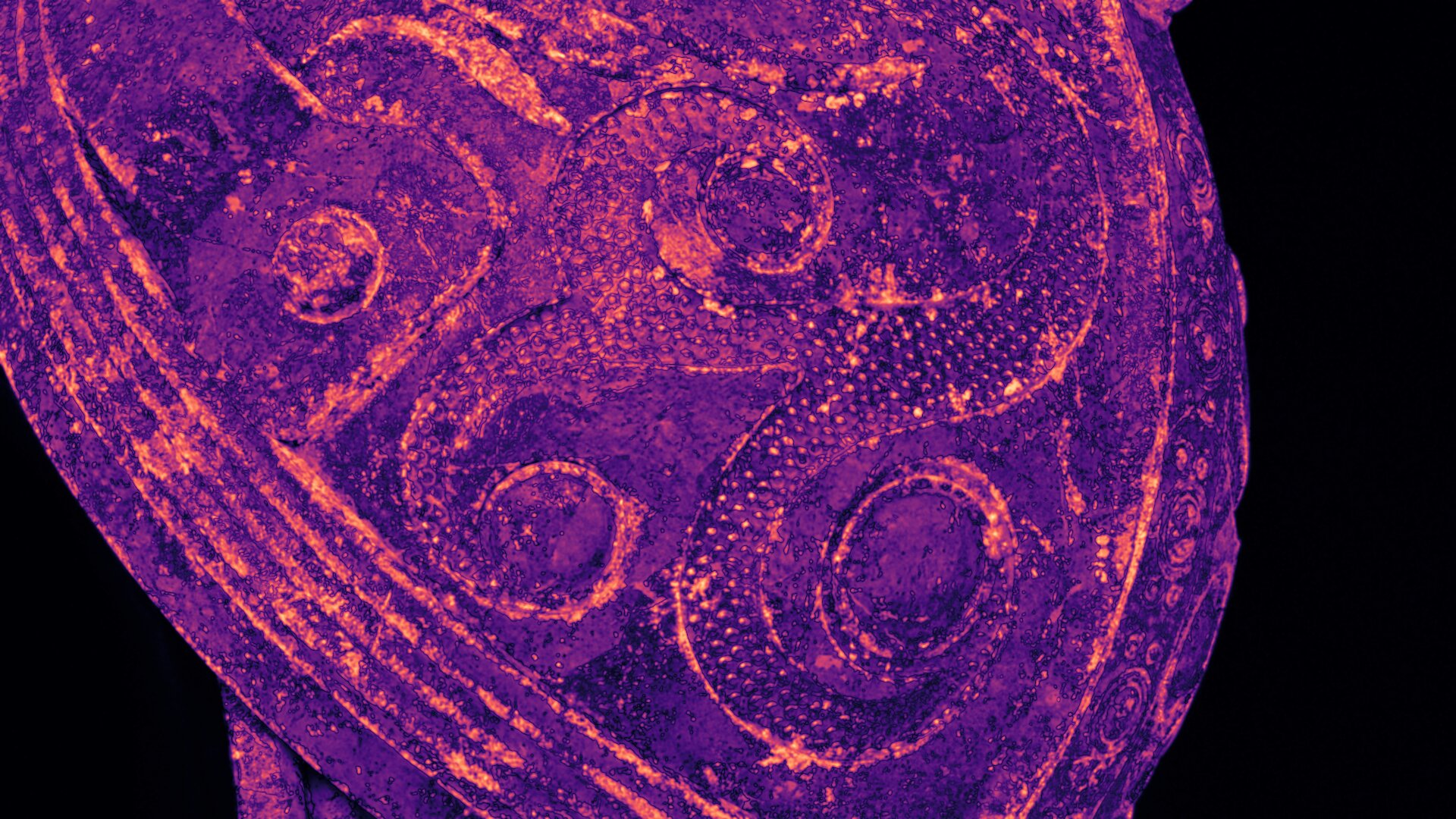}}
\put(-6.0,0){\rotatebox{90}{\hspace{8.5mm}\inkwell}}
\end{overpic}%
&\begin{overpic}[width=0.24575\textwidth,trim=270 200 330 138,clip]{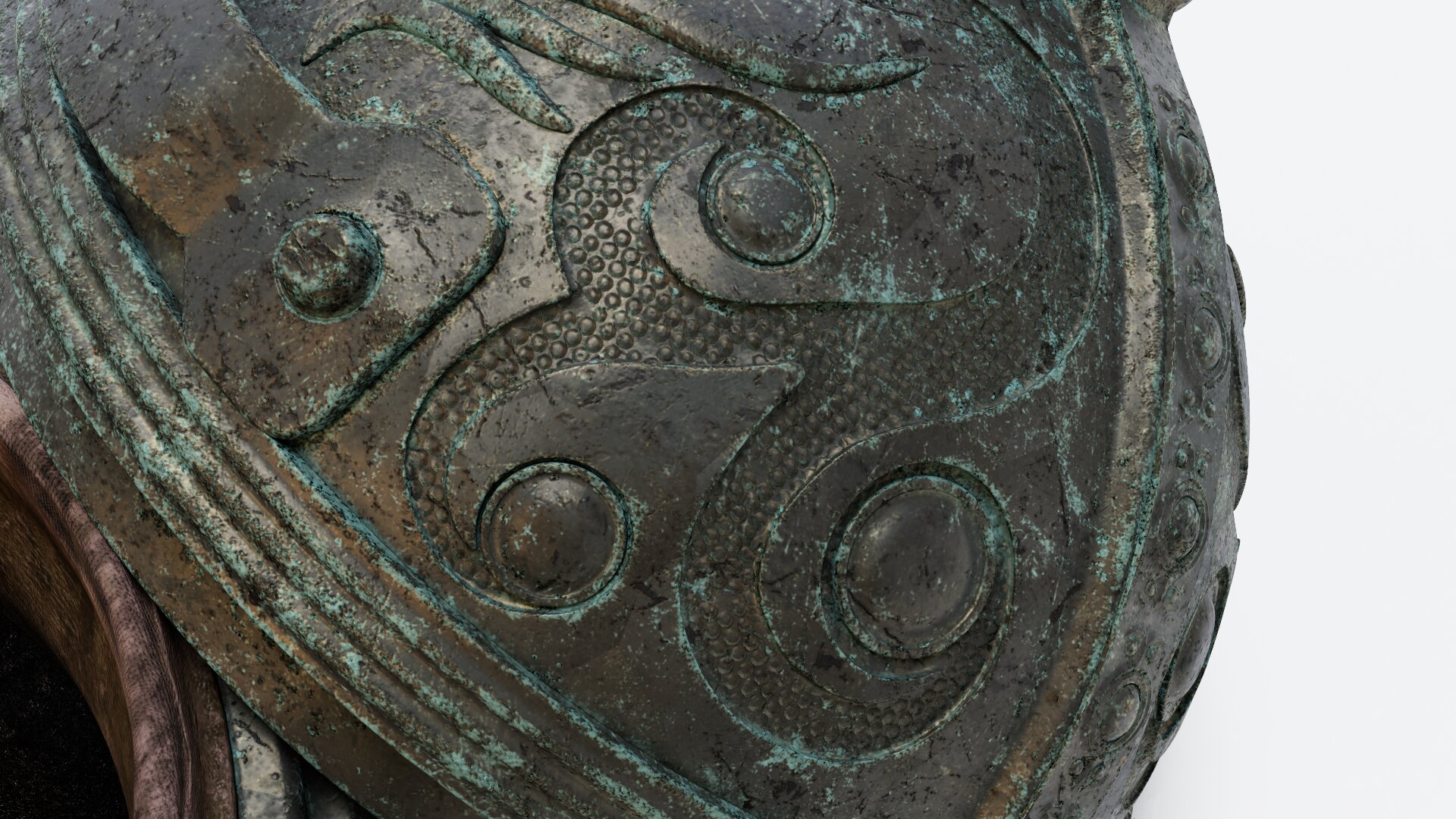}
\put(0,0){\includegraphics[width=0.07\linewidth,trim=270 200 330 138,clip]{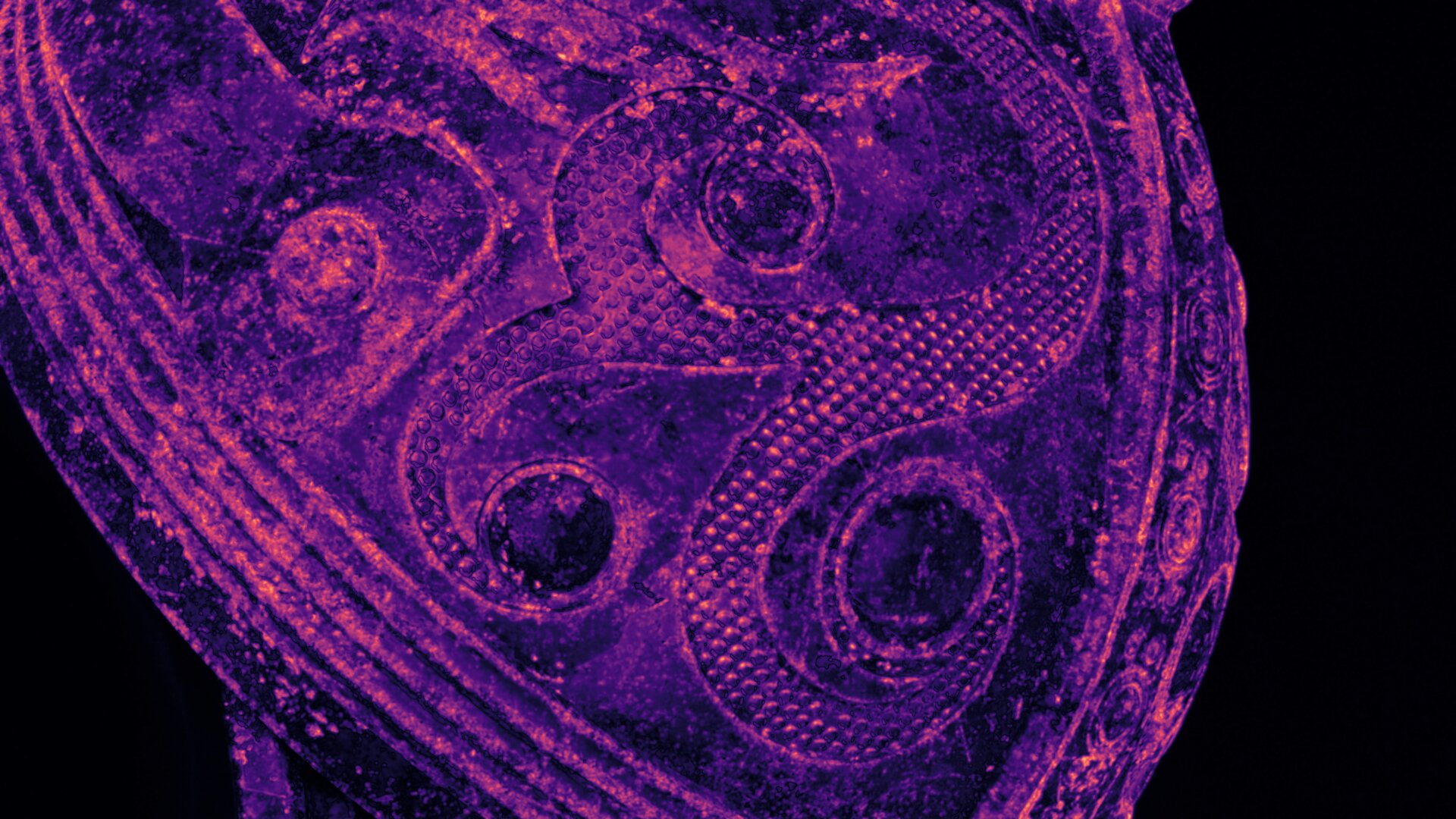}}
\end{overpic}%
&\begin{overpic}[width=0.24575\textwidth,trim=270 200 330 138,clip]{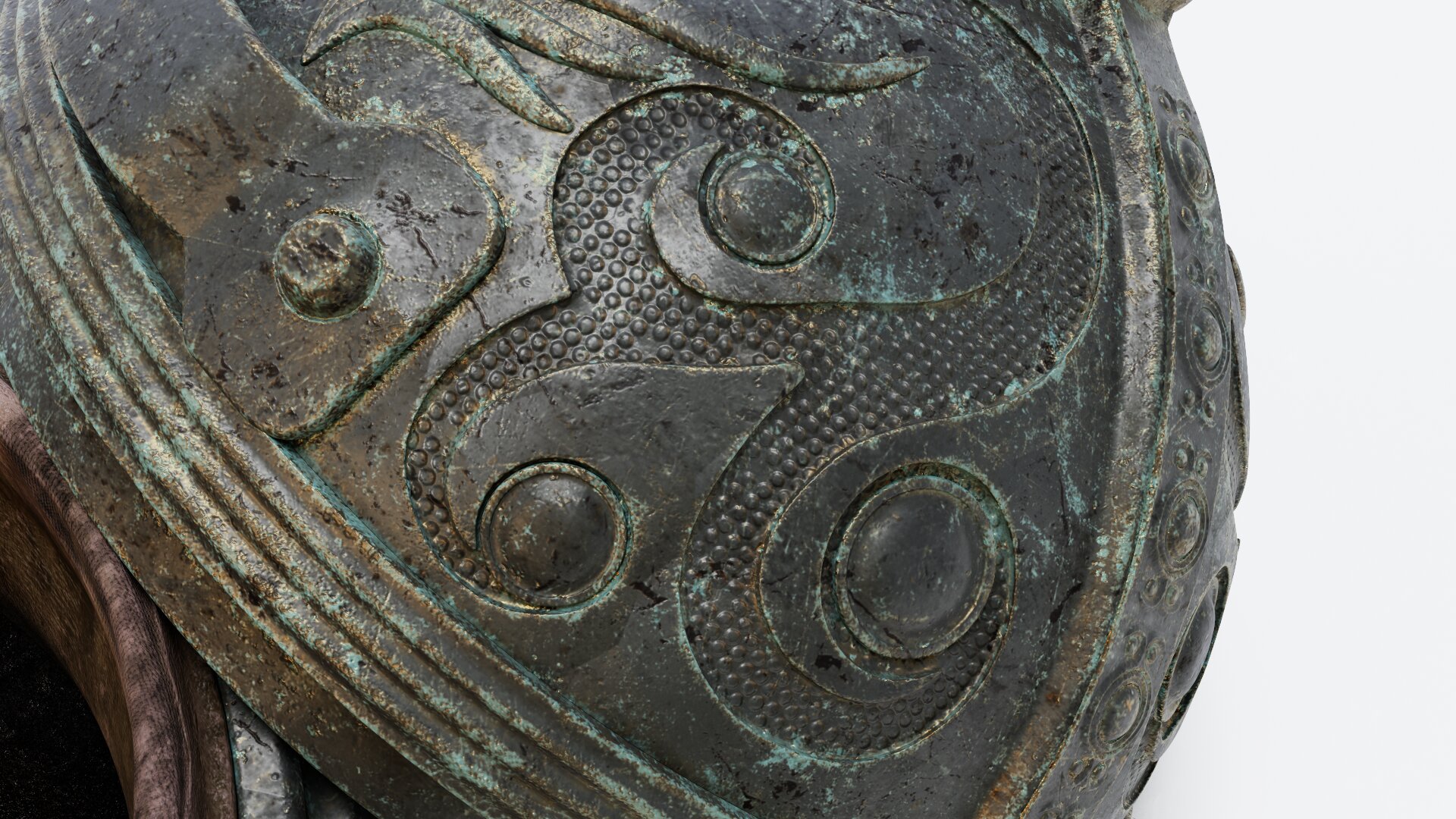}
\put(0,0){\includegraphics[width=0.07\linewidth,trim=270 200 330 138,clip]{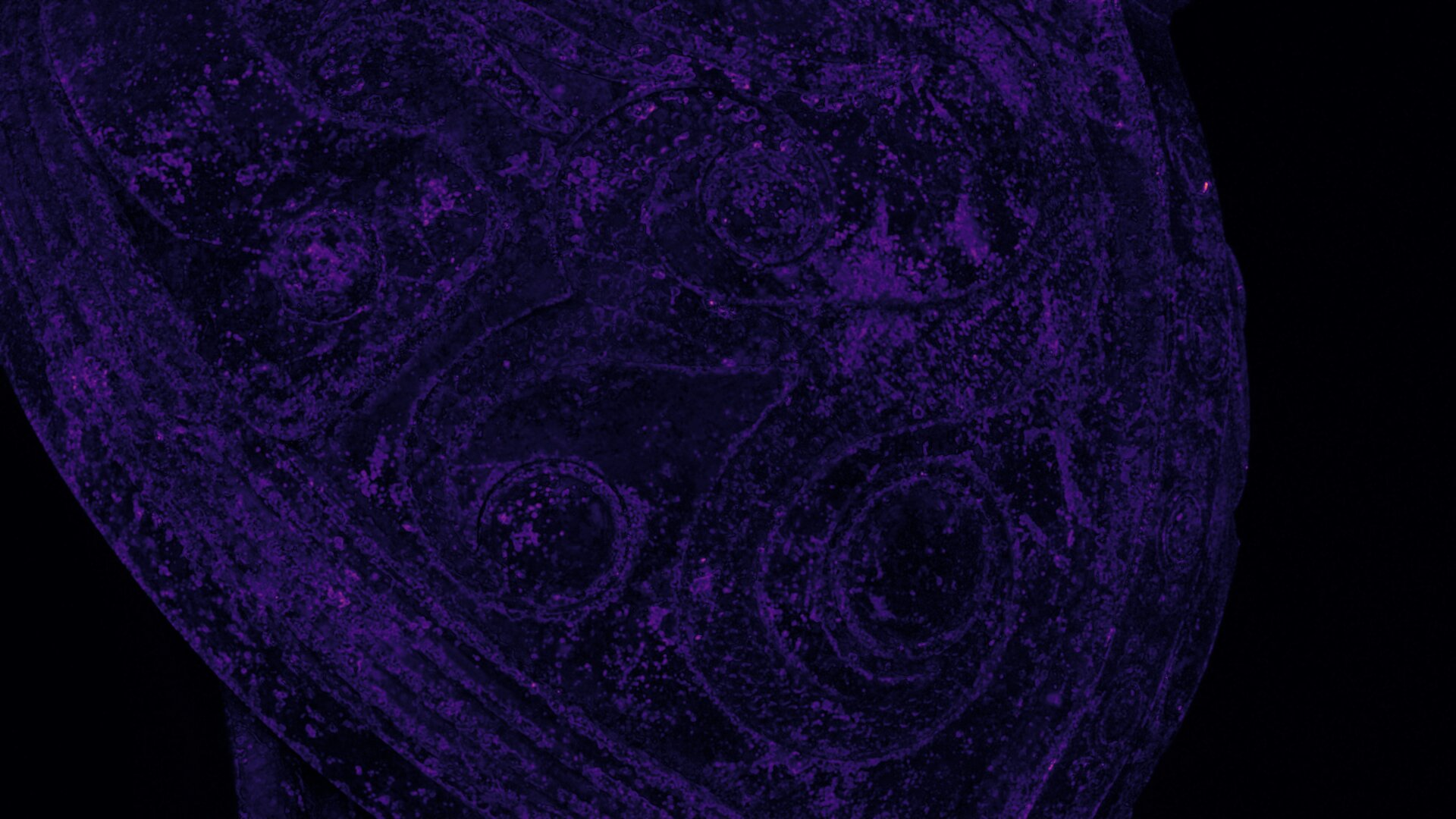}}
\end{overpic}%
&\begin{overpic}[width=0.24575\textwidth,trim=270 200 330 138,clip]{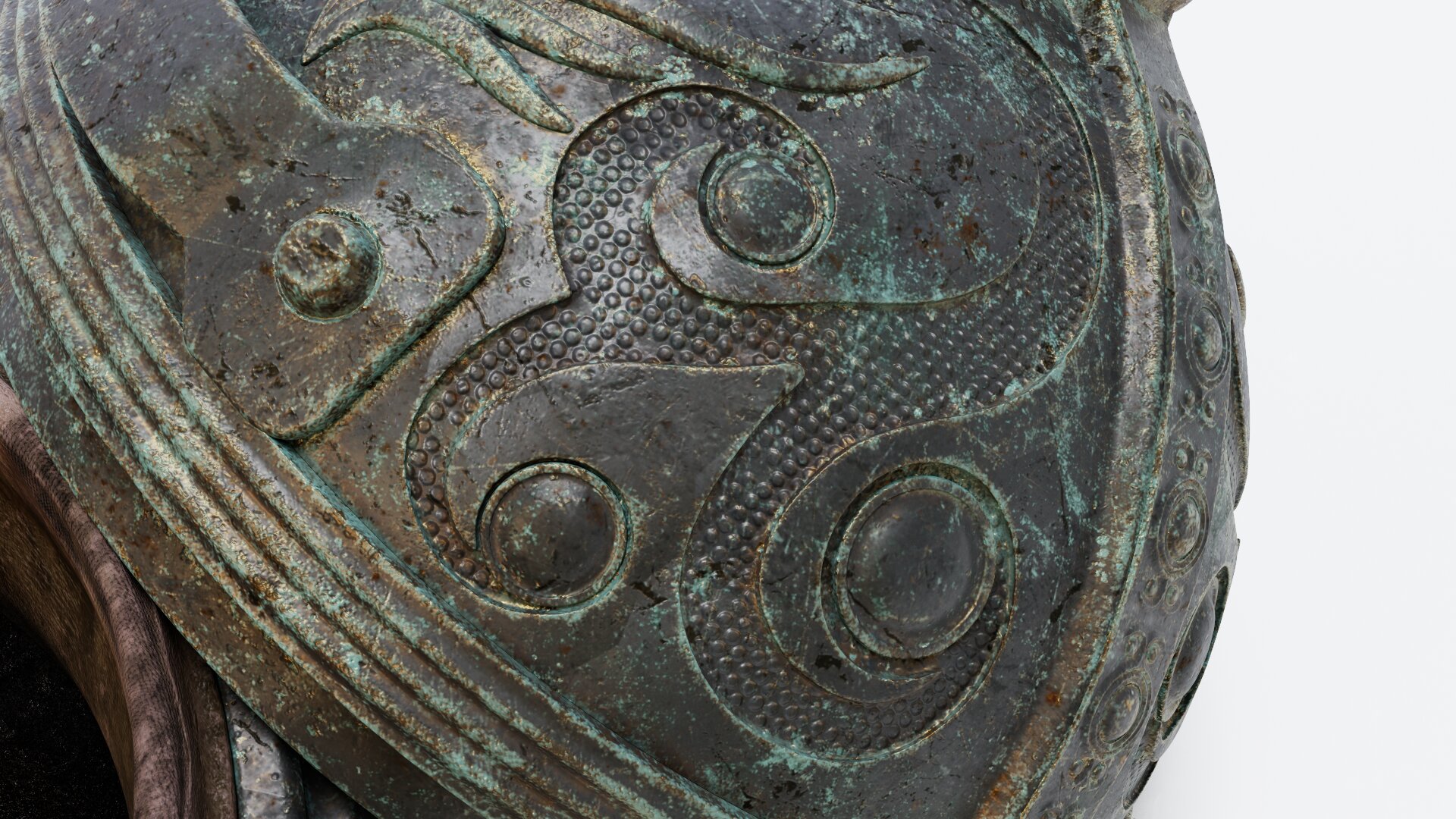}
\end{overpic}%
\\
\begin{overpic}[width=0.24575\textwidth,trim=270 200 330 138,clip]{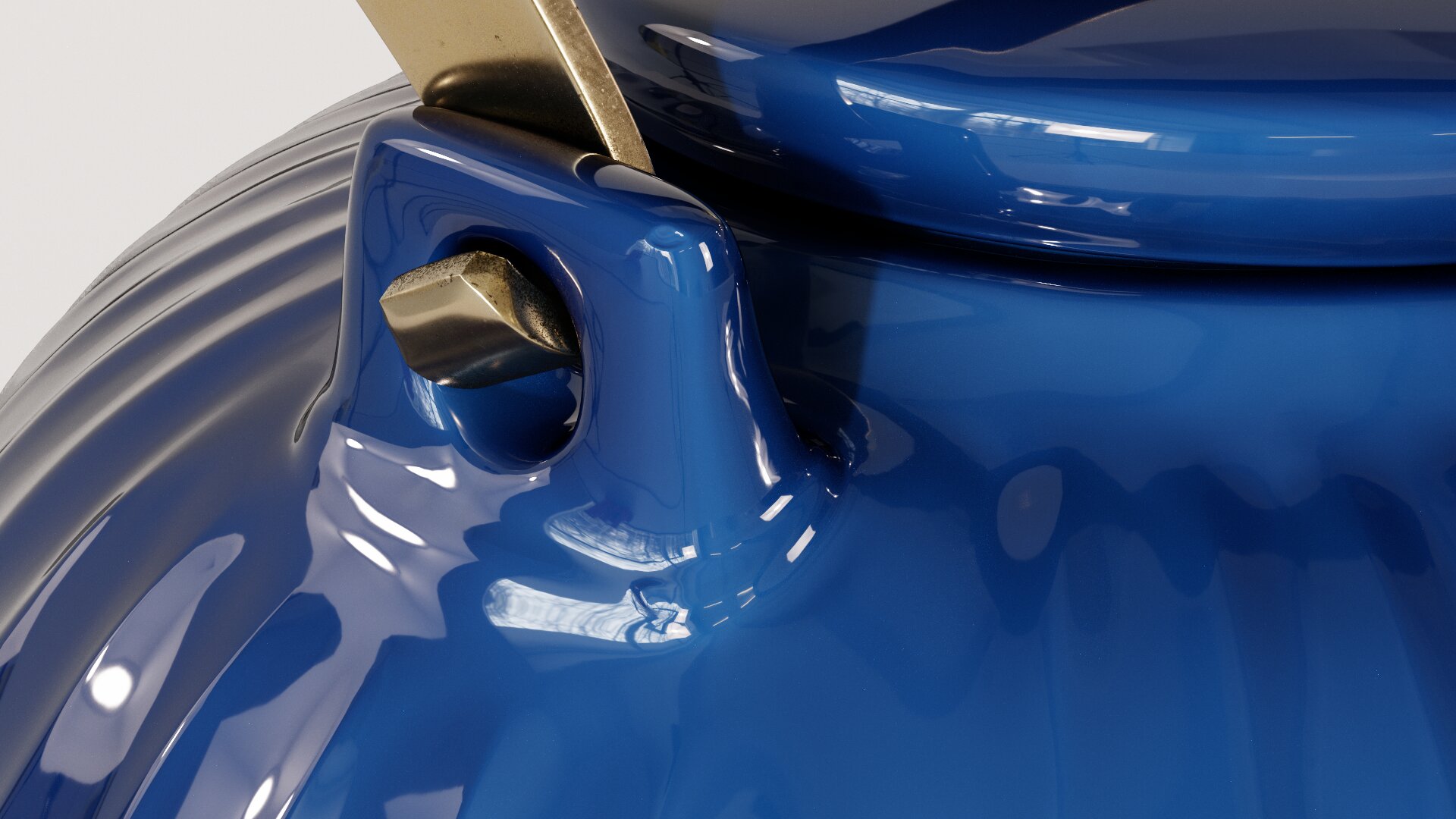}
\put(0,0){\includegraphics[width=0.07\linewidth,trim=270 200 330 138,clip]{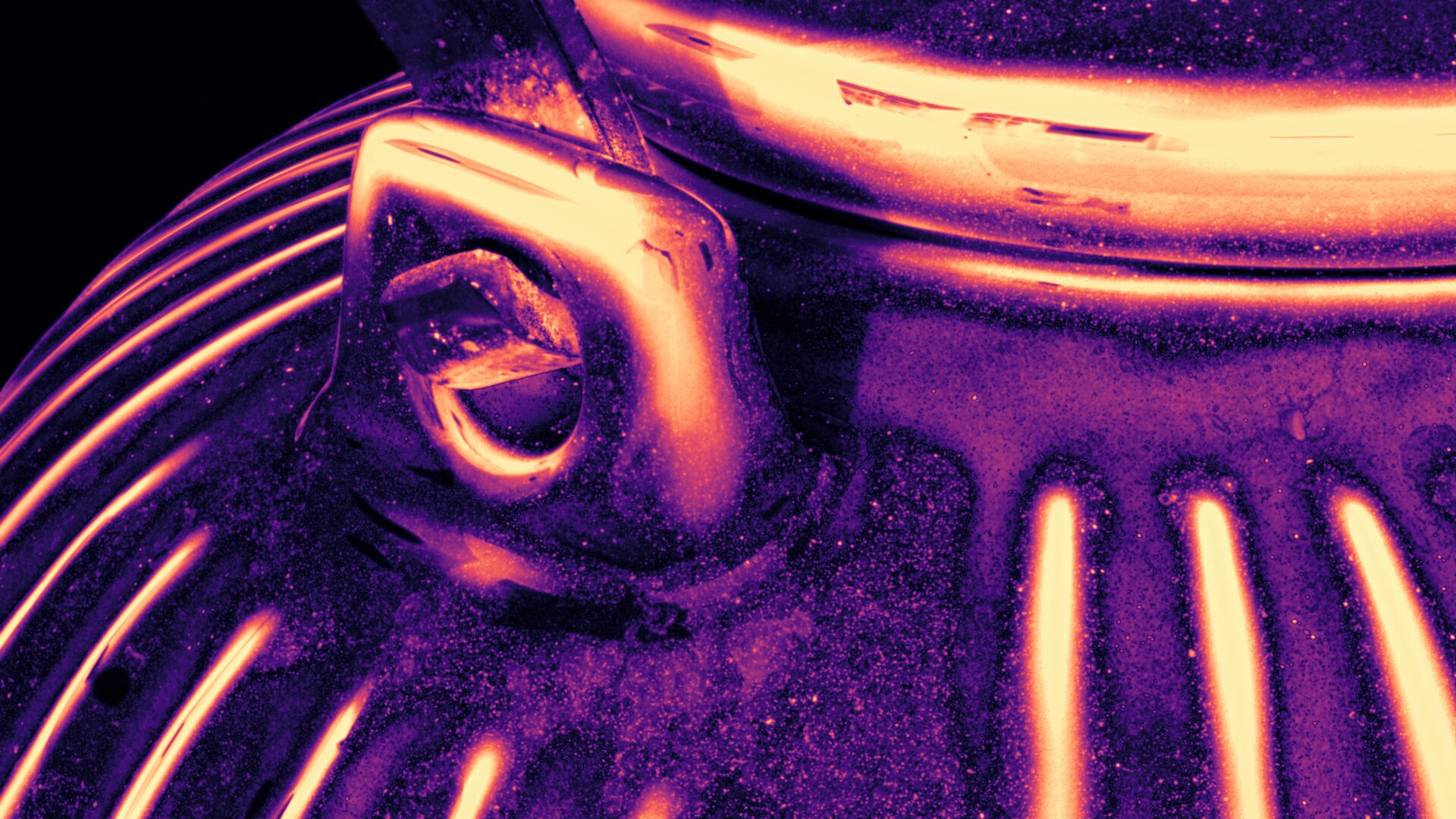}}
\put(-6.0,0){\rotatebox{90}{\hspace{9mm}\teapot}}
\end{overpic}%
&\begin{overpic}[width=0.24575\textwidth,trim=270 200 330 138,clip]{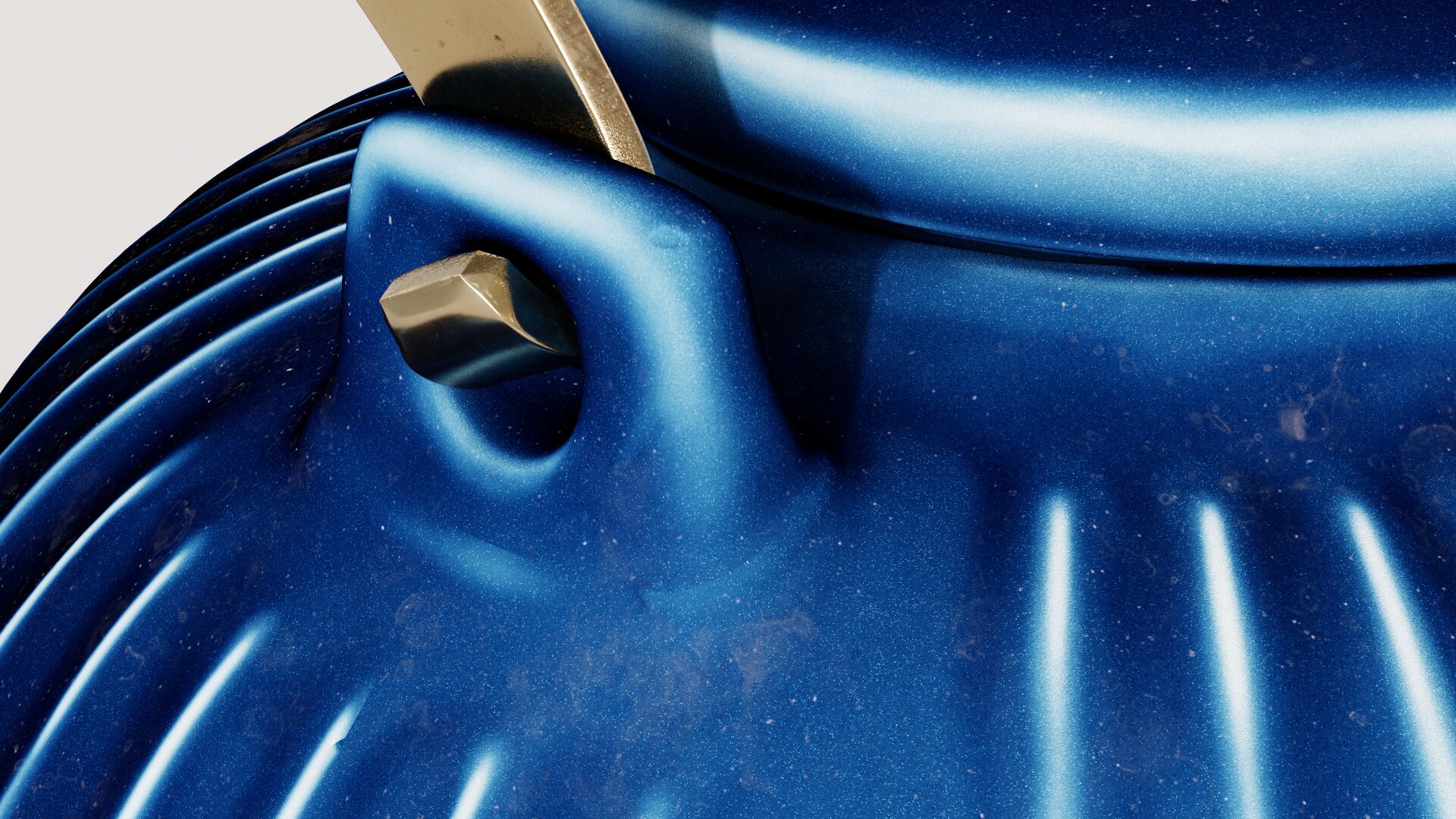}
\put(0,0){\includegraphics[width=0.07\linewidth,trim=270 200 330 138,clip]{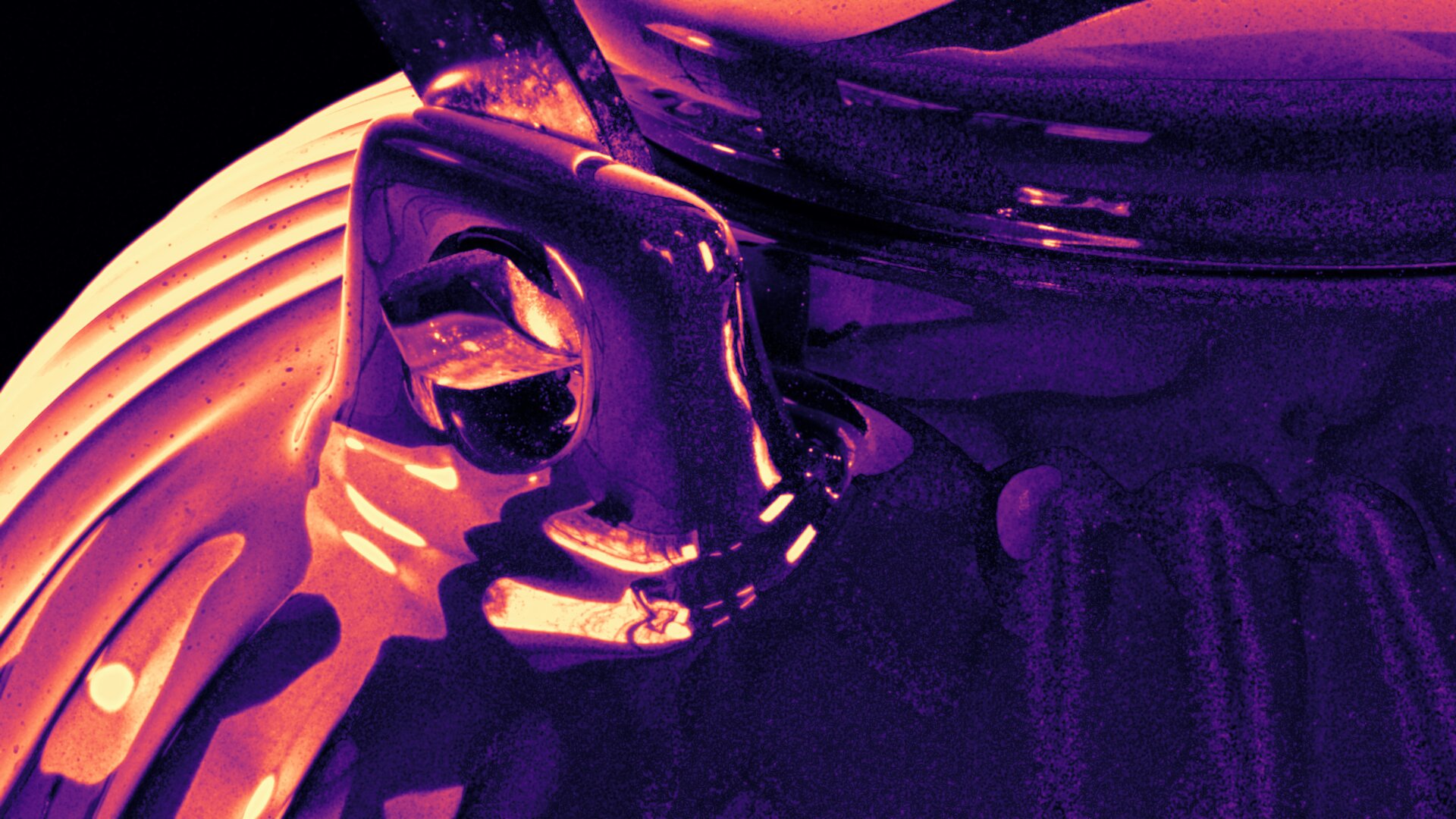}}
\end{overpic}%
&\begin{overpic}[width=0.24575\textwidth,trim=270 200 330 138,clip]{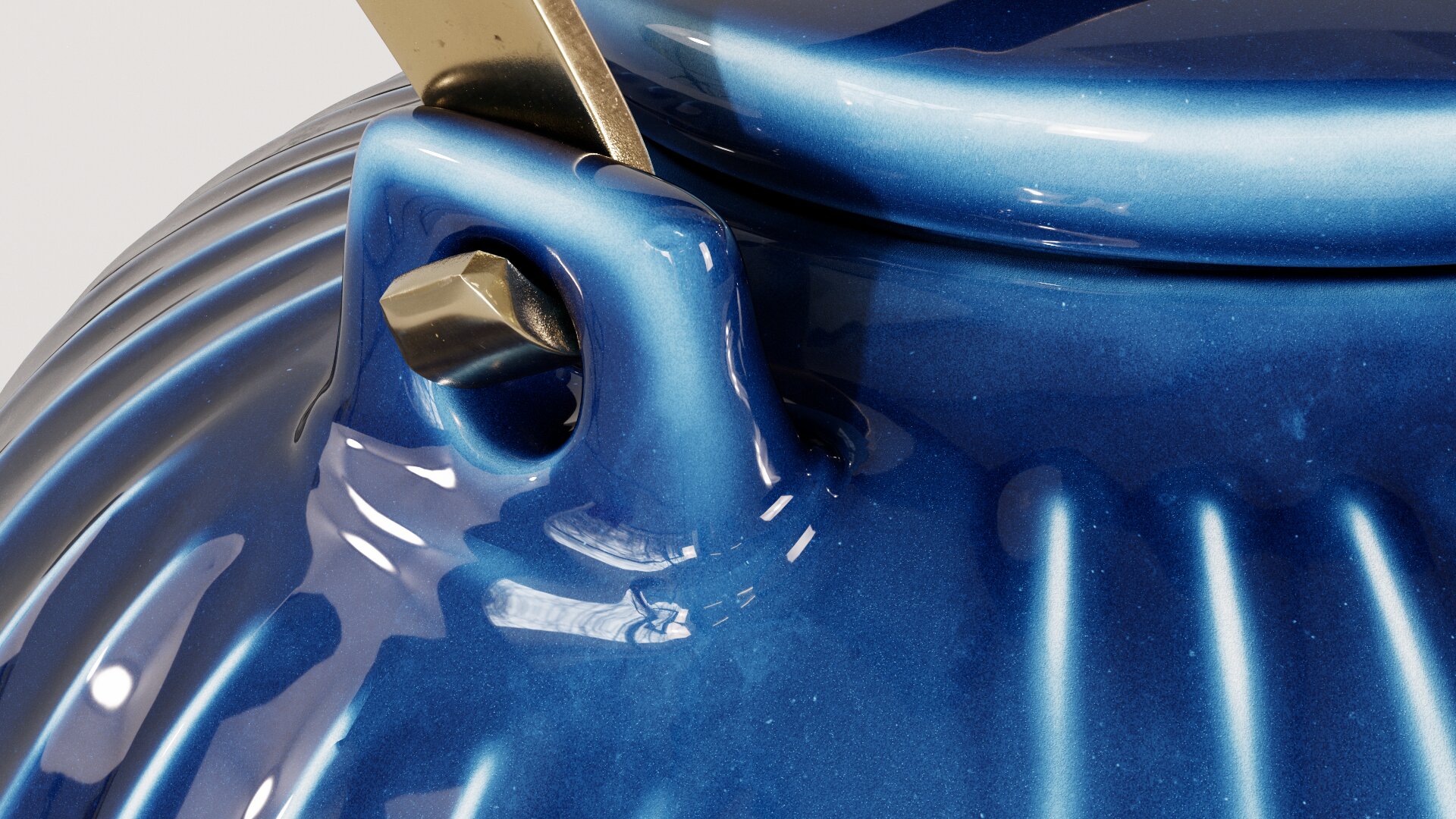}
\put(0,0){\includegraphics[width=0.07\linewidth,trim=270 200 330 138,clip]{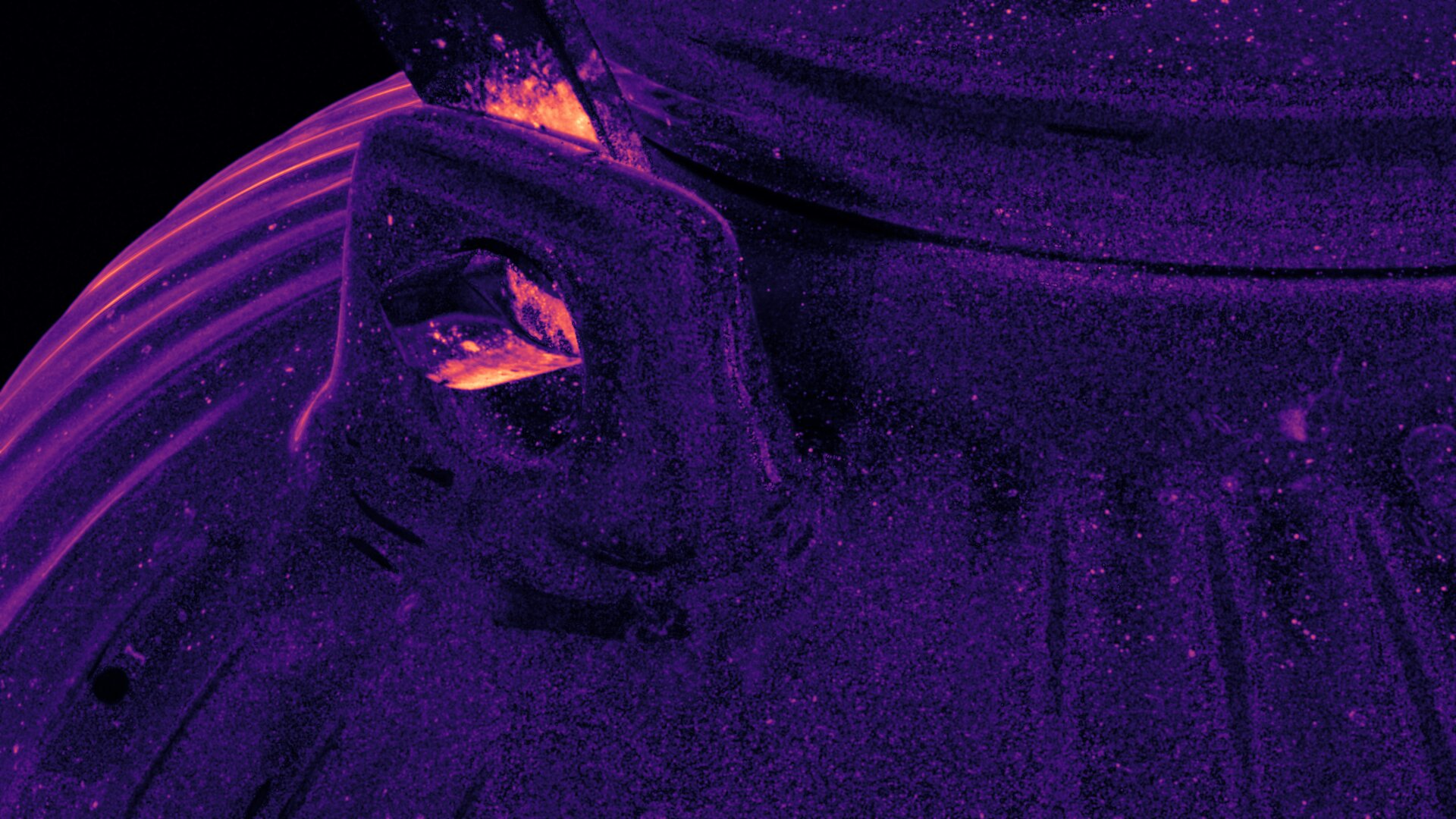}}
\end{overpic}%
&\begin{overpic}[width=0.24575\textwidth,trim=270 200 330 138,clip]{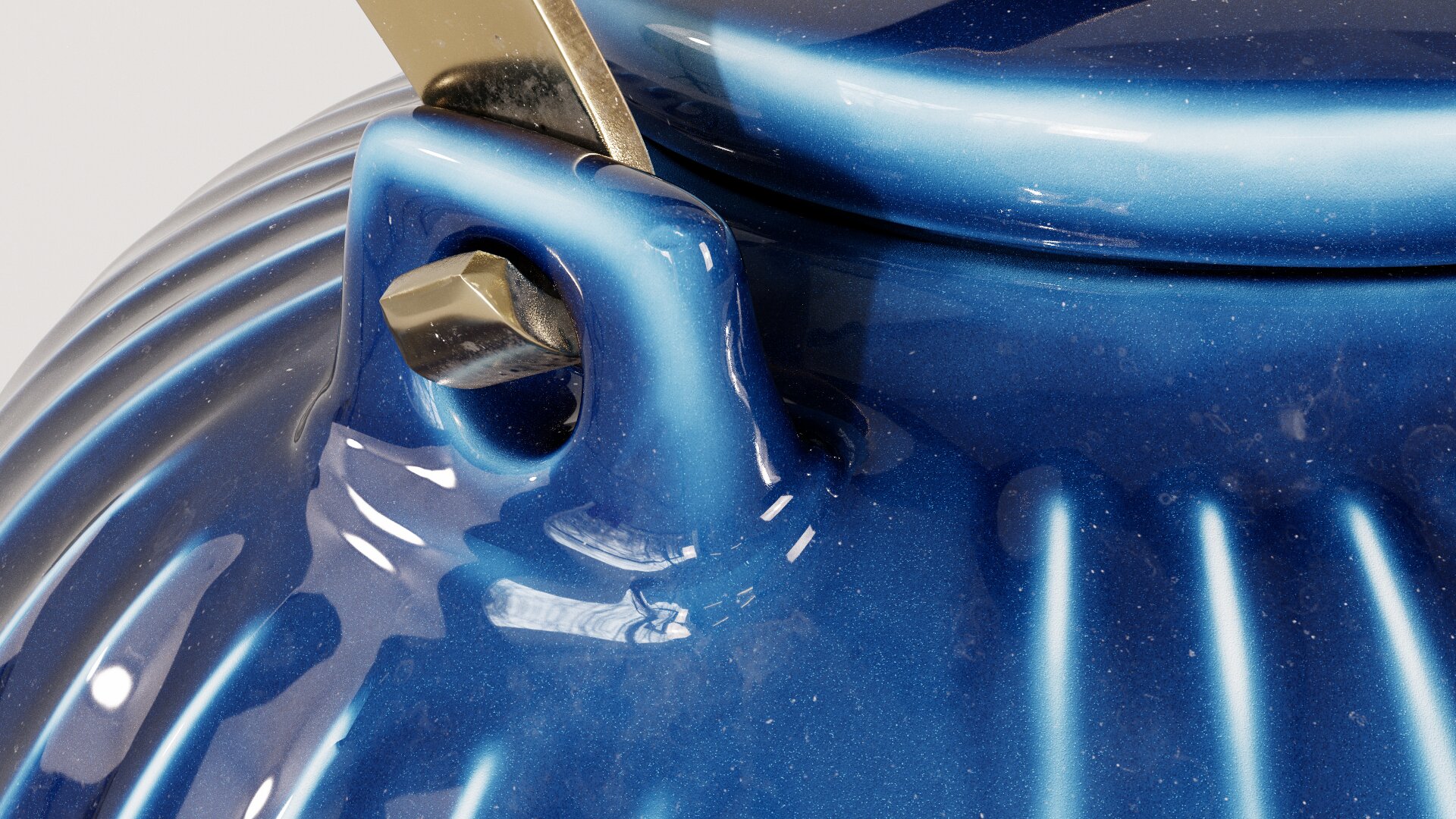}
\end{overpic}%
\\
\begin{overpic}[width=0.24575\textwidth,trim=270 200 330 138,clip]{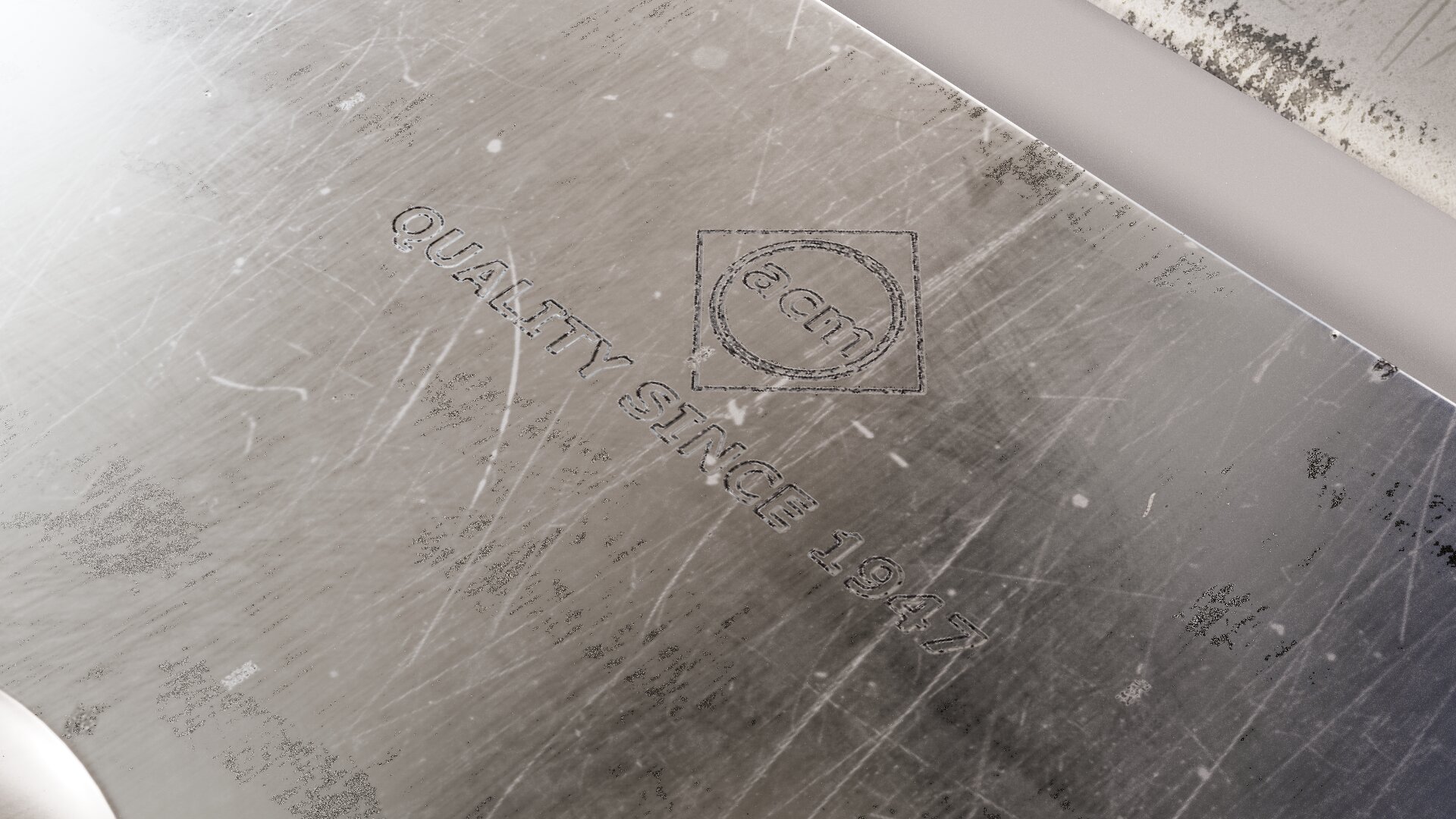}
\put(0,0){\includegraphics[width=0.07\linewidth,trim=270 200 330 138,clip]{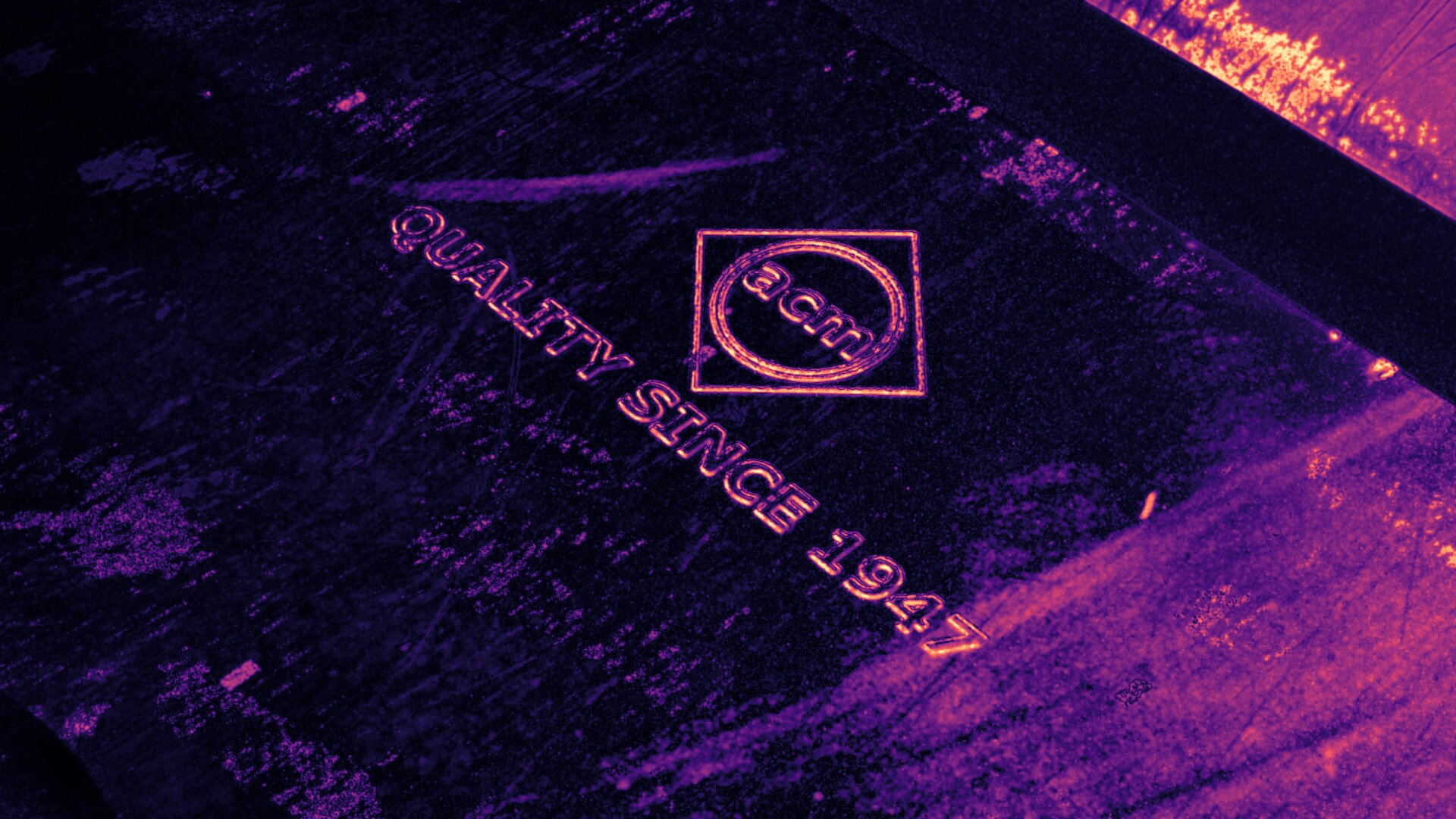}}
\put(-6.0,0){\rotatebox{90}{\hspace{2.5mm}\cheesegraterBlade}}
\end{overpic}%
&\begin{overpic}[width=0.24575\textwidth,trim=270 200 330 138,clip]{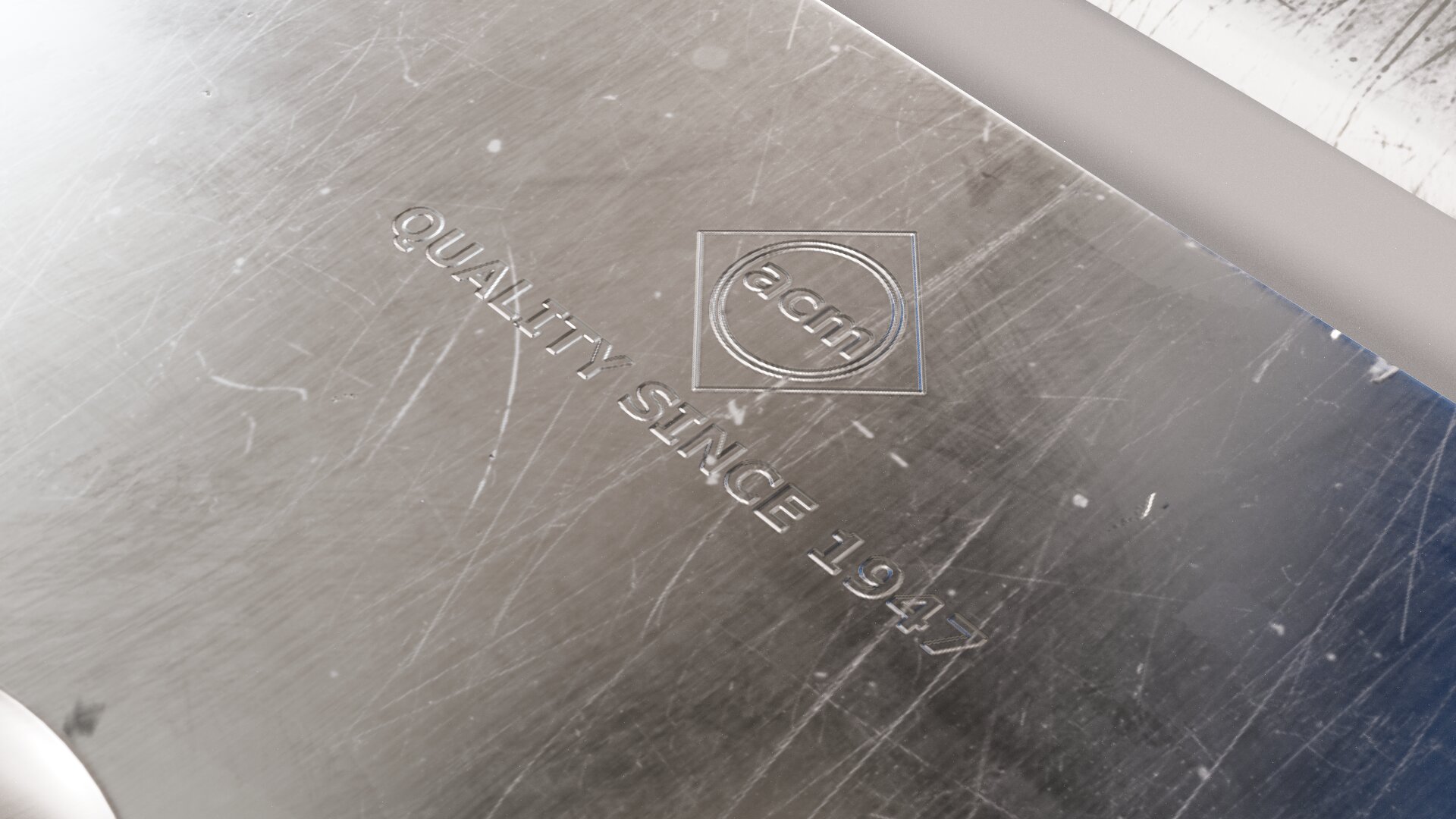}
\put(0,0){\includegraphics[width=0.07\linewidth,trim=270 200 330 138,clip]{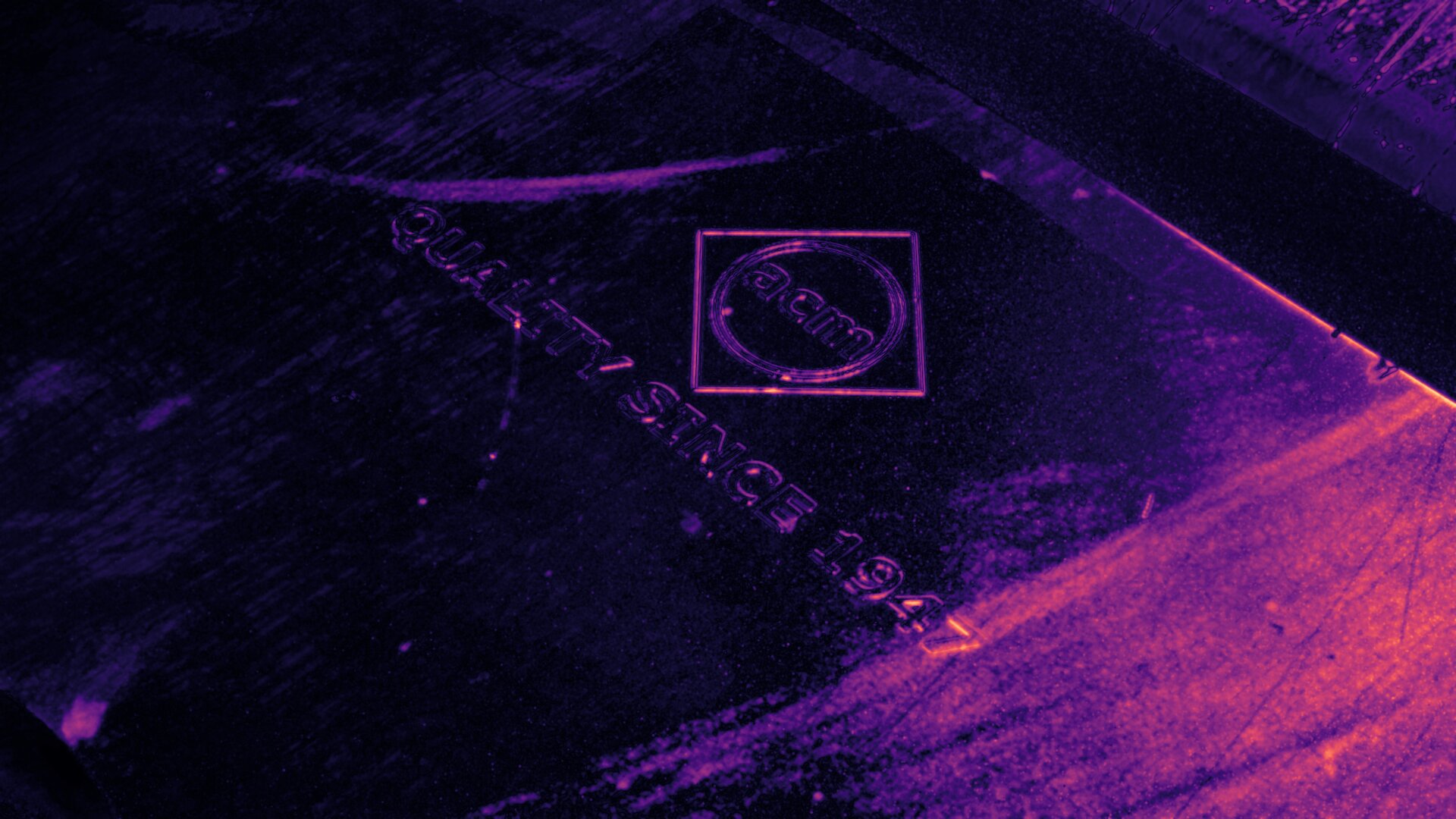}}
\end{overpic}%
&\begin{overpic}[width=0.24575\textwidth,trim=270 200 330 138,clip]{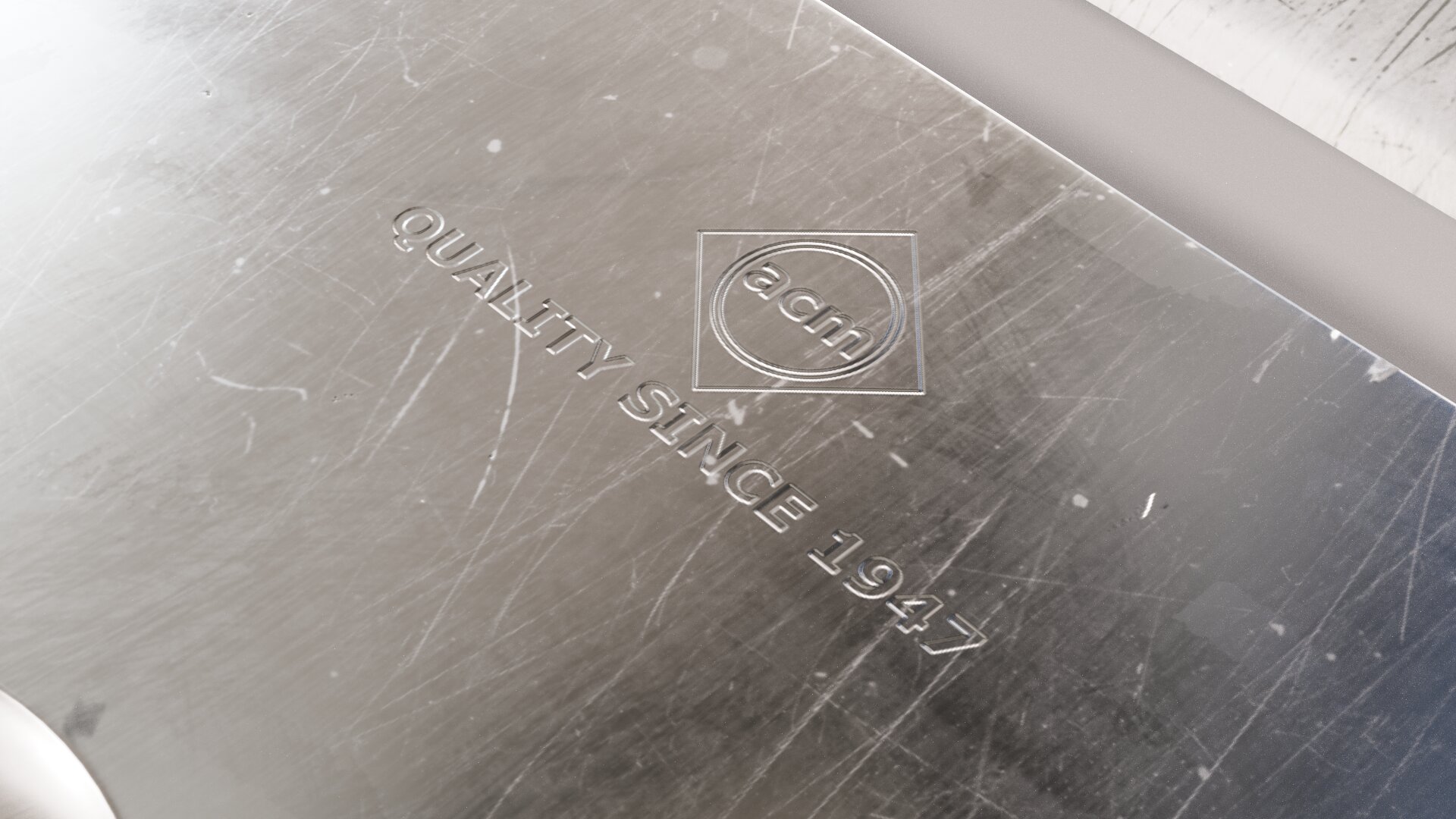}
\put(0,0){\includegraphics[width=0.07\linewidth,trim=270 200 330 138,clip]{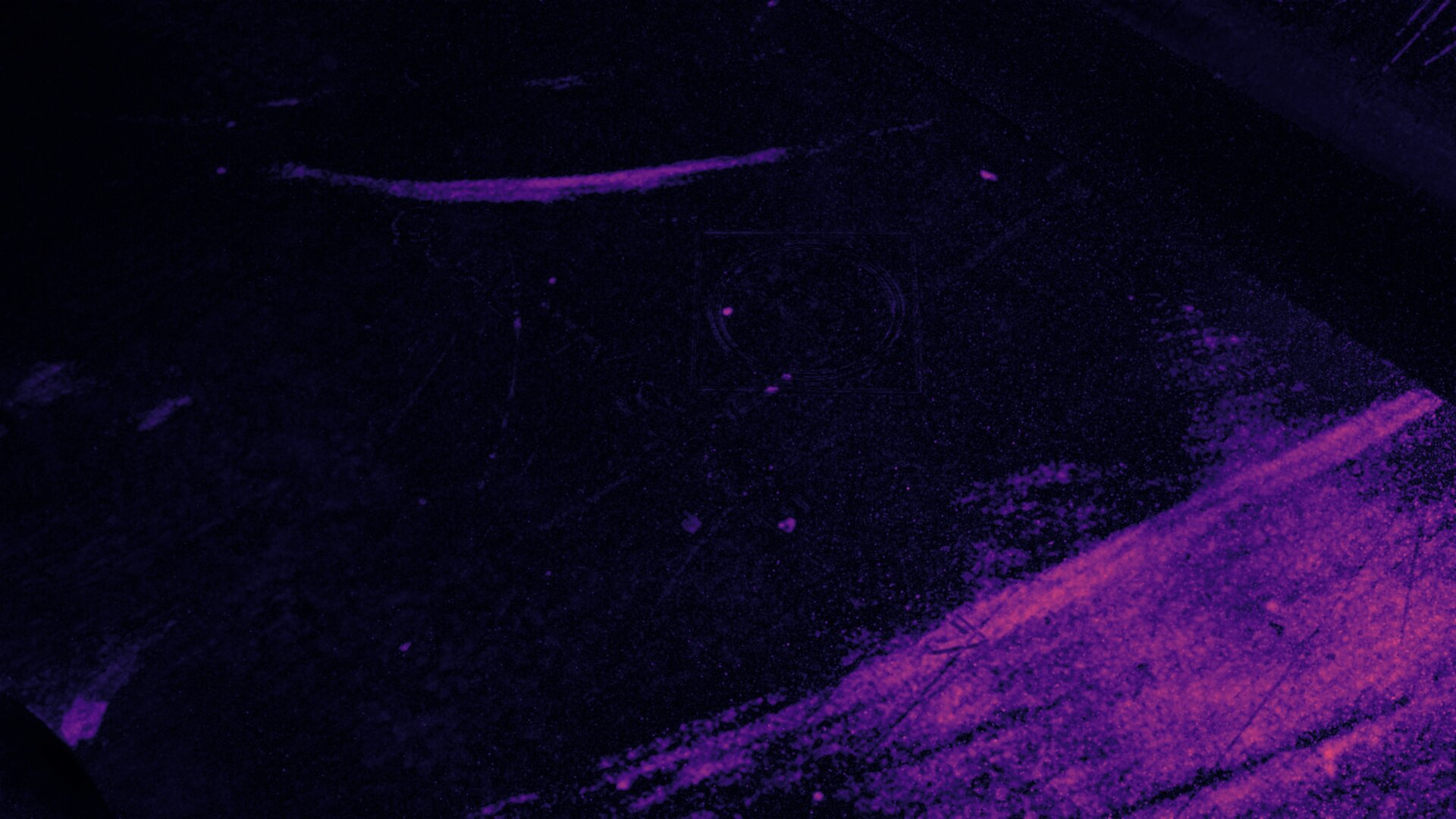}}
\end{overpic}%
&\begin{overpic}[width=0.24575\textwidth,trim=270 200 330 138,clip]{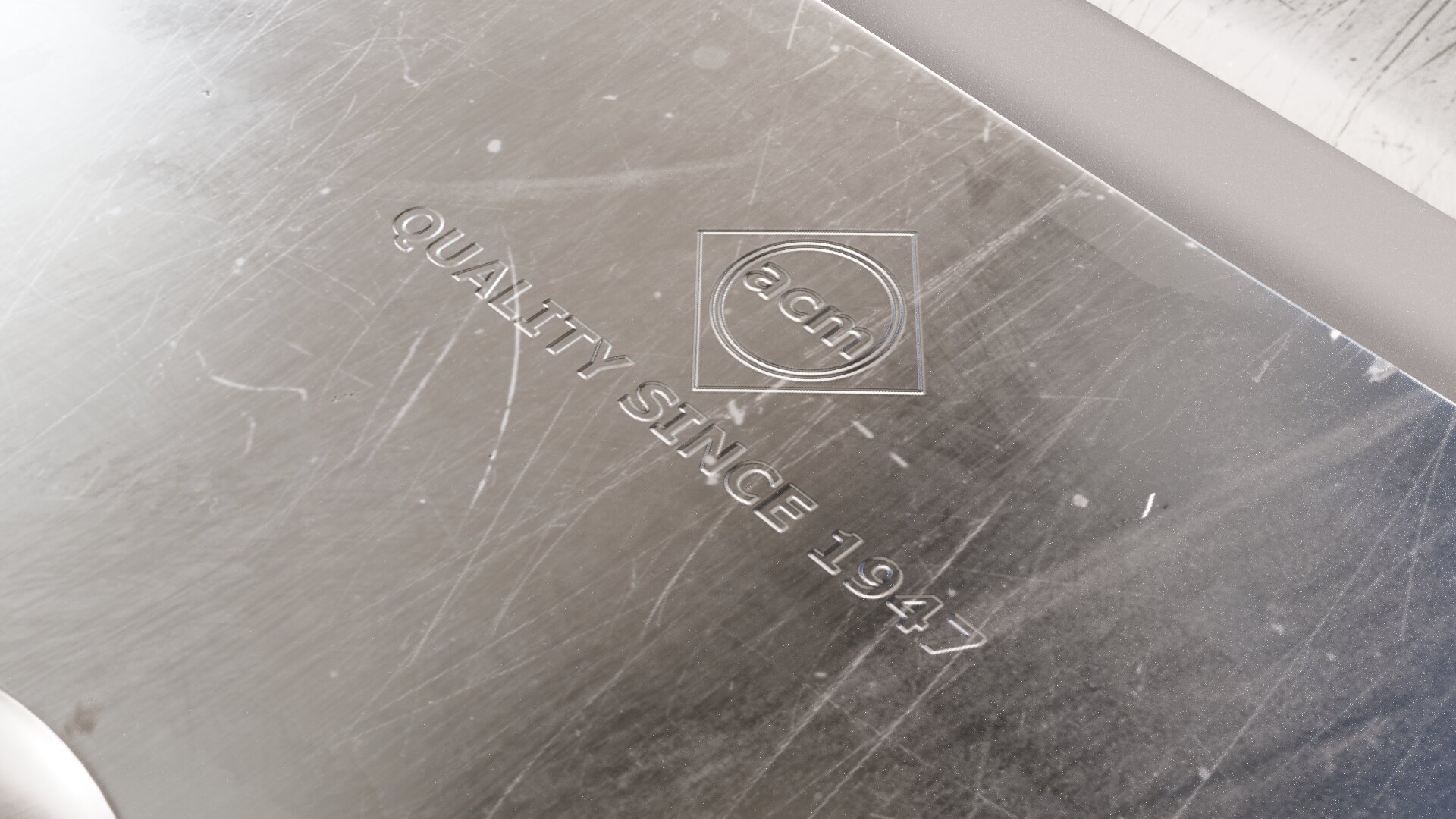}
\end{overpic}%
\\
\begin{overpic}[width=0.24575\textwidth,trim=270 280 330 58,clip]{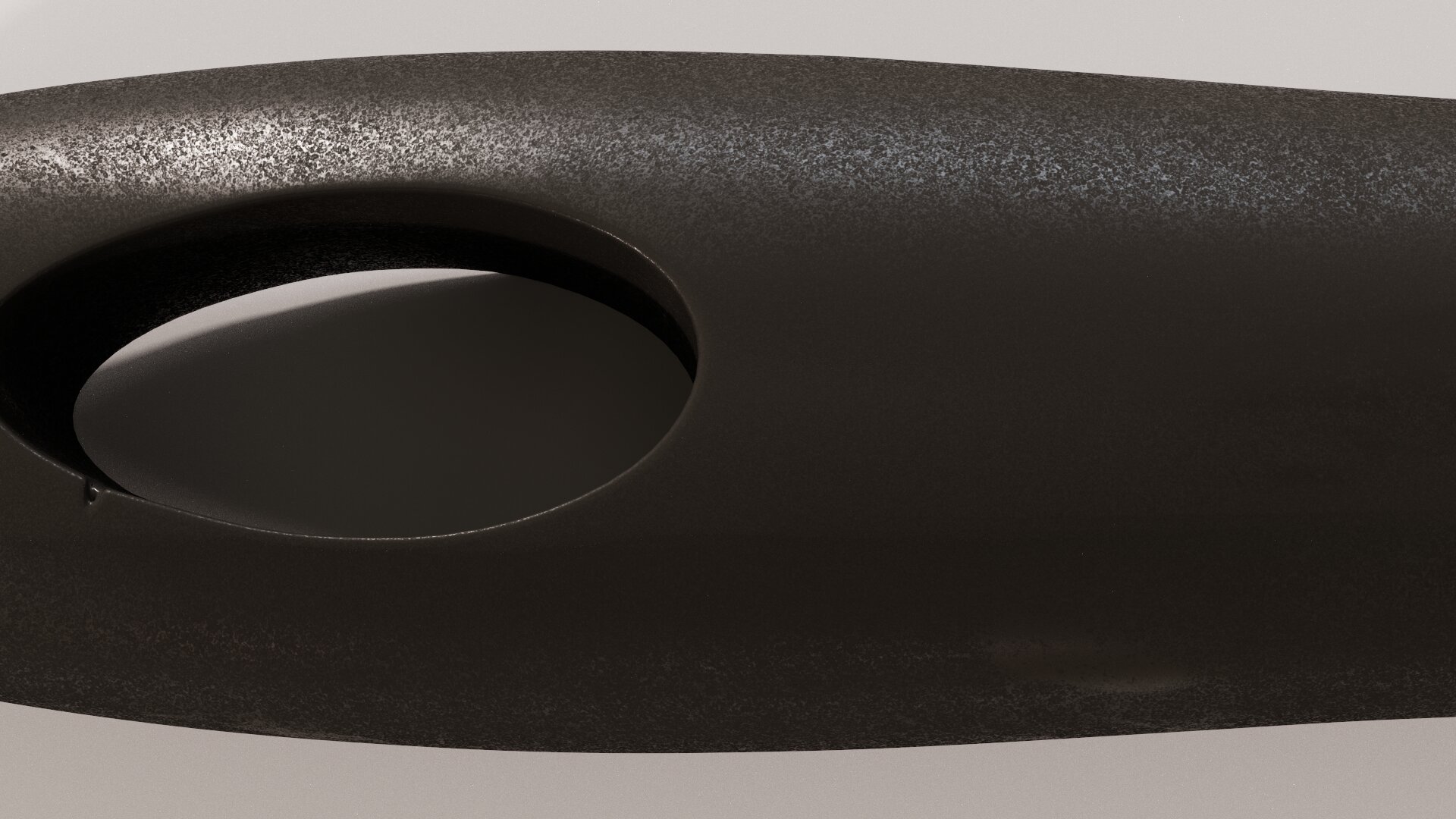}
\put(0,0){\includegraphics[width=0.07\linewidth,trim=270 280 330 58,clip]{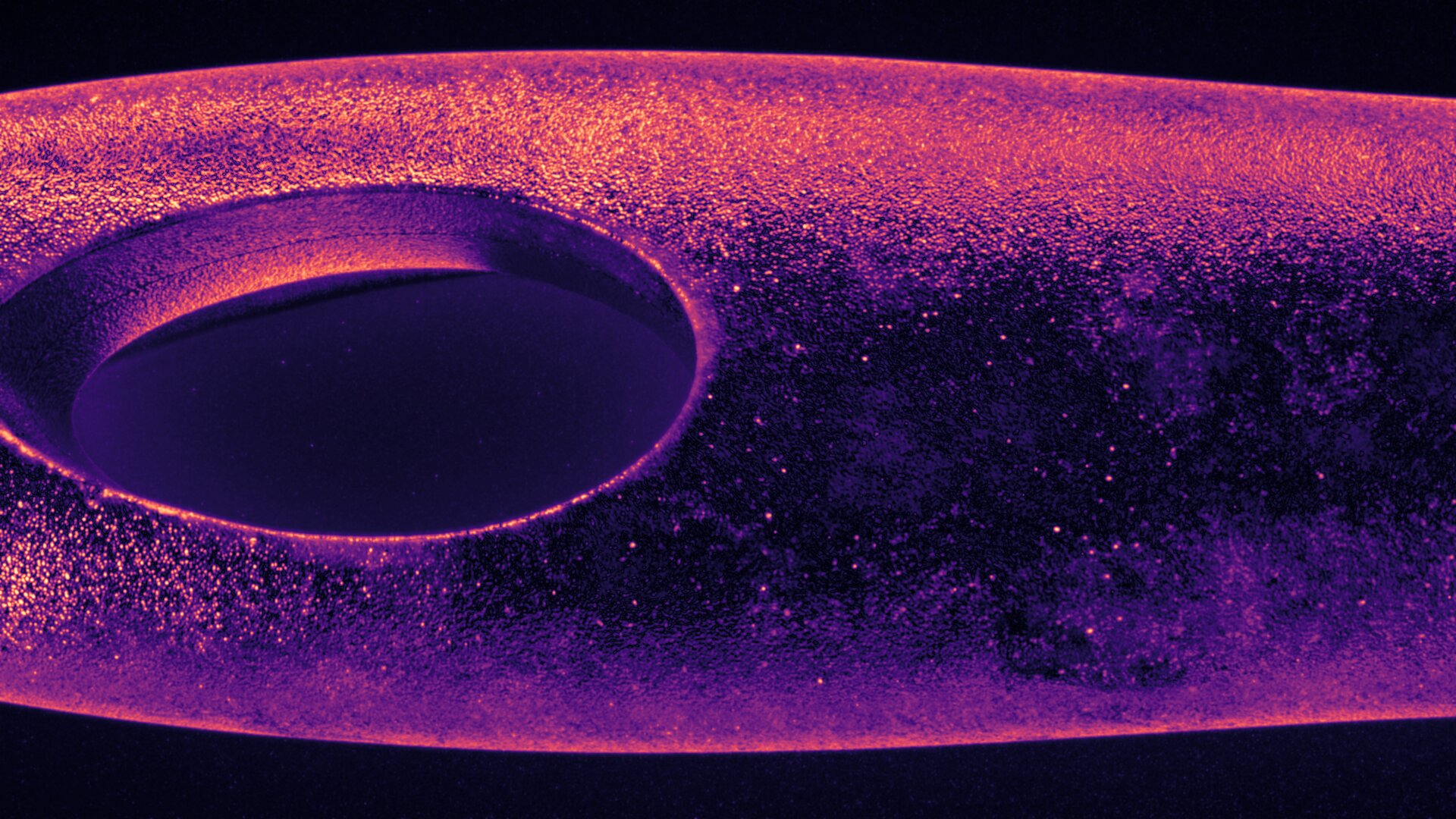}}
\put(-6.0,0){\rotatebox{90}{\hspace{1.5mm}\cheesegraterHandle}}
\end{overpic}%
&\begin{overpic}[width=0.24575\textwidth,trim=270 280 330 58,clip]{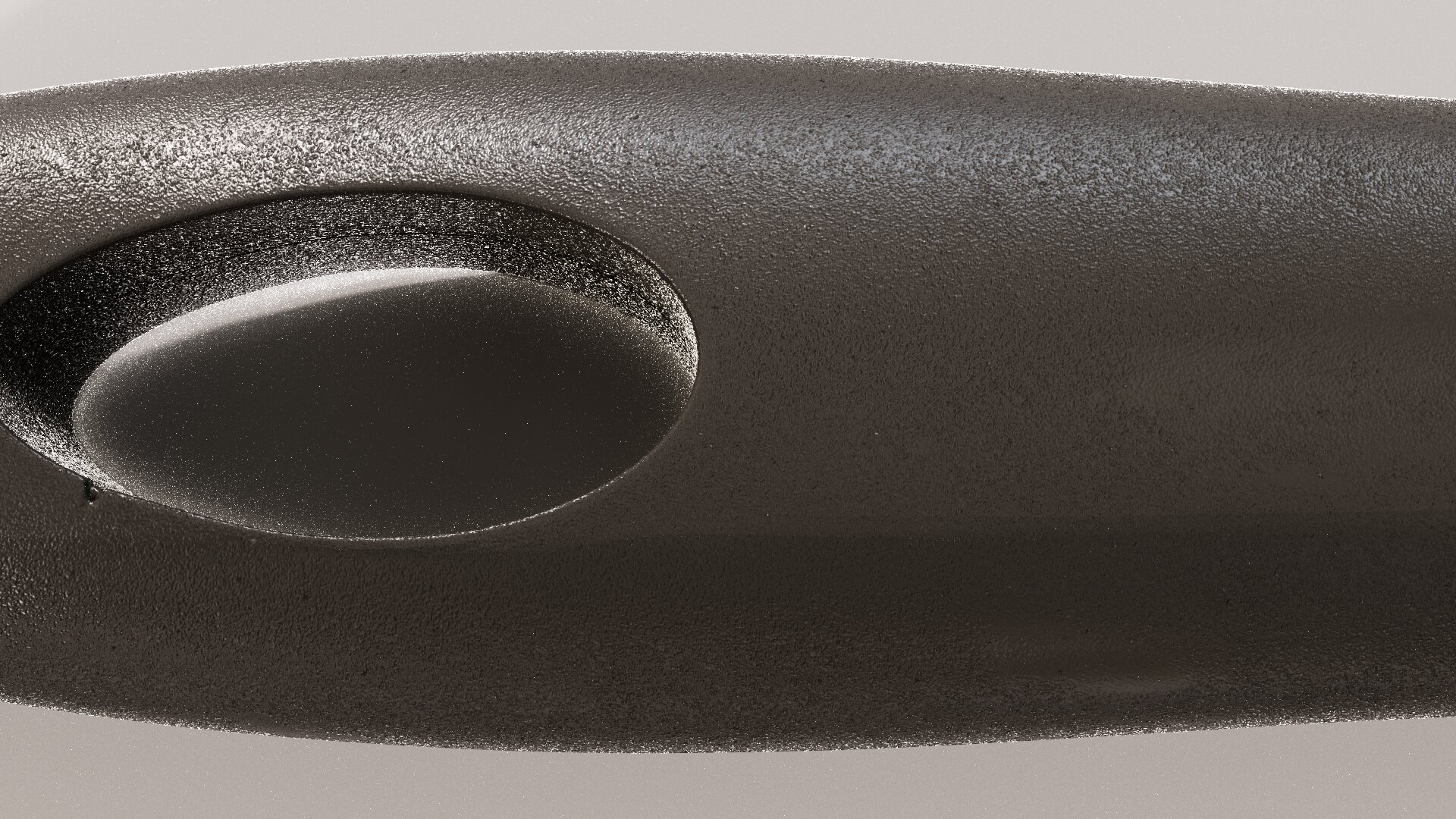}
\put(0,0){\includegraphics[width=0.07\linewidth,trim=270 280 330 58,clip]{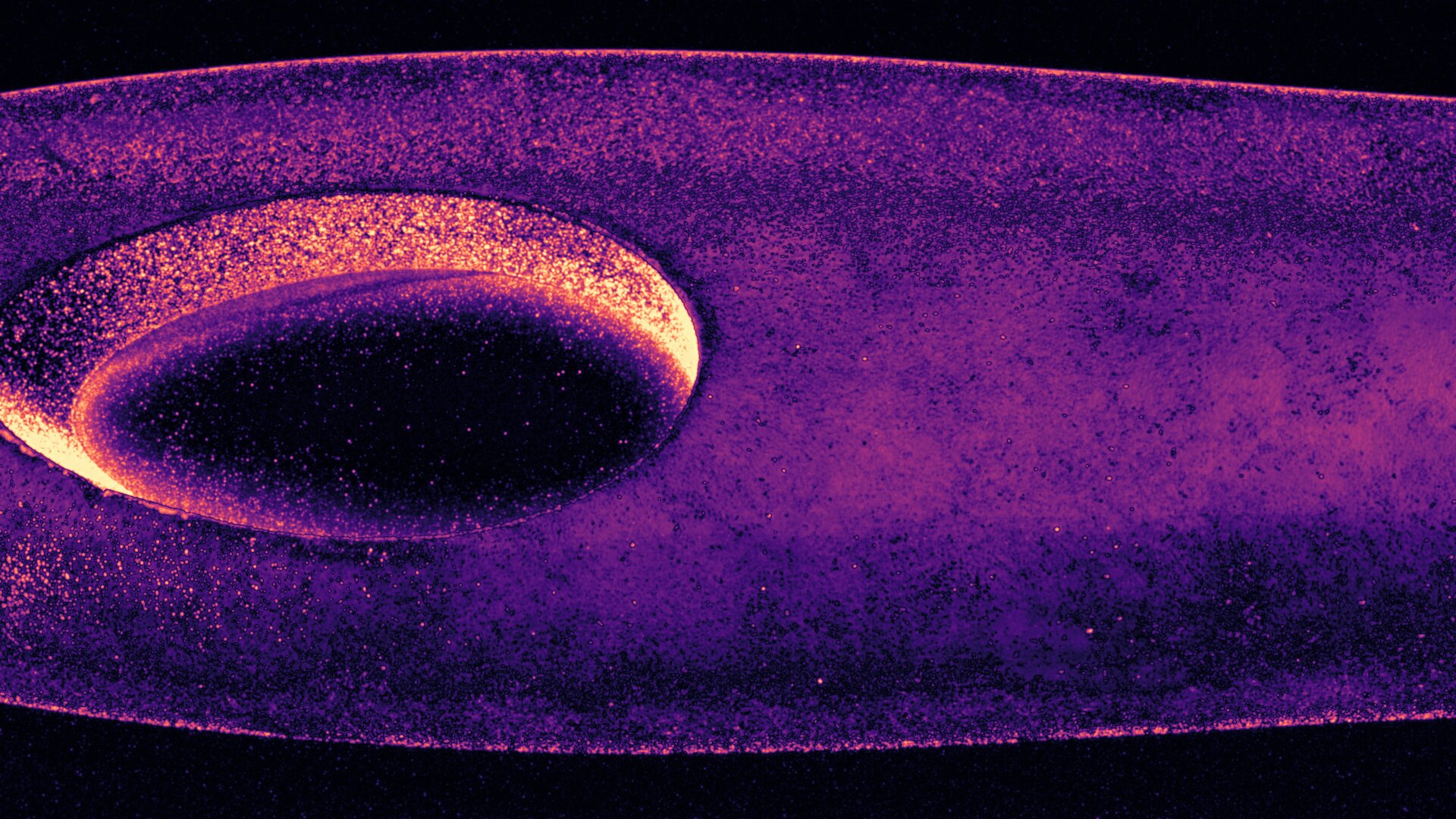}}
\end{overpic}%
&\begin{overpic}[width=0.24575\textwidth,trim=270 280 330 58,clip]{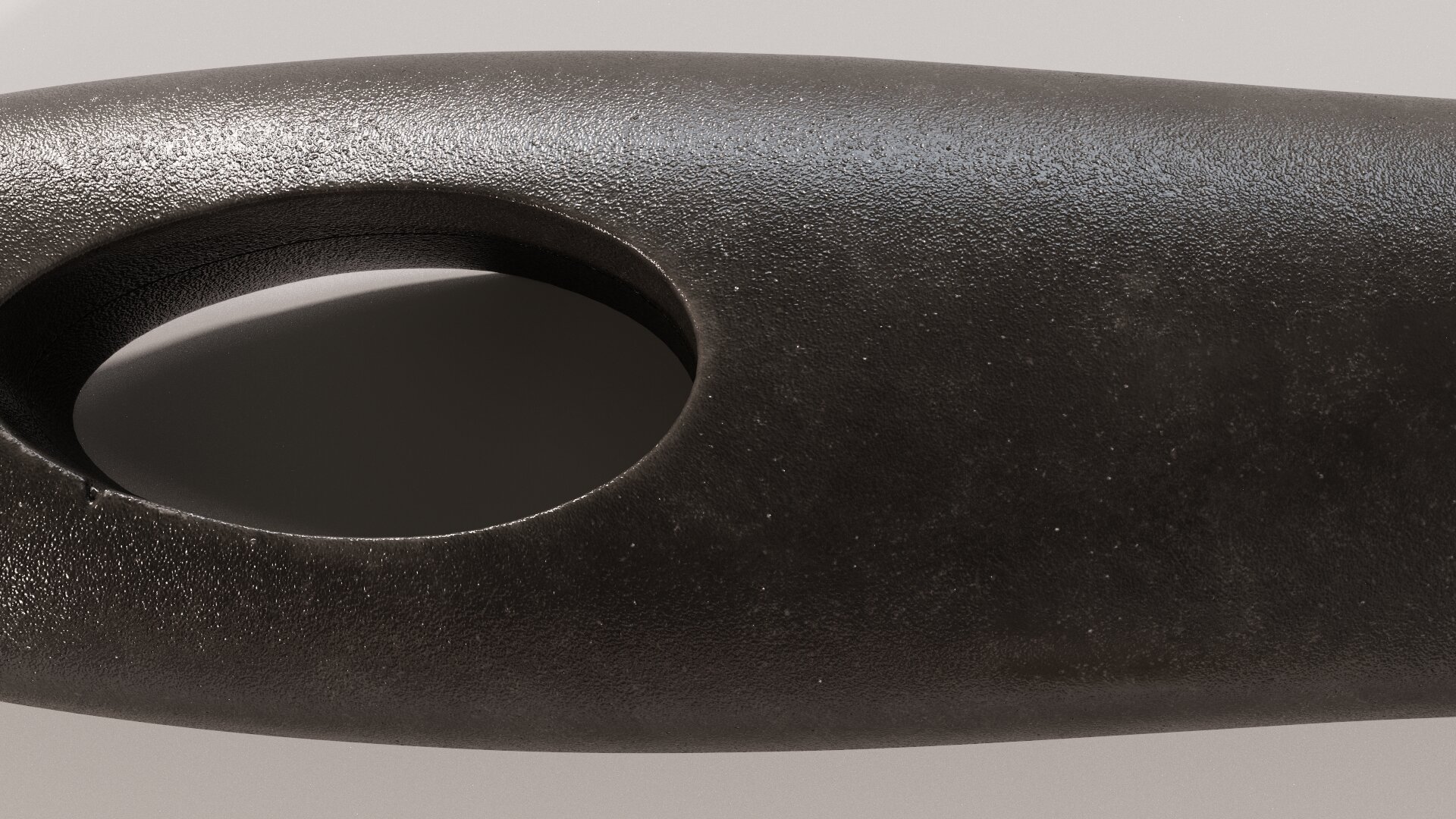}
\put(0,0){\includegraphics[width=0.07\linewidth,trim=270 280 330 58,clip]{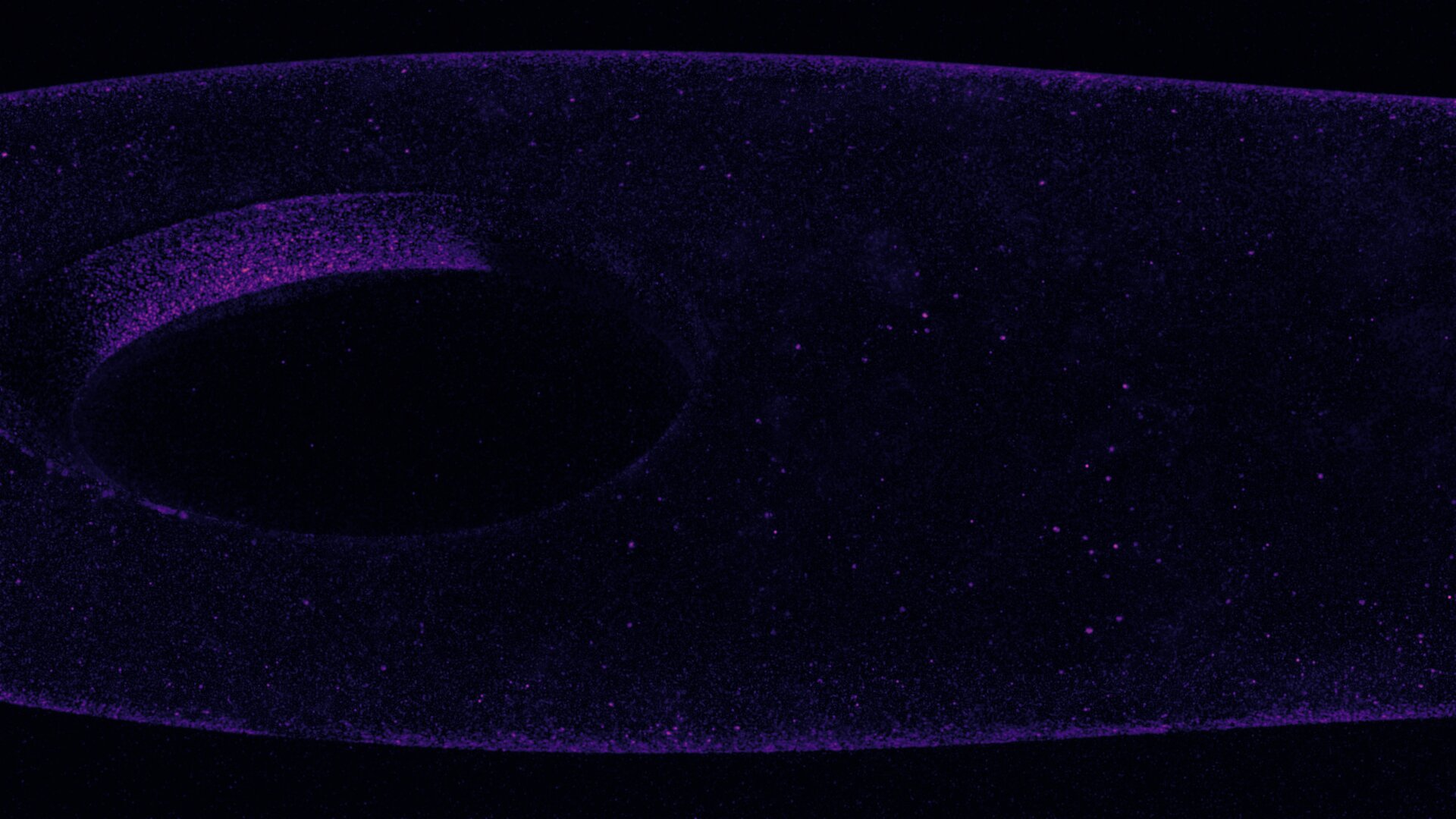}}
\end{overpic}%
&\begin{overpic}[width=0.24575\textwidth,trim=270 280 330 58,clip]{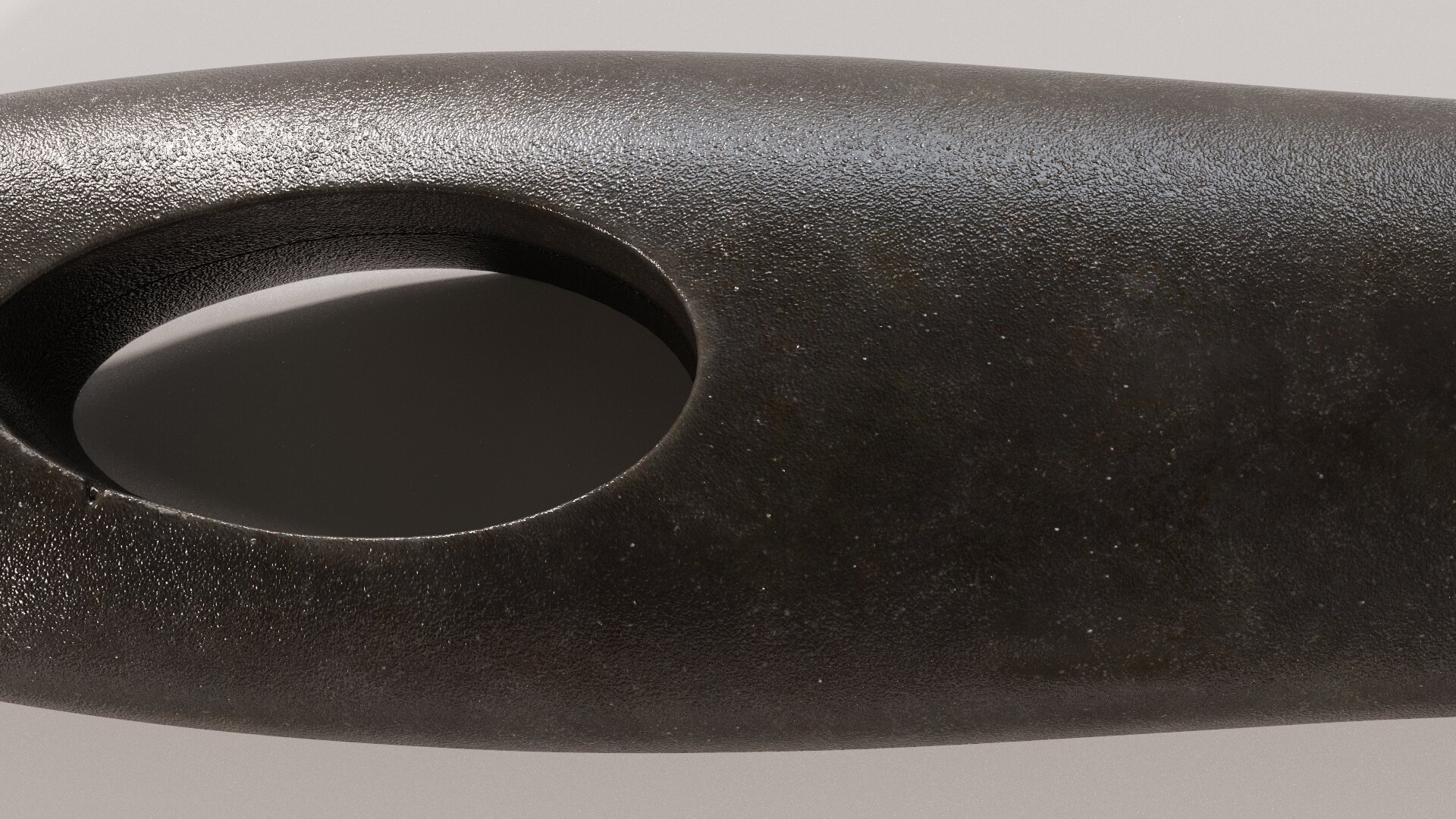}
\end{overpic}%
\\
\end{tabular}

    \caption{
    A qualitative comparison of two ablated variants and our full model at equal amount of training iterations.
    A vanilla MLP decoder with directly optimized latent texture (first column) provides limited quality.
    Training an encoder to produce the latent texture (second column) ensures that texels with identical appearance feature identical latent codes, easing the decoding to BRDF values.
    Augmenting the MLP decoder with an explicit transformation of directions to learned shading frames---our full model (third column)---further improves the reproduction of the reference image (last column).
    The bottom left corners show images of the \FLIP difference metric.
    The models without the shading frame extractor (first two columns) were equipped with an extra first layer with 8 neurons to roughly match the number of parameters of the full model.
    }\label{fig:ablation}
  \end{figure*}

Due to the difference in focus, it is hard to compare our work to previous approaches \emph{directly}.
Instead, we compare to two ablated variants of our model in \autoref{tab:metrics-ablation} and \autoref{fig:ablation}, and relate them to corresponding components in prior work.

\paragraph{Vanilla MLP decoder with latent texture.}
The basic variant utilizes only a hierarchical latent texture and a vanilla MLP decoder. As such, there is no explicit rotation to shading frames in the decoder, and the texels of the texture are optimized \emph{directly} via backpropagation.
This variant can be viewed as the decoder of \citet{Sztrajman2021} extended to handle spatial variations using a hierarchical neural texture~\cite{Thies2019}.
The model and the training procedure is also conceptually close to the NeuMIP model~\cite{Kuznetsov2021}, except that NeuMIP additionally features a UV-offsetting module for handling displaced surfaces.
The results of this variant (\autoref{fig:ablation}, first column) fail to correctly reproduce the spatial details of the reference material 
\revision{due to the vast number of latent texels that need to be optimized. We further analyze the scaling of latent-texture optimization with increasing resolution in \autoref{sec:latent-texture-optimization}.}
\paragraph{Latent texture encoder.}
The second column in \autoref{fig:ablation} shows the benefits of adding the encoder (\autoref{sec:encoder}).
The texture detail is reproduced more faithfully due to two main reasons.
First, the encoder prevents situations where multiple texels with identical BRDF end up with different latent codes after optimization. Such surjective mapping of latents to BRDF values often occurs in the basic model (first column) depleting the modeling capacity of the decoder.
Second, the encoder amortizes each training record over many latent texels instead of optimizing a single latent texel.

While the spatial variations are captured well \revision{in this particular example}, the decoder is unable to \revision{additionally} capture the narrow reflection lobe of the \teapot{} ceramic \revision{even though it was correctly captured by the vanilla MLP decoder}. This suggests that the model has insufficient modelling capacity to accurately reproduce both the spatial variations and the high-frequency reflections. This can be alleviated by increasing the size of the decoder.

Our encoder-decoder architecture is reminiscent of the auto-encoder used by \citet{Rainer2019} for compressing BTFs, with the key distinction that we chose to encode the material parameters (albedo, roughness, normal, etc.) instead of encoding the reflectance measurements.
This allows our system to further improve scaling to very high-resolution textures, since the encoder can exploit the redundancy in the material parameterization.

\paragraph{Transformation to learned shading frames.}
In the third column of \autoref{fig:ablation}, we prepend the MLP decoder with the transformation of directions to two learned shading frames, which are 
extracted from the latent code using an extra trainable layer with 12 neurons.
This constitutes our complete model.
As discussed in \autoref{sec:rotation-op}, performing a multiplicative operation on the inputs explicitly spares the MLP from approximating it using its non-linear layers.
The quality of the results improves, including effects that are not necessarily related to normal mapping.
This suggests that modeling capacity retained by the explicit shading frame transformation is ``invested'' in better capturing the shape and spatial variations of the BRDF.

\begin{table}[t]
    \centering
    \small
    \caption{
      Image error metrics averaged over the four images in \autoref{fig:ablation} for each of the three compared variants. Material-specific statistics are included in the supplemental material.
    }\label{tab:metrics-ablation}
    \begin{tabular}{lrrr}
\toprule
{} & \shortstack[r]{Vanilla \\ MLP} & \shortstack[r]{with \\ encoder} & \shortstack[r]{with frame \\ transform} \\
\midrule
Mean \FLIP & 0.2390 & 0.1956 & 0.0815 \\
Mean abs. error & 0.0769 & 0.0652 & 0.0183 \\
Mean sqr. error & 0.0682 & 10.1933 & 0.0057 \\
Mean rel. abs. error & 0.2177 & 0.3439 & 0.0656 \\
Mean rel. sqr. error & 0.0798 & 265.4018 & 0.0090 \\
SMAPE & 0.2670 & 0.2397 & 0.0713 \\
\bottomrule
\end{tabular}

    \vspace{-2mm}
\end{table}

\subsection{Filtering} 
\begin{figure*}[t]
    \setlength{\tabcolsep}{0.002\textwidth}%
\renewcommand{\arraystretch}{1}%
\footnotesize%
\begin{tabular}{ccccccc}
Reference & Footprint-based & Level 0 & Level 1 & Level 2 & \hspace{-1.5mm}$\cdots$\hspace{-1.5mm} & Level 5 \\
\includegraphics[width=0.1625\textwidth]{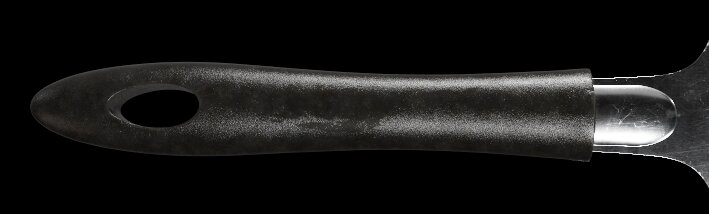}%
&\includegraphics[width=0.1625\textwidth]{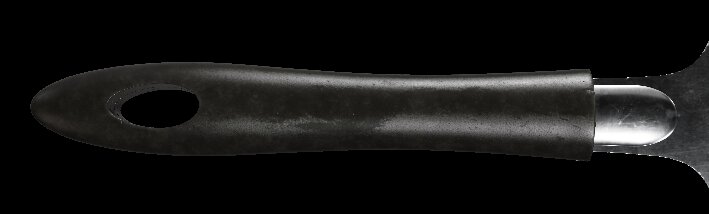}%
&\includegraphics[width=0.1625\textwidth]{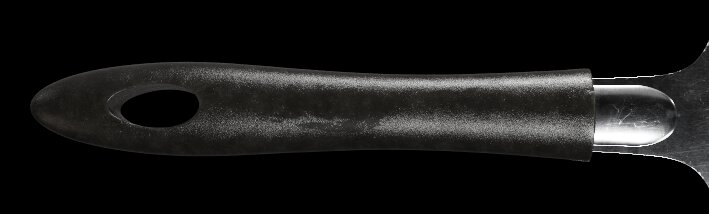}%
&\includegraphics[width=0.1625\textwidth]{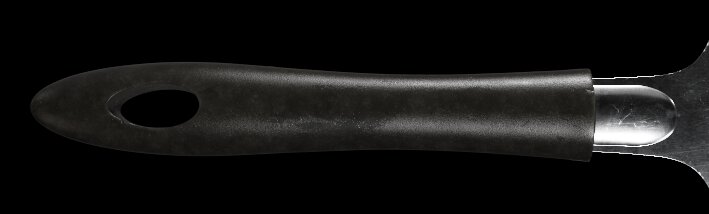}%
&\includegraphics[width=0.1625\textwidth]{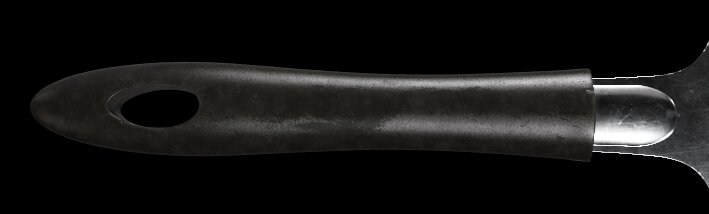}%
&&\includegraphics[width=0.1625\textwidth]{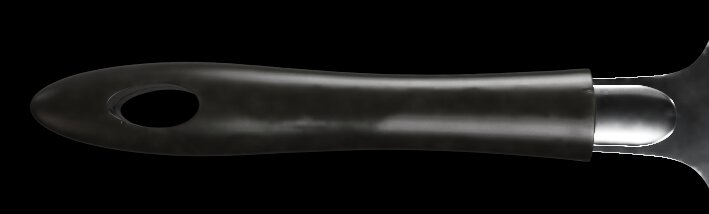}%
\\
\includegraphics[width=0.1625\textwidth]{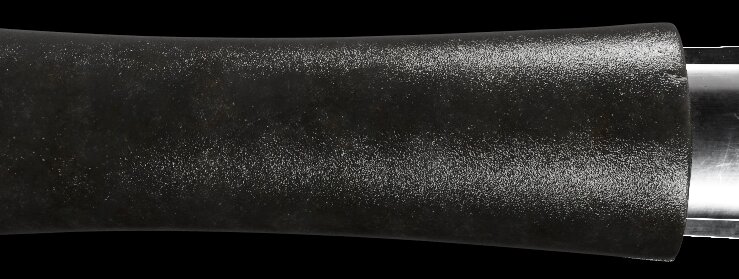}%
&\includegraphics[width=0.1625\textwidth]{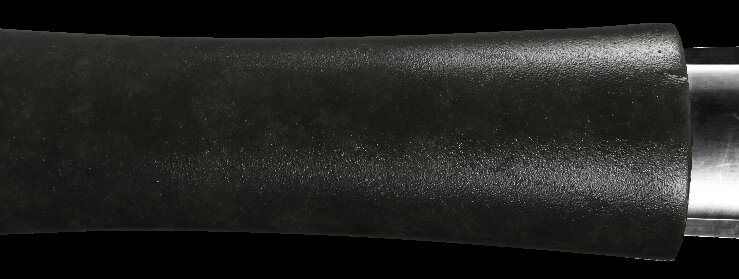}%
&\includegraphics[width=0.1625\textwidth]{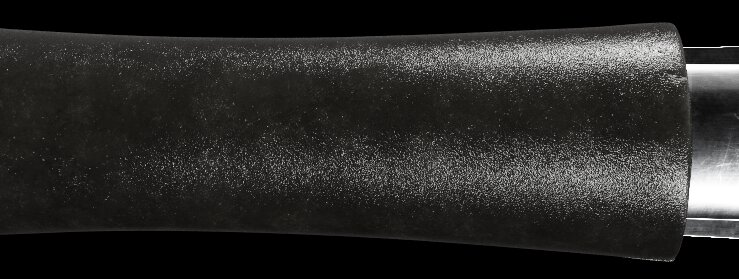}%
&\includegraphics[width=0.1625\textwidth]{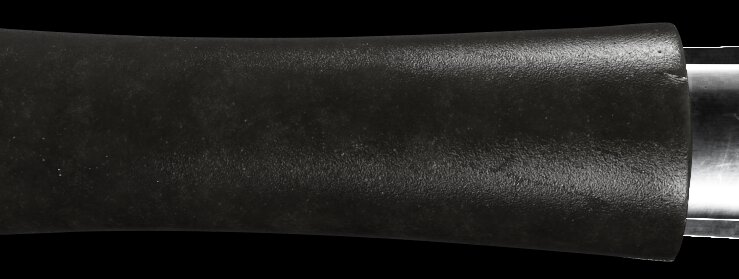}%
&\includegraphics[width=0.1625\textwidth]{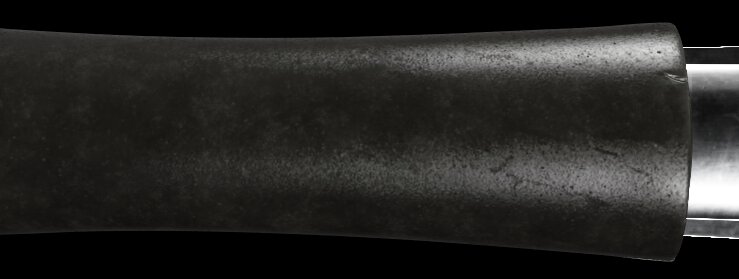}%
&&\includegraphics[width=0.1625\textwidth]{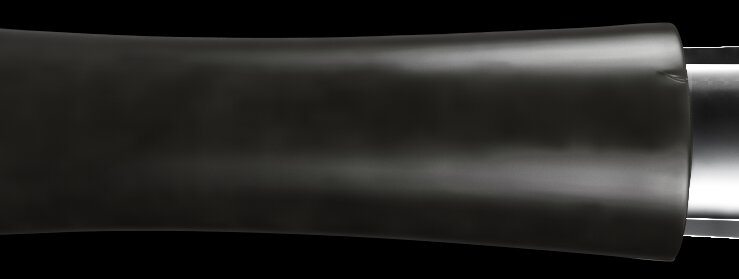}%
\\
\includegraphics[width=0.1625\textwidth]{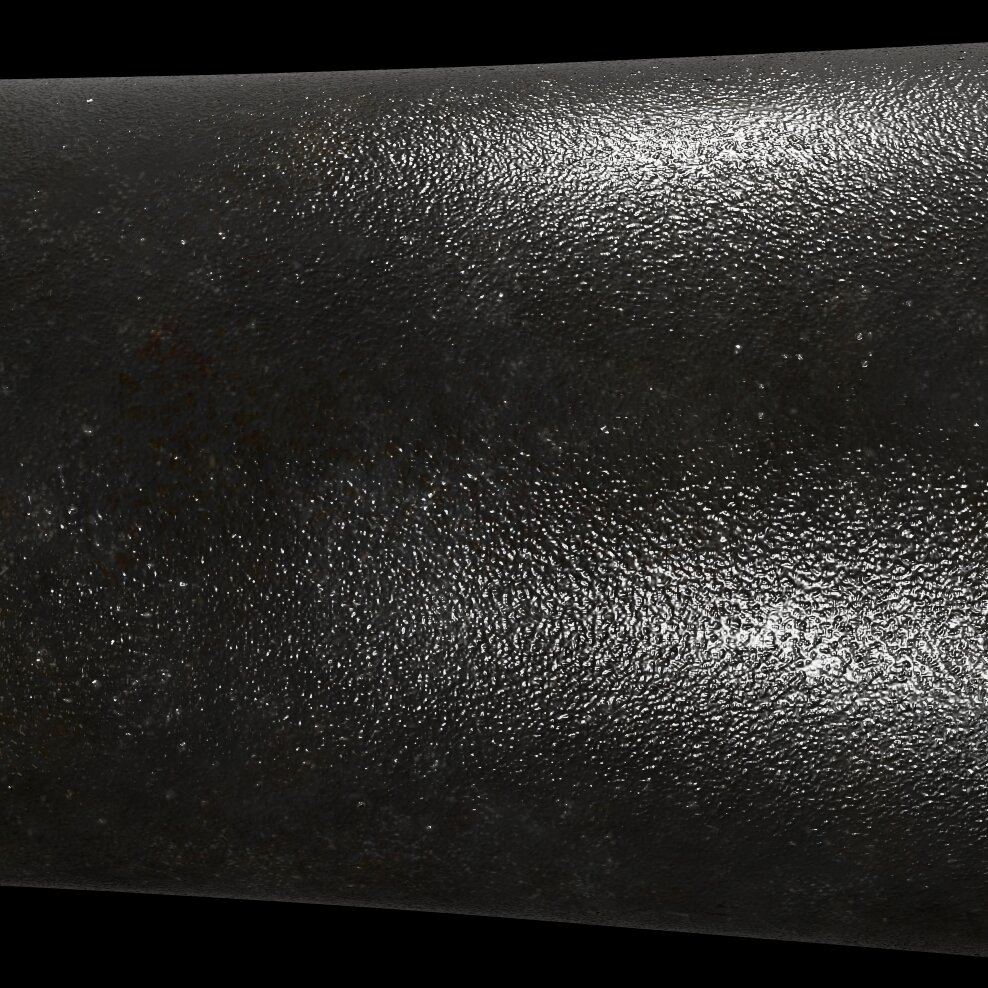}%
&\includegraphics[width=0.1625\textwidth]{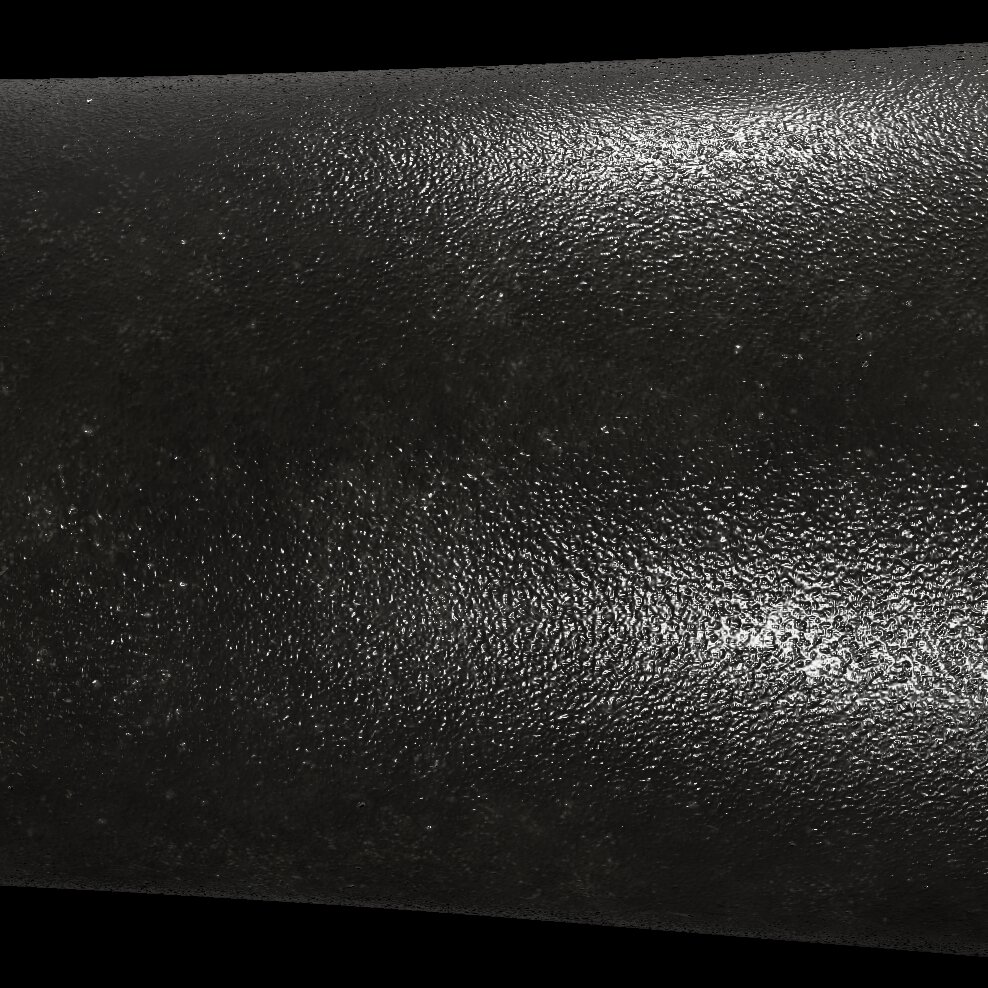}%
&\includegraphics[width=0.1625\textwidth]{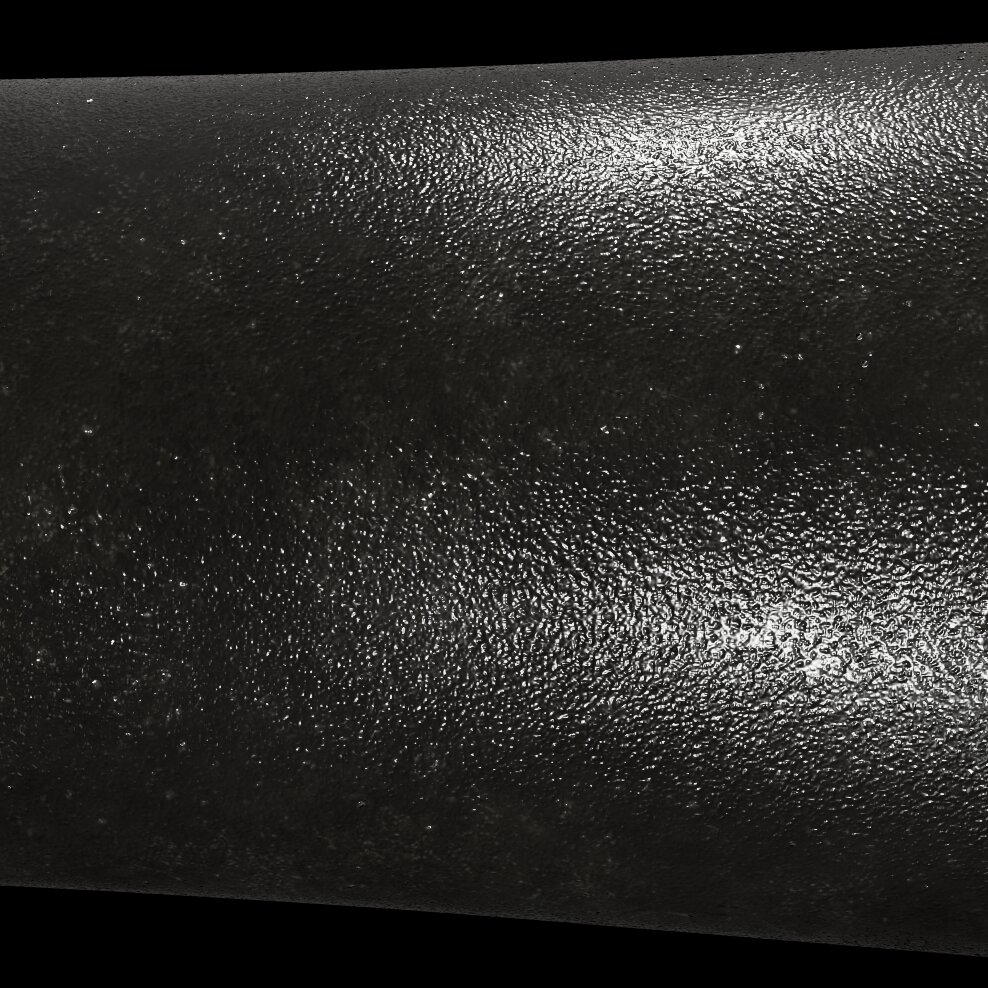}%
&\includegraphics[width=0.1625\textwidth]{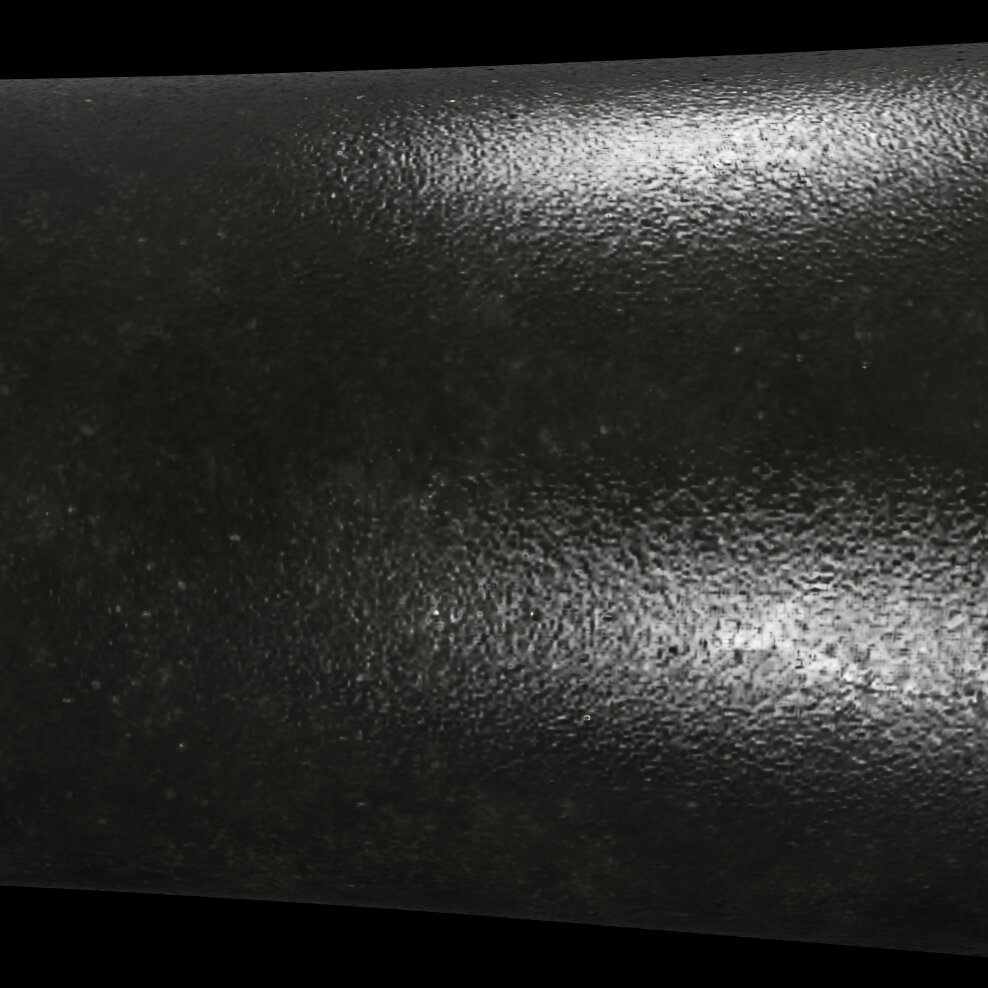}%
&\includegraphics[width=0.1625\textwidth]{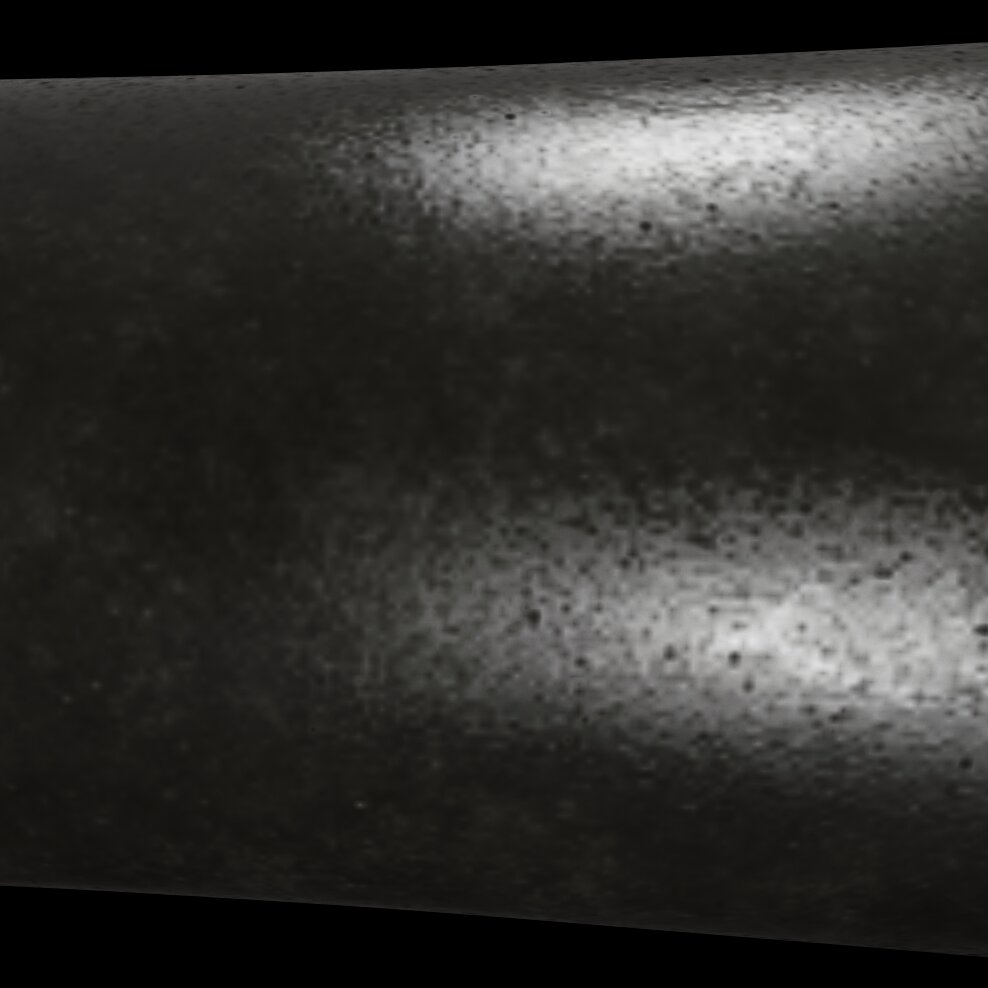}%
&&\includegraphics[width=0.1625\textwidth]{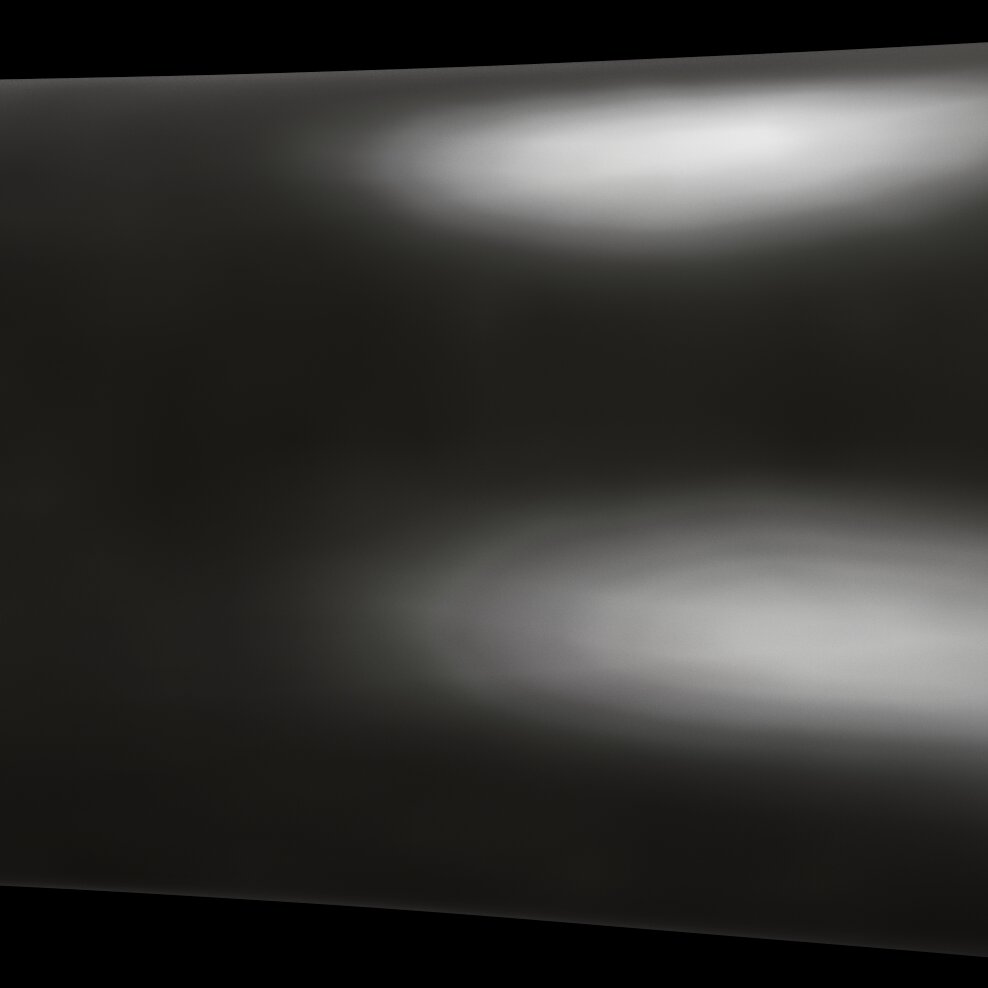}%
\end{tabular}

    \caption{We evaluate the quality of our filtering by comparing footprint-based level selection to fixed latent pyramid levels (rendered with supersampling) on the \cheesegrater\ asset at different distances. Up close, coarser levels show loss of small detail such as glints, which reflects in our filtered result. This is not the case for level 0, which is a near perfect match to the ground truth (at the cost of aliasing). From afar, all levels average to visually similar appearance.
    }\label{fig:fixed-lod}
\end{figure*}

We evaluate the quality of our filtering in \autoref{fig:fixed-lod} by comparing individual levels of the latent pyramid to ground truth rendered with supersampling. Our filtered model is a good match up close, but loses small details from a medium distance. This is because latent optimization does not work as well for coarser levels as it does for level 0 and slightly overblurs the result. This may be compensated by biasing our level selection towards finer MIP levels, at the cost of some aliasing. From afar, all levels have a similar appearance.

\subsection{Latent texture optimization}
\label{sec:latent-texture-optimization}

\begin{figure}[t]
    \setlength{\tabcolsep}{1pt}
    \footnotesize
    \centering
    \begin{overpic}[width=1\linewidth]{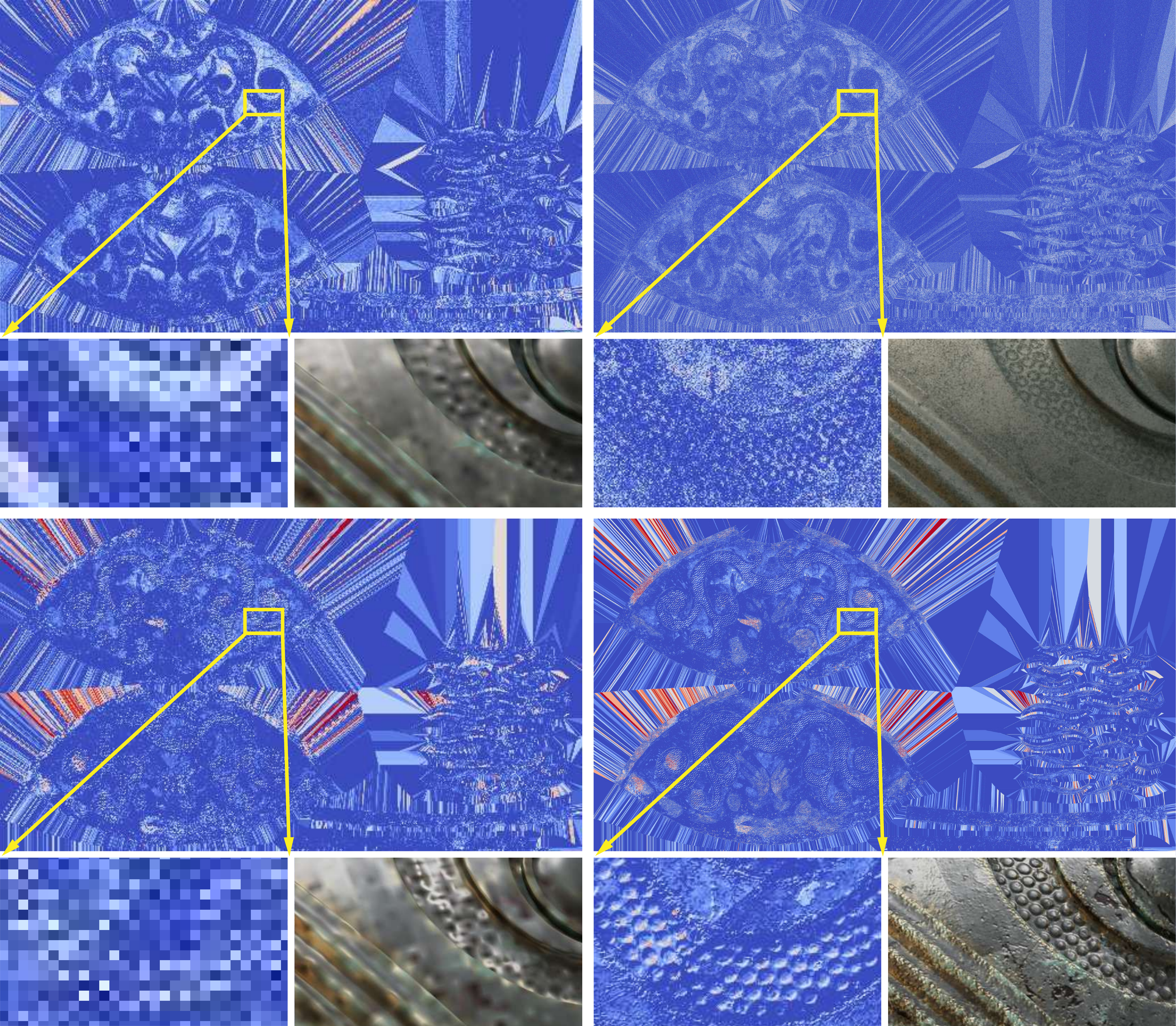}
        \put(19,88){512 $\times$ 512}
        \put(71,88){4k $\times$ 4k}
        \put(-3,24){\rotatebox{90}{Encoder}}
        \put(-3,61){\rotatebox{90}{Direct optimization}}
        \put(1,1){\textcolor{white}{zoom-in}}
        \put(1,45){\textcolor{white}{zoom-in}}
        \put(51,1){\textcolor{white}{zoom-in}}
        \put(51,45){\textcolor{white}{zoom-in}}
        \put(26,1){\textcolor{white}{render}}
        \put(26,45){\textcolor{white}{render}}
        \put(76,1){\textcolor{white}{render}}
        \put(76,45){\textcolor{white}{render}}
    \end{overpic}
    \caption{
        Latent textures of the \inkwell{} asset. Direct optimization (top row) works well for small textures (left) but struggles with high resolutions (right) as independently optimizing texels is computationally inefficient; the latent texture still contains a large amount of initialization noise after many iterations. Therefore, we train an encoder (bottom row) that transforms PBR surface attributes into latent codes, and can be executed at any resolution. All analyzed configurations were optimized using the same amount of data. The left inset zooms-in on a small part of the texture that is partly visible in the rendered inset on the right.
    }\label{fig:latent-texture}
\end{figure}

We further analyze the benefits of using the encoder in \autoref{fig:latent-texture}, in which we compare the latent textures of different configurations at MIP level 0.
We visualize latent textures obtained via direct optimization (top row) and using the encoder at small (512~$\times$~512, left) and large (4k~$\times$~4k, right) resolutions.
The bottom insets show a close-up of the learned texture and the rendered appearance of this area.
While direct optimization and the encoder perform comparably at small resolutions (as used for instance in NeuMIP~\cite{Kuznetsov2021}),
the difference becomes apparent at high resolutions.
At resolution 4k$~\times$~4k, the directly optimized texels receive roughly 64$\times$ fewer gradient updates than texels of the $512~\times~512$ latent texture.
This results in the decoder having to map vastly different latent codes (due to random initialization) to the same BRDF value, hindering its performance.
Much of the initialization noise is still visible in the converged model.
On the other hand, the encoder provides a more data- and compute-efficient approach, yielding high-fidelity visuals.
All models were trained using the same amount of training data. Despite being computationally less intense during training, the models with direct optimization nearly doubled the training times (up to 10 hours) due to their higher memory requirements.

\subsection{Importance sampling}

\begin{figure*}[t]
    \begin{overpic}[width=1\linewidth]{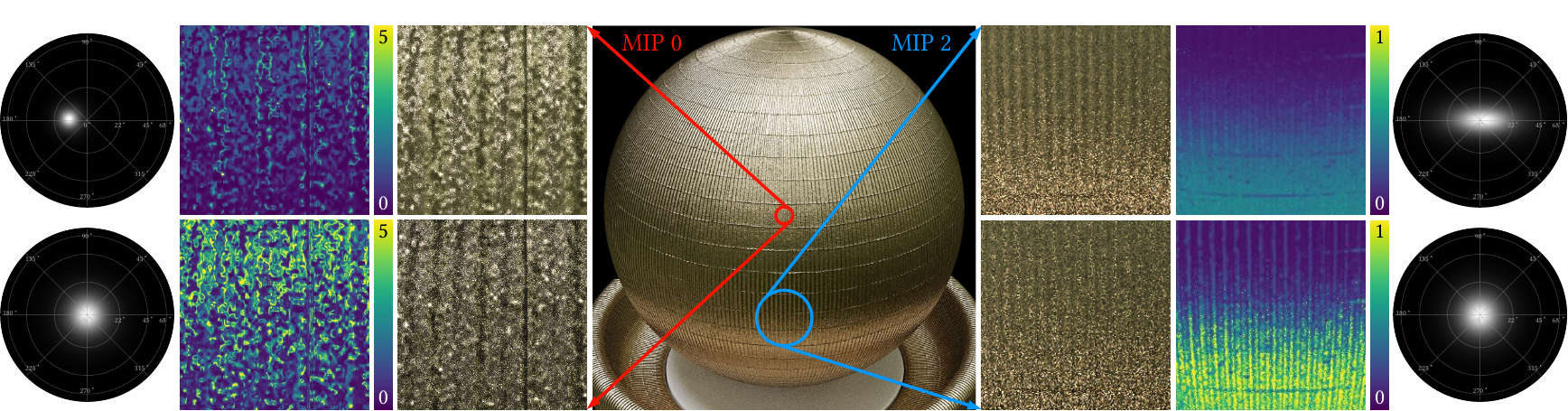}
        \footnotesize
        \put(-1.7,2.5){\rotatebox{90}{Isotropic PDF}}
        \put(-1.7,16){\rotatebox{90}{Our PDF}}
        \put(1.8,25){Example NDF}
        \put(14,25){Std. deviation}
        \put(27.8,25){Zoomed view}
        \put(46,25){MIP 0 reference}
        \put(64.8,25){Zoomed view}
        \put(77,25){Std. deviation}
        \put(90.7,25){Example NDF}
        \put(16.8,13){\textcolor{white}{Mean: 0.70}}
        \put(16.8,0.5){\textcolor{white}{Mean: 1.73}}
        \put(80.4,13){\textcolor{black}{Mean: 0.22}}
        \put(80.4,0.5){\textcolor{black}{Mean: 0.42}}
    \end{overpic}
    \caption{
        The importance sampler (top row) reduces noise levels compared to a simpler variant only supporting isotropic specular reflections (bottom row), in the spirit of \citet{Sztrajman2021} and \citet{Fan2022}. Left: Fine details of a normal map are captured using a non-centered microfacet NDF. Right: At coarser MIP levels, the filtered distribution is strongly anisotropic. The zoomed views are rendered using 4 SPP. False-color images show the pixel-wise standard deviation and its mean across the entire inset.
    }\label{fig:lod-importance-sampling}
\end{figure*}

We compare the importance sampler described in \autoref{sec:importance-sampling} against a simplified variant resembling that from \citet{Sztrajman2021} and \citet{Fan2022}. This variant is trained to only produce two outputs: an isotropic roughness parameter and a relative weight for mixing the specular and diffuse components. \autoref{fig:lod-importance-sampling} shows the benefit of the more general approach in the context of level-of-detail rendering, where it is useful to sample both non-centered and anisotropic NDFs for normal mapped and filtered BRDFs.

\begin{figure*}[t]
  \vspace{1mm}
  \begin{overpic}[width=1\linewidth]{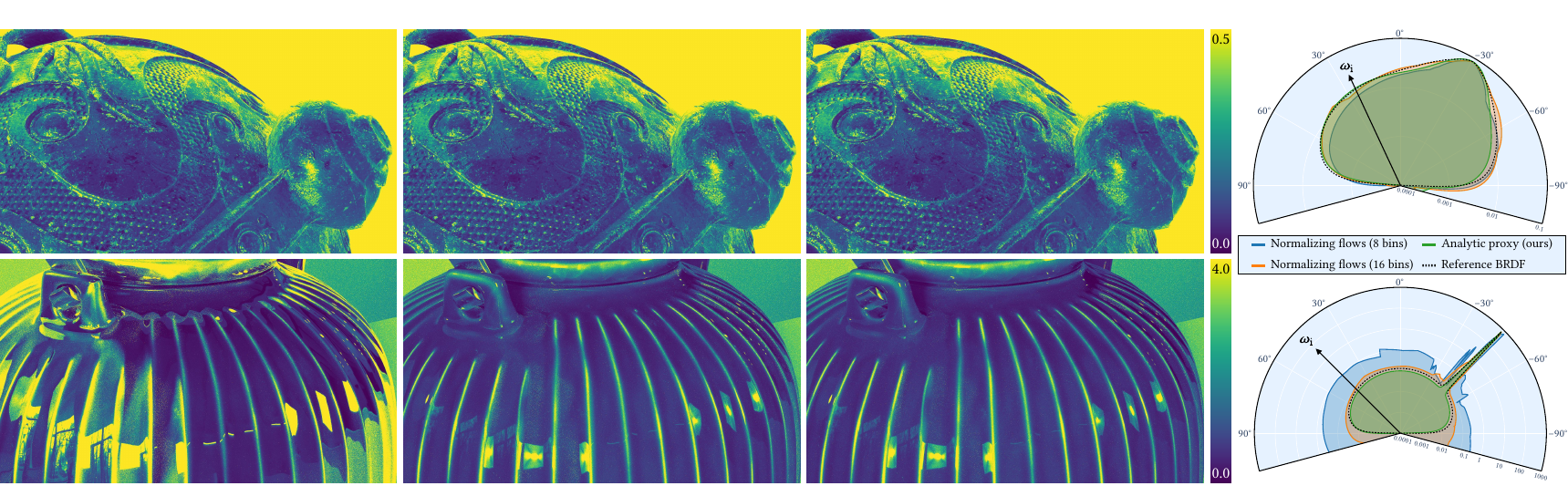}
    \footnotesize
    \put(-1.5,5.8){\rotatebox{90}{\teapot}}
    \put(-1.5,19.4){\rotatebox{90}{\inkwell}}
    \put(5.5,29.7){Normalizing flows (8 bins)}
    \put(31.5,29.7){Normalizing flows (16 bins)}
    \put(58.5,29.7){Analytic proxy (ours)}
    \put(85.5,29.7){Example PDF}
    \put(16.9,27.8){Time: 7.93 ms}
    \put(15.5,26.4){TTUV: 12.26 ms}
    \put(41.9,27.8){Time: 14.31 ms}
    \put(41.2,26.4){TTUV: 22.12 ms}
    \put(68.2,27.8){Time: 3.06 ms}
    \put(67.5,26.4){TTUV: 4.73 ms}
    \put(16.2,13){\textcolor{white}{Time: 10.59 ms}}
    \put(14.8,11.6){\textcolor{white}{TTUV: 366.00 ms}}
    \put(41.9,13){\textcolor{white}{Time: 17.69 ms}}
    \put(40.5,11.6){\textcolor{white}{TTUV: 330.06 ms}}
    \put(68.2,13){\textcolor{white}{Time: 4.55 ms}}
    \put(66.9,11.6){\textcolor{white}{TTUV: 70.34 ms}}
  \end{overpic}
    \caption{
        Pixel-wise standard deviation images of our importance sampler against an alternative implementation based on normalizing flows. The sampler architecture in the first column (using warps with 8 bins, matching that of \citet{Zheng2021}), is adequate for the glossy \inkwell{} metal it struggles with the highly specular peak of the \teapot{} ceramic. The second column (using a higher-quality warp with 16 bins) captures the peak and roughly matches the variance of our sampler based on the analytic proxy (third column). The last column shows corresponding (log scale) polar plots of the learned densities. The overlaid numbers report rendering time (for the full frame at 1 SPP) and the \emph{time to unit variance} (TTUV), i.e. the product of mean variance and render time. This reveals a significant runtime overhead of normalizing flows. The size of the evaluation network is fixed at 2 layers with 32 neurons in all cases.
    }\label{fig:flows-importance-sampling}
\end{figure*}

We also considered using samplers based on normalizing flows \citep{Dinh2017} in our system. In particular, the variant described by~\citet{Zheng2021} where the distribution of half-vectors is represented by two piecewise quadratic warps~\citep{Mueller2019}, each parameterized by an MLP (3 layers with 16 neurons). We found this to yield comparable sampling quality to our chosen approach, but it increases the total frame render time by a factor of $2$--$3.8\times$ (see \autoref{fig:flows-importance-sampling}), making it less viable in our real-time context. This is explained by the additional overhead of the warps and the need to evaluate a larger number of MLPs at shading time. Normalizing flows generally run 4 MLPs at each hit: 2 when sampling an outgoing direction and 2 when evaluating the associated PDF, e.g. for computing multiple importance sampling (MIS) weights~\cite{Veach1995}. In contrast, our method only needs to query the sampling network once per hit and caches the resulting analytic proxy parameters for subsequent sampling and PDF evaluation steps.


\subsection{Albedo inference}
\begin{figure}[t]
    \vspace{1mm}
    \setlength{\tabcolsep}{2pt}
    \footnotesize
    \begin{tabular}{cc}
        Rendering & Visualization of learned albedo\\[0.5mm]
        \includegraphics[width=0.49\linewidth,trim=300 0 200 0,clip]{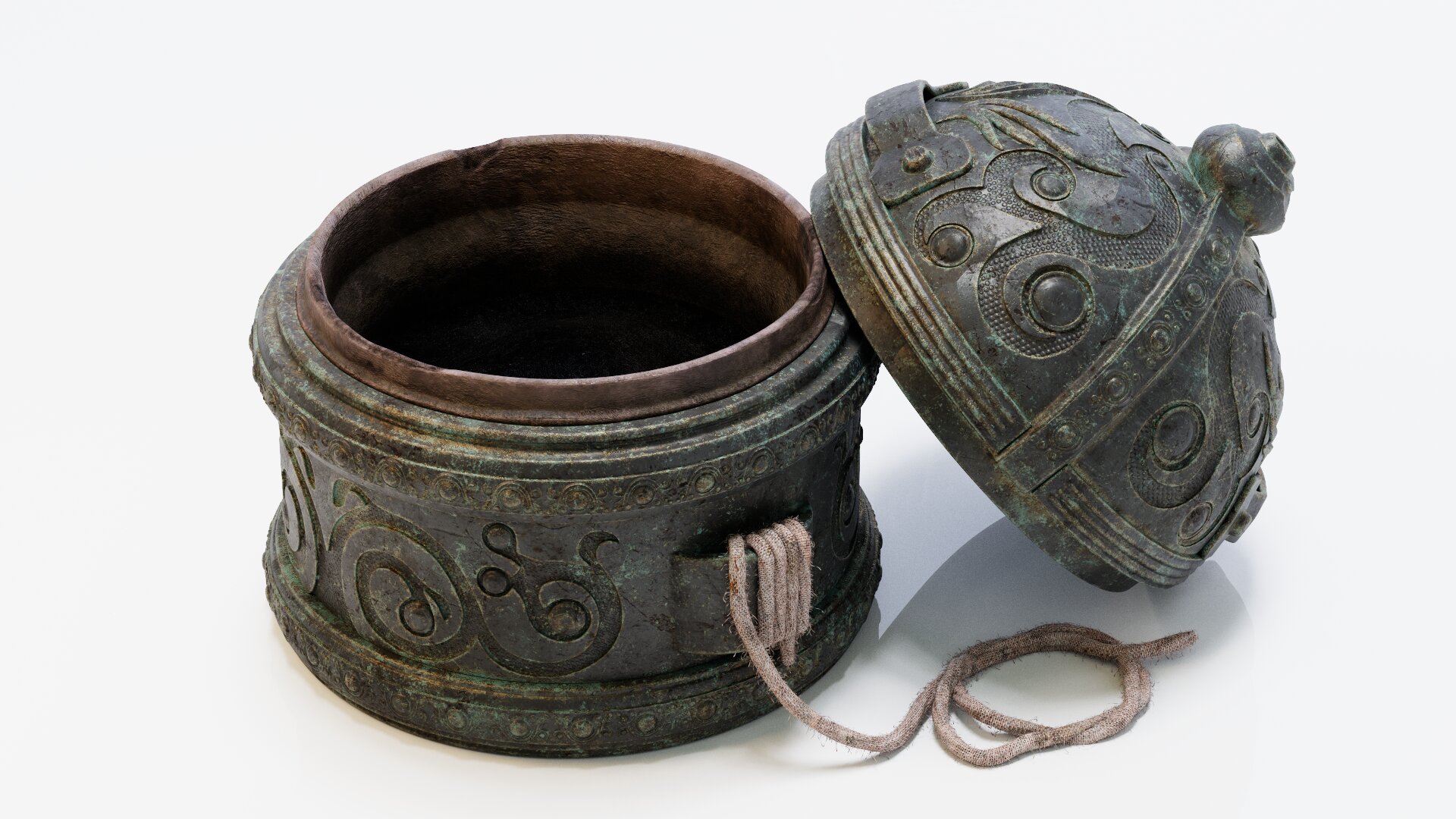}
        &\includegraphics[width=0.49\linewidth,trim=300 0 200 0,clip]{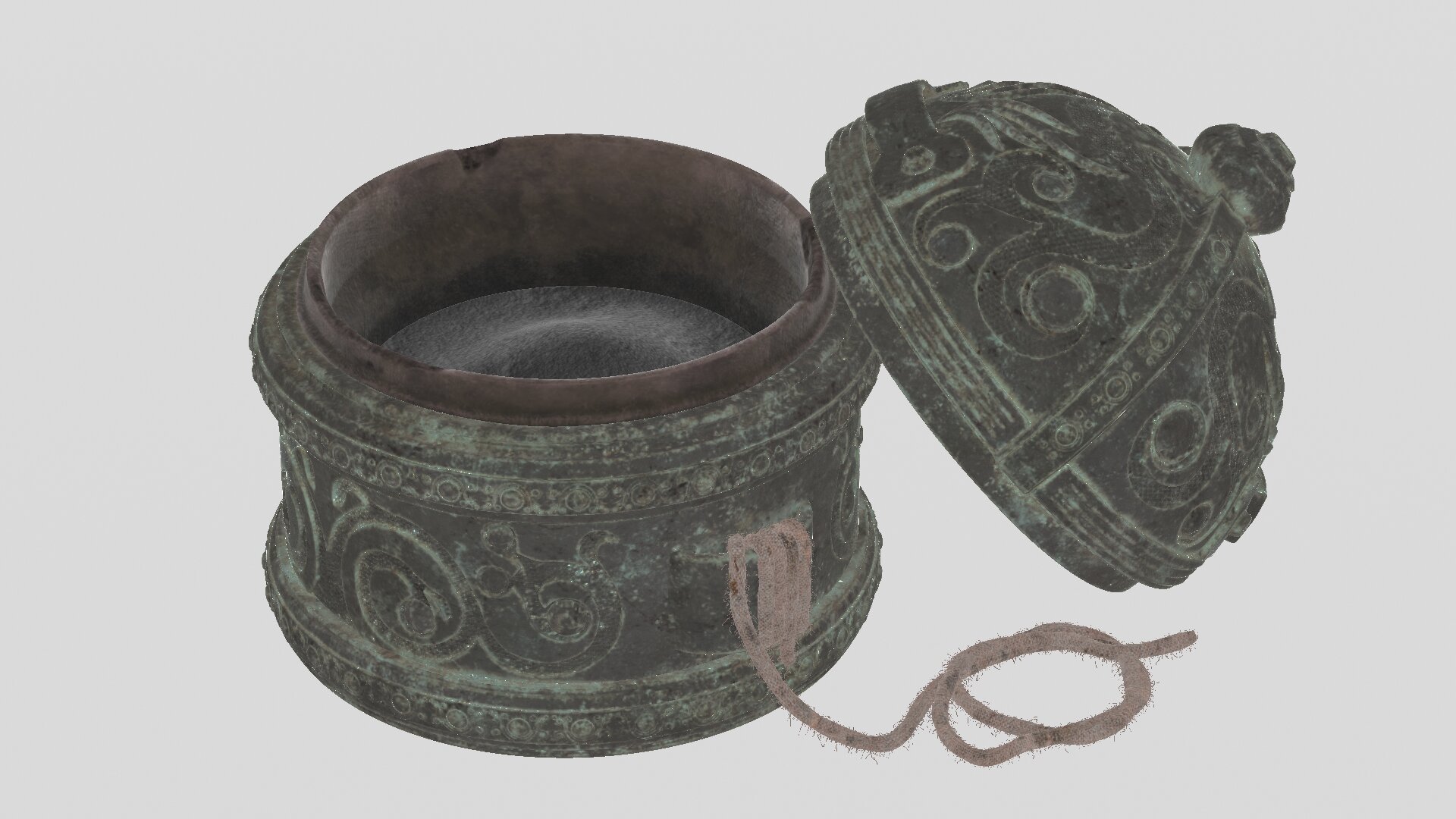}
    \end{tabular}
    \caption{
        The BRDF decoder can be trained to additionally infer the albedo of the material by optimizing its additional RGB output against a Monte Carlo estimate of the albedo of the reference material.
    }\label{fig:albedo-inference}
\end{figure}
\autoref{fig:albedo-inference} demonstrates the ability of a data-driven BRDF model to learn additional material characteristics.
The BRDF decoder outputs an extra RGB triplet approximating the albedo of the multilayer material.
We optimize the triplet against (one-sample) estimates of the true albedo during training using the $L_2$ loss, which ensures convergence towards the mean.
The ability to predict albedo gives our approach an edge over complex materials composed of analytical models, that can only output texture values of \emph{individual} components, since numerical albedo estimation is typically infeasible in a path tracer.
The albedo value can be used, e.g., to guide a denoiser.

\section{Inline Neural Materials}
\label{sec:system}

In this section, we describe the runtime system for inlining our neural appearance model in ray tracing shaders.
Similar to recent work on real-time NeRFs~\cite{Mueller2022}, we implement fully fused neural networks from scratch on the GPU. Instead of hand-written kernels however, we use run-time code generation to evaluate the neural model \emph{inline} with rendering code.
This allows fine-grained execution of neural networks at every hit point in a ray tracing shader program,
intermixed with hand-written code.
There are several technical challenges in making this possible.

First, existing machine learning frameworks, such as PyTorch and TensorFlow,
are built for coherent execution of neural networks in large batches.
Tools for integrating neural networks in real-time shading languages such as GLSL or HLSL
with potentially divergent execution, are largely non-existent.
Second, we want to leverage hardware accelerated matrix multiply-accumulate (MMA) operations
in recent GPU architectures by 
AMD,\footnote{\url{https://gpuopen.com/learn/wmma_on_rdna3}} 
Intel,\footnote{\url{https://www.intel.com/content/www/us/en/developer/articles/technical/introduction-to-the-xe-hpg-architecture.html}}
and NVIDIA,\footnote{\url{https://developer.nvidia.com/tensor-cores}}
but these instructions are not exposed in current shading languages.
Last, the execution and data divergence in a renderer are challenging for neural networks,
which load large amounts of parameter data from memory.

In the following, we discuss how we address each of these challenges in order to reach real-time performance.

\subsection{Neural material shaders}
\label{sec:neural_material_shaders}

Our neural model consists of several small MLPs, interconnected by blocks of non-neural operations.
We train materials offline and export a description of the final model along with its learned 
hierarchical latent textures, stored as mipmapped 16-bit RGBA images.
Texture compression of the latents is an interesting avenue for future work.
In particular, neural texture compression~\cite{Vaidyanathan2023} may be very fruitful as the 
compression and neural material model could be trained end-to-end.

The runtime system compiles the neural material description into optimized shader code.
We target the open source Slang shading language~\cite{He2018},
which has backends for a variety of targets including Vulkan, Direct3D~12, and CUDA.
Slang supports shader modules and interfaces for logically modularizing code. 
We generate one shader module per neural material, implementing the same interface as hand-written materials. In other words, neural materials are executed by the renderer no differently than classical ones.
\revision{See the supplemental material for implementation details and pseudocode examples 
for functional reproducibility of our work.}

\paragraph{Code Generation}

GPUs use a \emph{single instruction, multiple threads} (SIMT) execution model,
where batches (\emph{wavefronts} or \emph{warps}) of threads execute in lockstep.
Threads may be terminated
or masked out due to control flow.
Because each thread may process a different hit point and material,
there is no guarantee that all threads in a warp evaluate the same network.

We handle this by generating two code paths, optimized for divergent and coherent execution respectively.
The shader selects dynamically per warp which path to take.
In the divergent case, we rely on the hardware SIMT model to handle divergence and generate an unrolled sequence
of arithmetic and load instructions.
A majority of the instructions evaluate the large matrix multiplies in the MLP feedforward layers.
We use fused multiply-add (FMA) instructions to operate on
two packed 16-bit weights at a time.
The weights are laid out in memory in order of access, and special care is taken to generate
128-bit vectorized loads. %

\subsection{Tensor core acceleration}
\label{sec:tensor_cores}

Some recent GPU architectures offer hardware units for accelerating general 
matrix multiplication. %
While implementation details vary, core functionality is similar.
We focus on NVIDIA's \emph{tensor cores} which provide many flavors of matrix multiply instructions,
although the same idea applies to other architectures.

These instructions are currently limited to compute APIs and are not exposed in shaders.
To address this, we modified an open source LLVM-based DirectX shader compiler\footnote{\url{https://github.com/microsoft/DirectXShaderCompiler}}
to add custom intrinsics for low-level access.
This mechanism allows us to generate Slang shader code evaluating neural networks %
very efficiently using tensor cores, which operate on 16~$\times$~16 blocks of the weight matrix simultaneously.

MMA instructions require cooperation across the warp, which limits this fast path
to coherent warps where all threads evaluate the same material.
Additionally, loading network parameters also benefits from coherent access,
requiring careful consideration of how to construct coherent warps, which we
discuss next.

\begin{figure}[t]
	\setlength{\tabcolsep}{0.004\textwidth}%
\renewcommand{\arraystretch}{1}
\begin{center}
    \footnotesize
    \begin{tabular}{cc}
        \cakebox{} scene & Ratio of coherent warps per path length\\
        \raisebox{1.8mm}{\includegraphics[width=0.455\columnwidth]{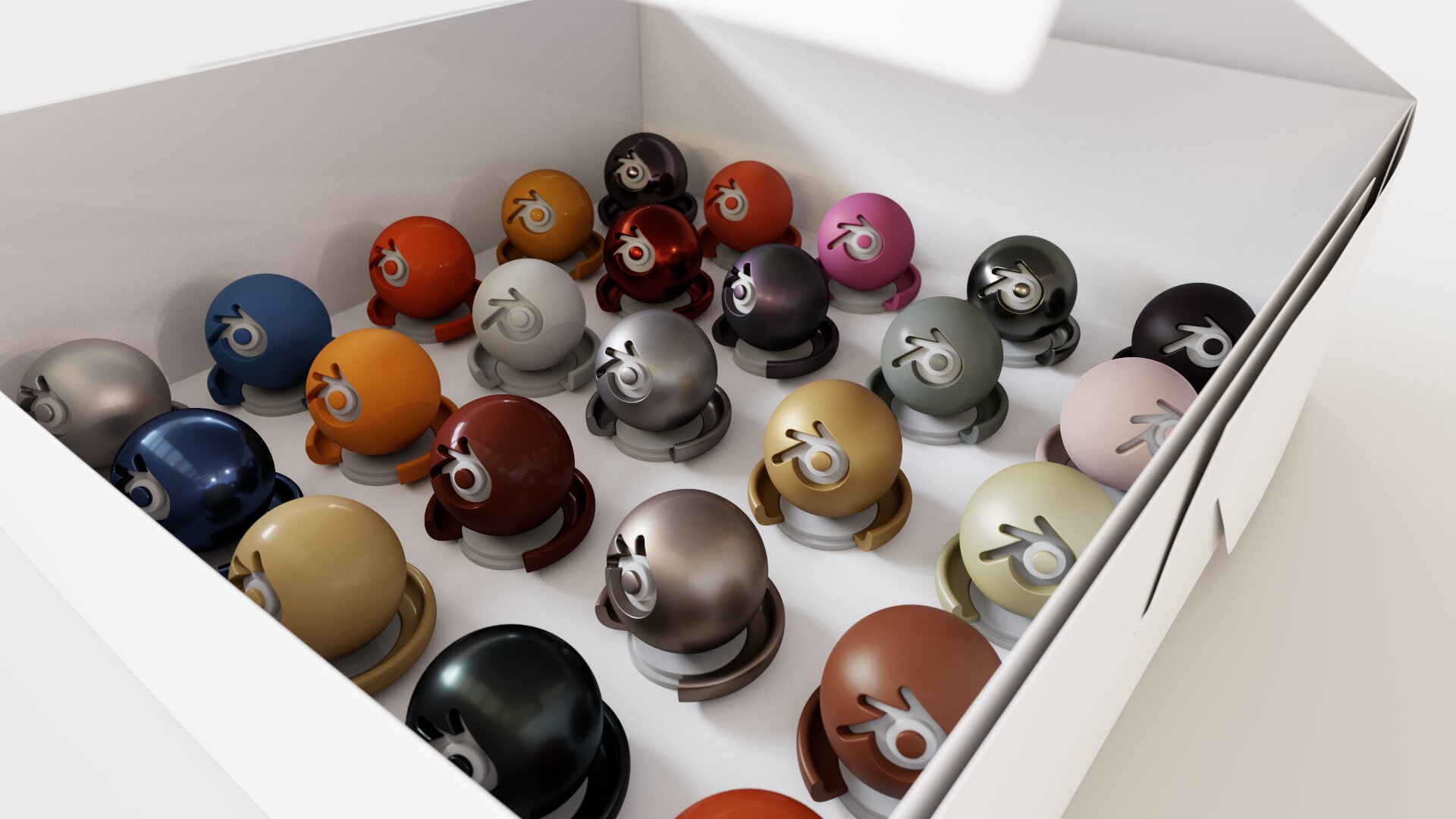}}&
        \includegraphics[width=0.49\columnwidth]{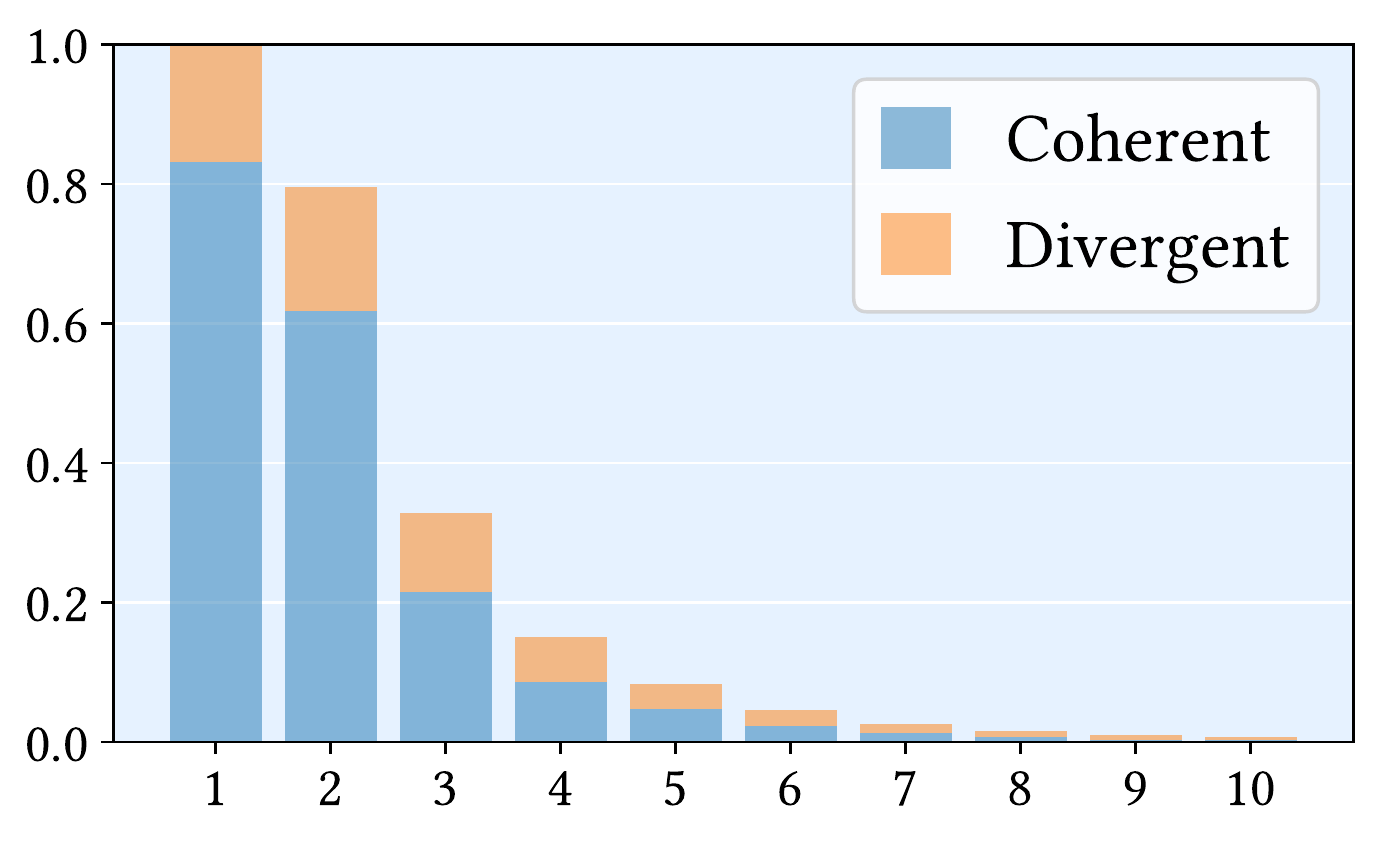}
    \end{tabular}
\end{center}

	\vspace{-4mm}
	\caption{
		This partially open \cakebox{} is filled with 25 different neural materials. The statistics show that our megakernel path tracer achieves a high degree of shading coherency using shader execution reordering (SER) over all vertices along long light paths.
	}\label{fig:system_merl_scene}
\end{figure}

\subsection{Shading coherency}

Neural materials allow us to reproduce a variety of materials using the same shader code,
simply by swapping out network weights and latent textures. This improves warp utilization 
(and thus performance) even for workloads with traditionally high execution divergence,
such as path tracing.

However, the increase in \emph{data divergence} puts pressure on the memory system,
and we can extract additional performance by increasing shading coherence.
Classical coherent approaches like wavefront path tracing~\cite{VanAntwerpen2011,Laine2013}
store hits to memory and globally reorder them after each bounce, but the high bandwidth requirements
fundamentally limit their performance.
Recent hardware features such as Intel's thread sorting unit 
(TSU)\footnote{\url{https://www.intel.com/content/www/us/en/developer/articles/guide/real-time-ray-tracing-in-games.html}}
and NVIDIA's shader execution reordering 
(SER),\footnote{\url{https://developer.nvidia.com/sites/default/files/akamai/gameworks/ser-whitepaper.pdf}}
instead reorder work \emph{locally}. We use a megakernel path tracer
to keep paths on-chip, and benefit from the increased data coherence provided by SER.
\autoref{fig:system_merl_scene} shows that the majority of warps are fully coherent 
(shading the same material with all threads active) with our path tracing architecture.

\begin{figure*}[t!]
	\setlength{\tabcolsep}{0.002\textwidth}%
\renewcommand{\arraystretch}{1}%
\footnotesize%
\begin{tabular}{cccc}
2 layers with 16 neurons & 2 layers with 32 neurons & 3 layers with 64 neurons & Reference\\
\begin{overpic}[width=0.24575\textwidth]{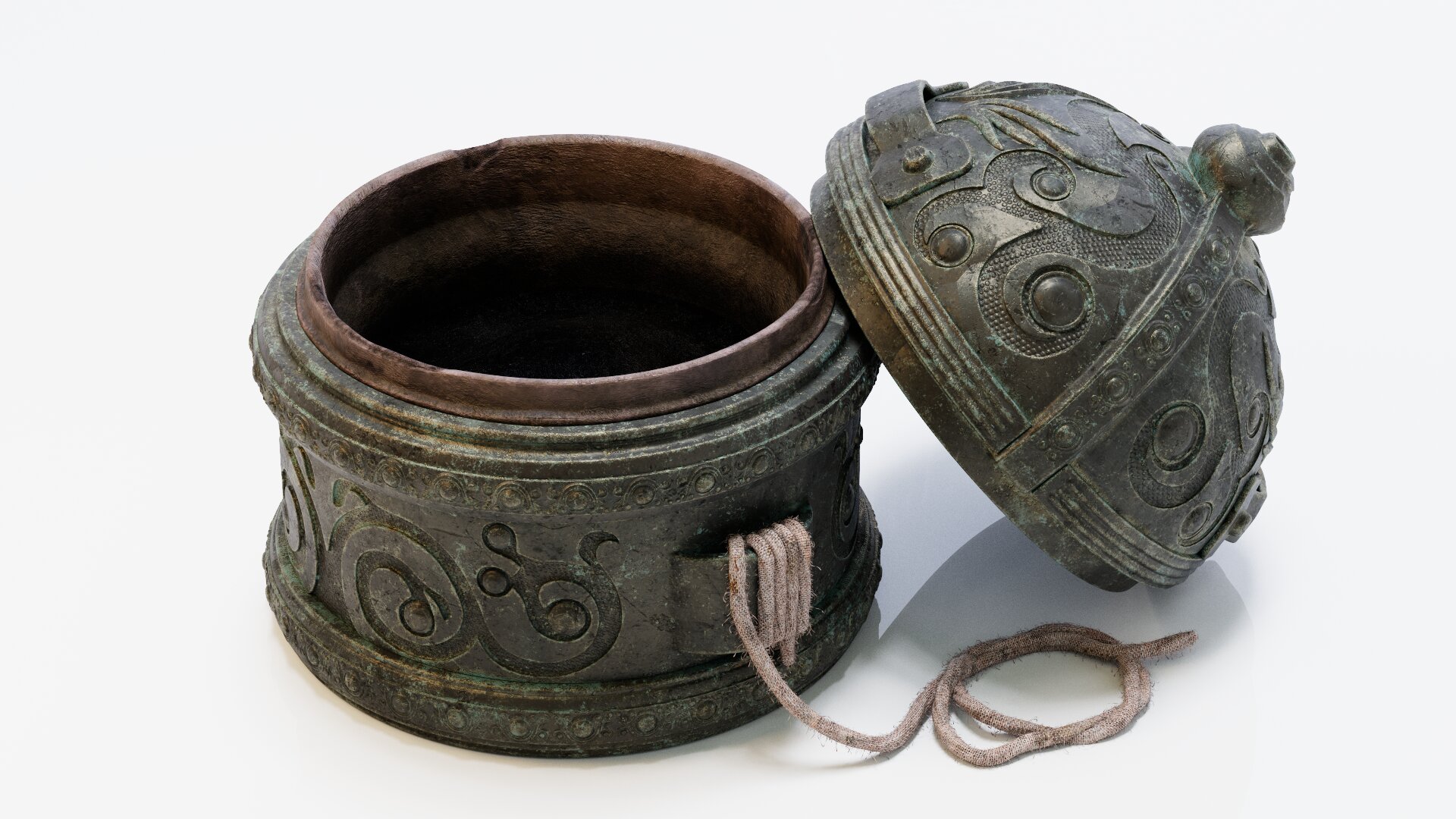}
\put(78,2){\textcolor{black}{$3.64$ ms}}
\put(0,0){\includegraphics[width=0.05\linewidth]{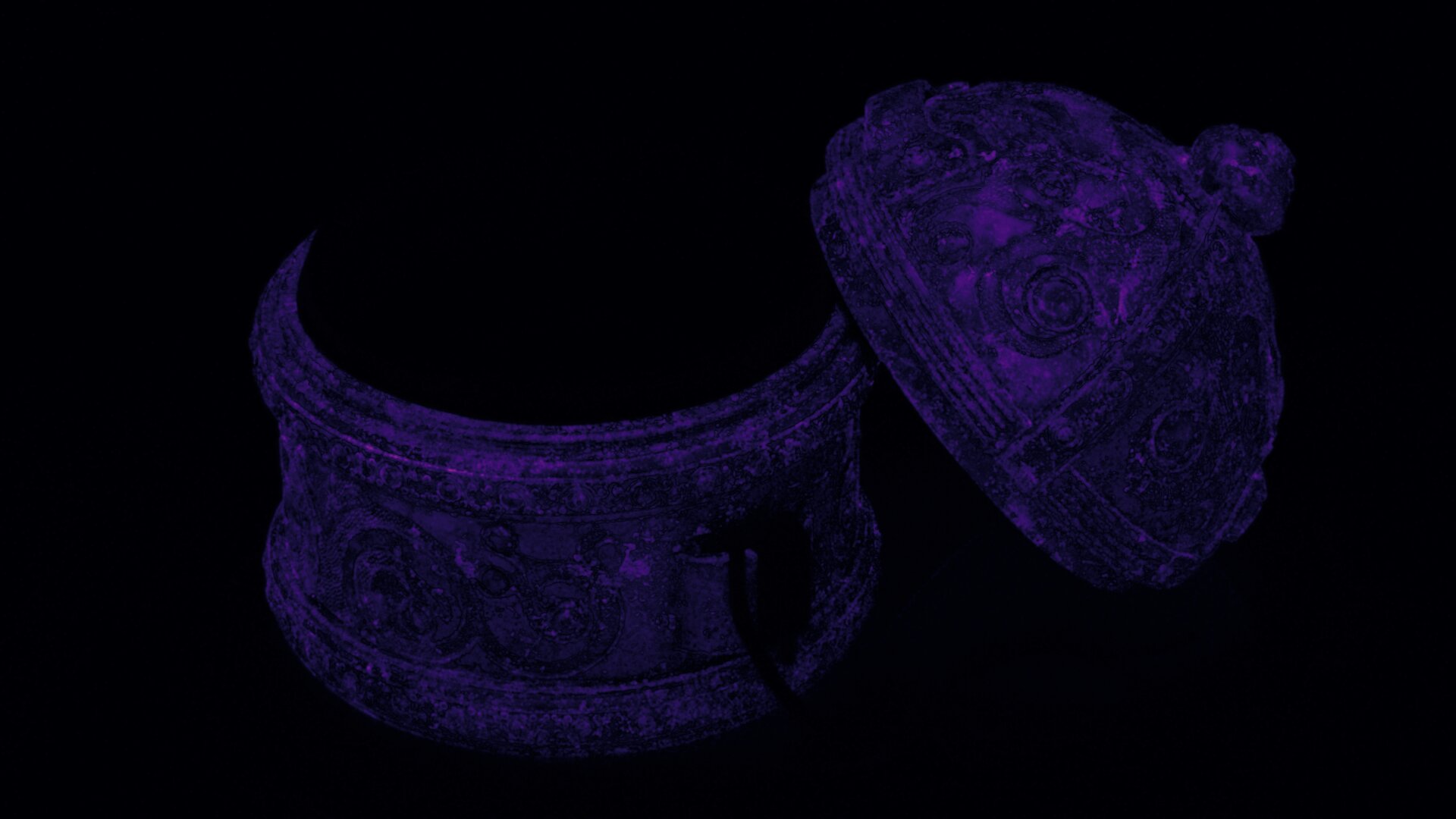}}
\put(-6.0,0){\rotatebox{90}{\hspace{9mm}View 1}}
\end{overpic}%
&\begin{overpic}[width=0.24575\textwidth]{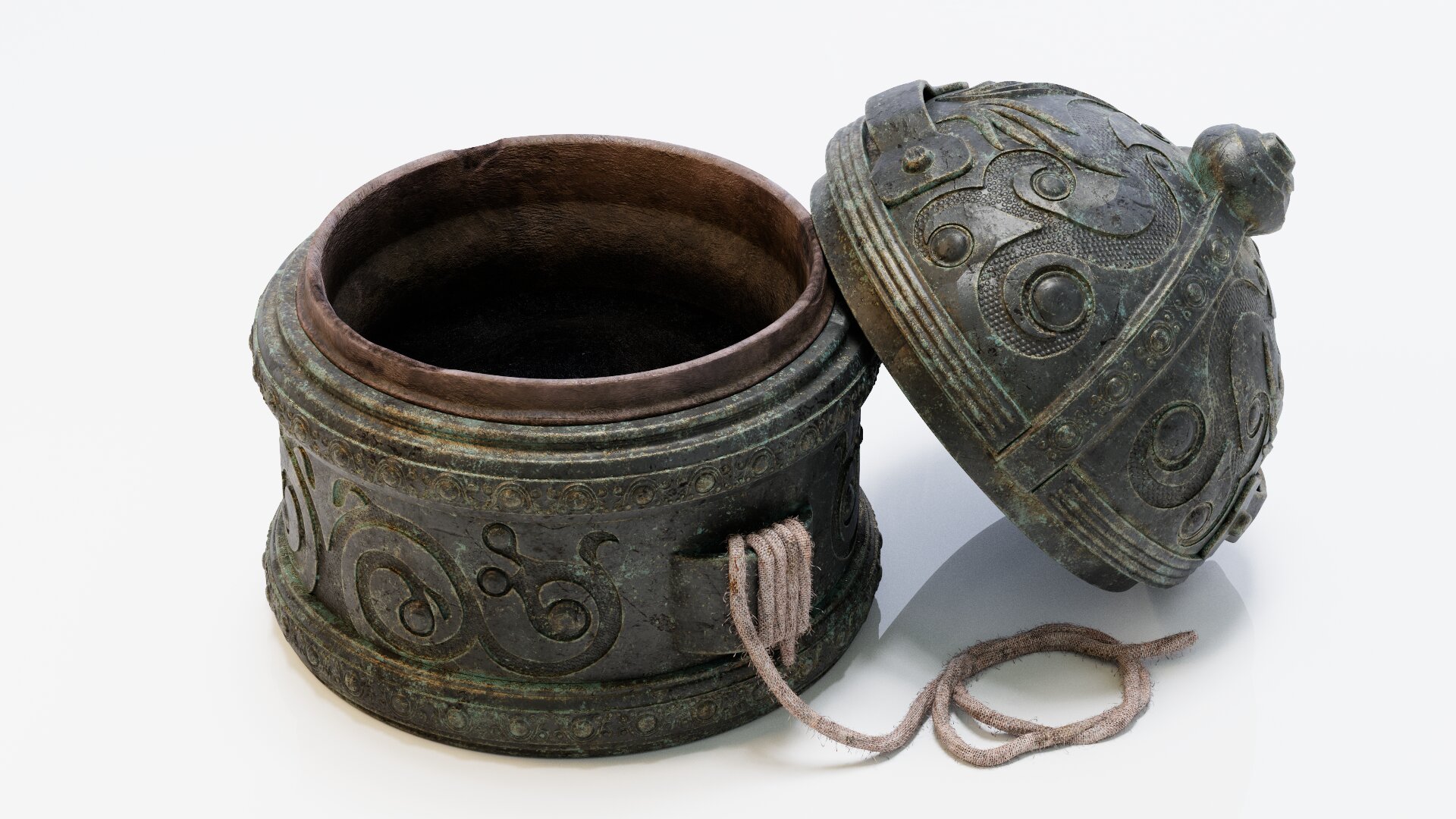}
\put(78,2){\textcolor{black}{$4.36$ ms}}
\put(0,0){\includegraphics[width=0.05\linewidth]{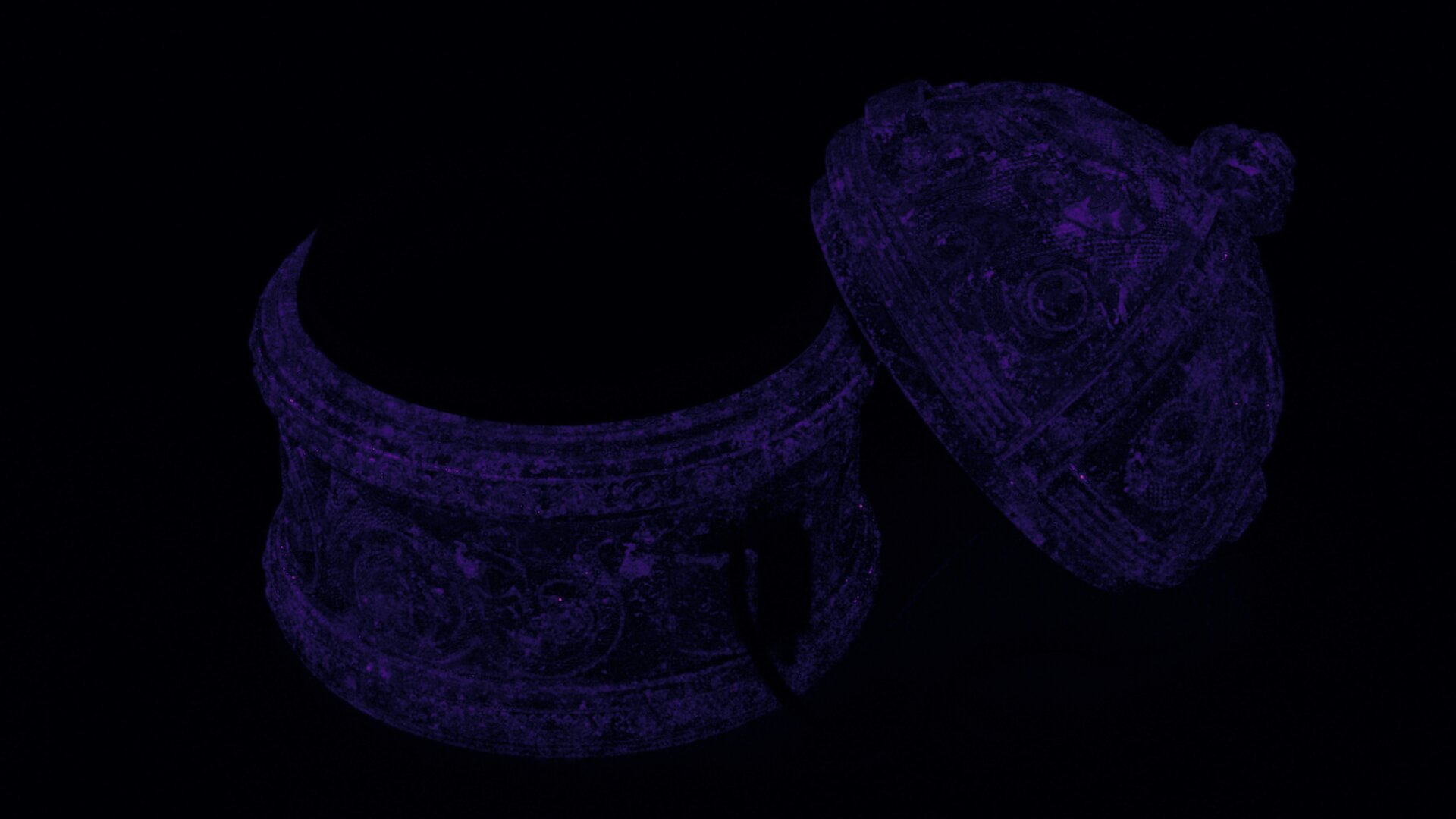}}
\end{overpic}%
&\begin{overpic}[width=0.24575\textwidth]{images/performance_inkwell/3x64_camera_0_C_Cam_A_spp_8192_tonemap.jpg}
\put(78,2){\textcolor{black}{$9.94$ ms}}
\put(0,0){\includegraphics[width=0.05\linewidth]{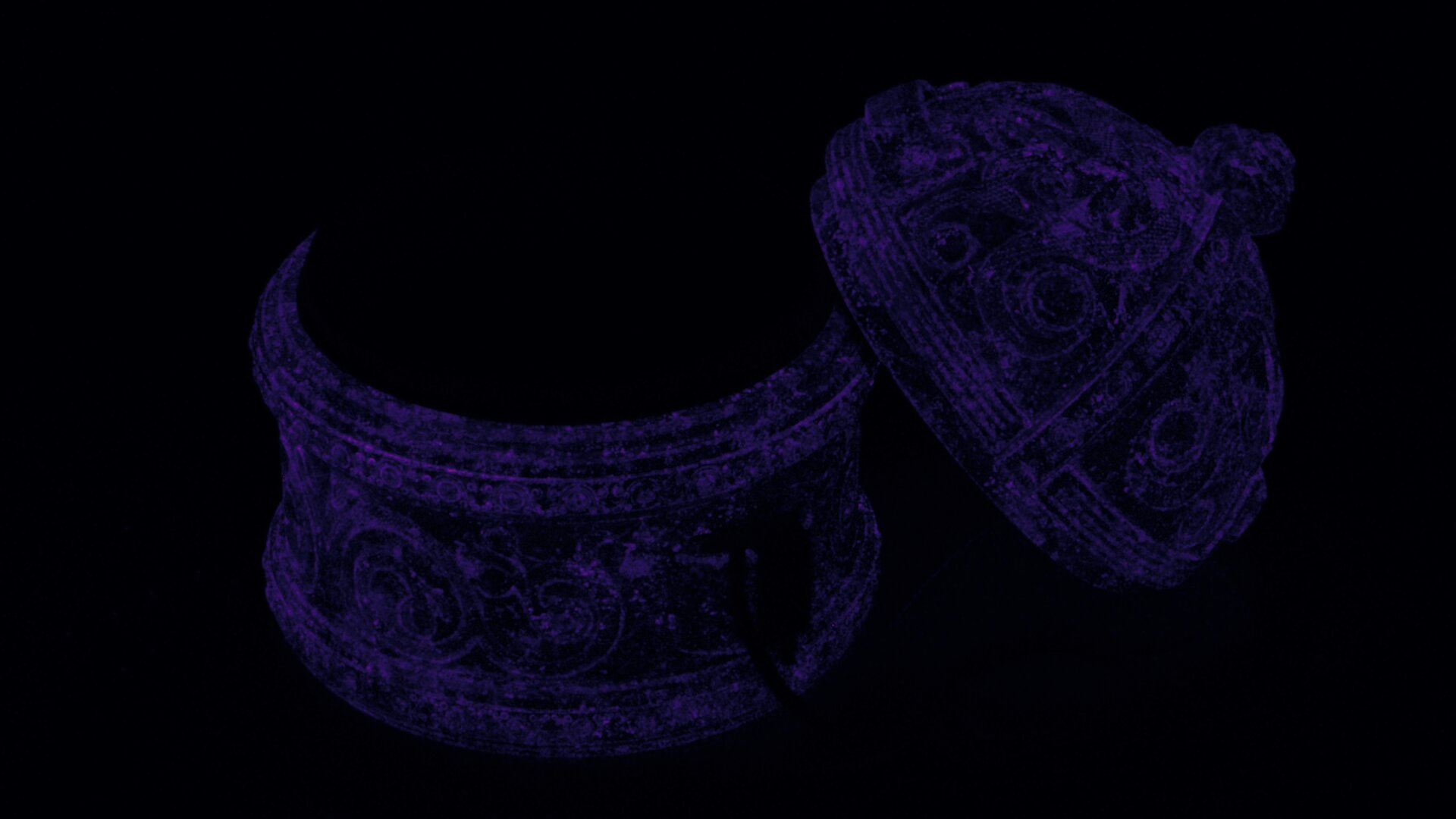}}
\end{overpic}%
&\begin{overpic}[width=0.24575\textwidth]{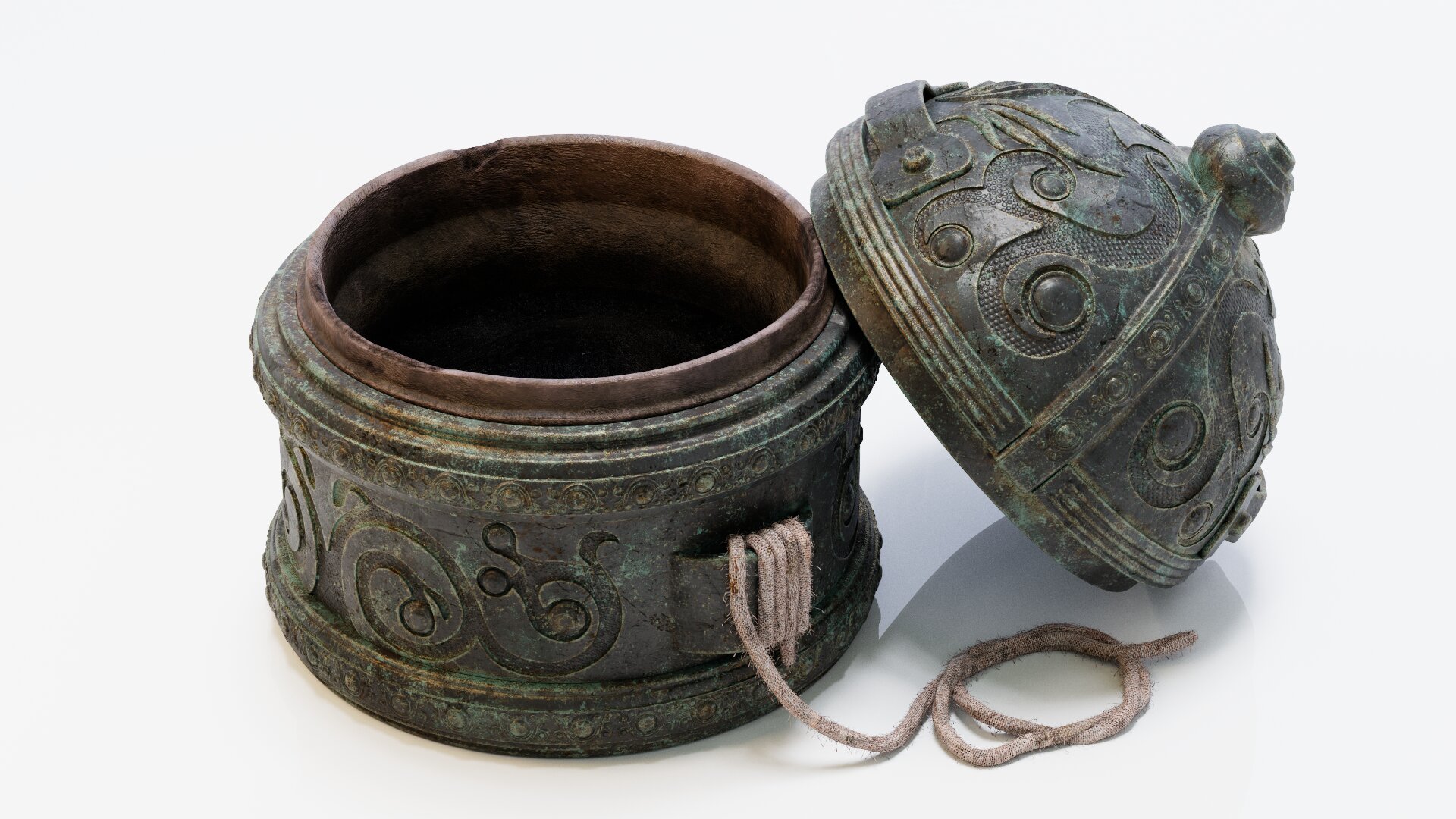}
\put(78,2){\textcolor{black}{$14.58$ ms}}
\end{overpic}%
\\
\begin{overpic}[width=0.24575\textwidth]{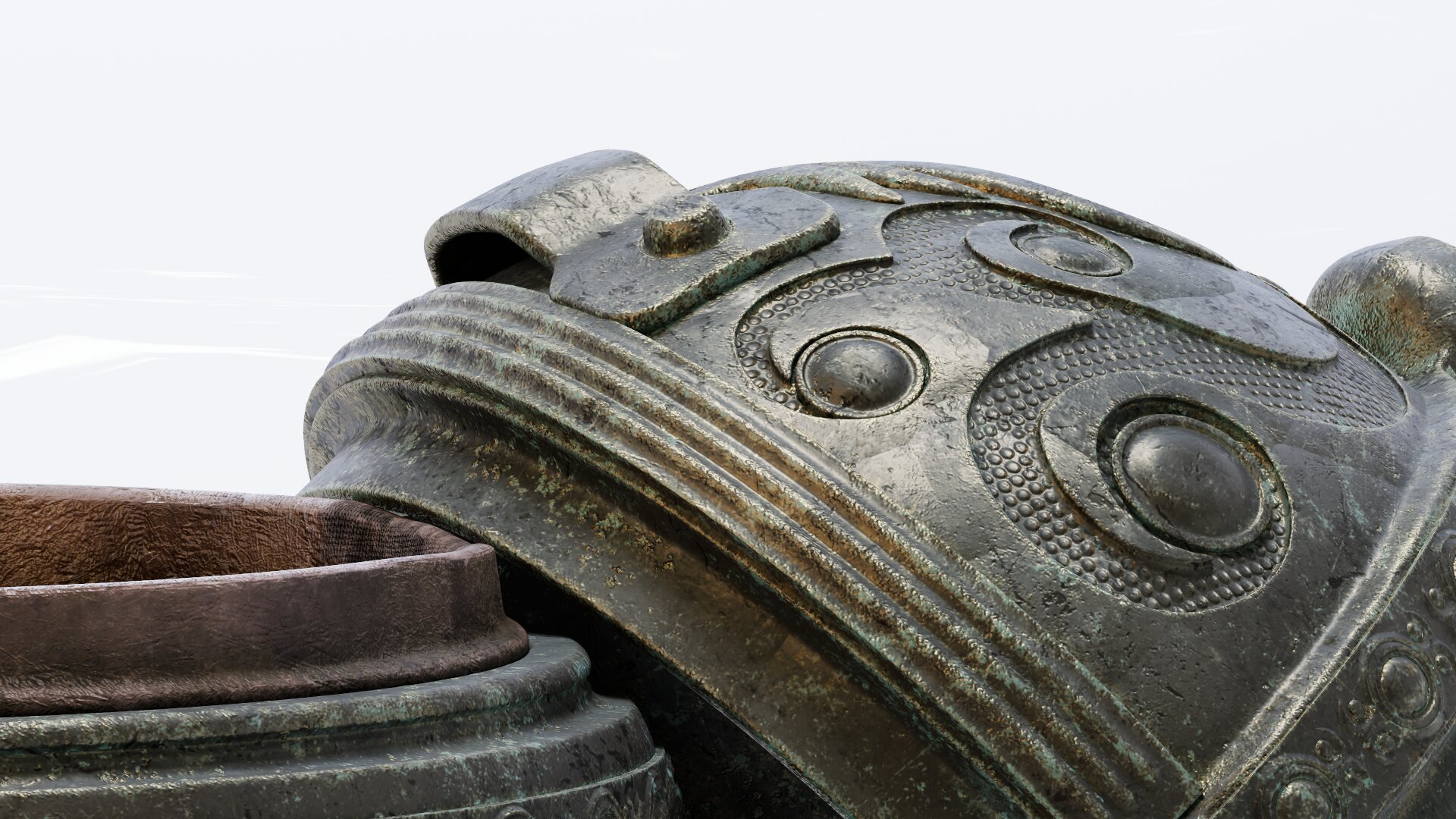}
\put(78,2){\textcolor{white}{$3.26$ ms}}
\put(0,0){\includegraphics[width=0.05\linewidth]{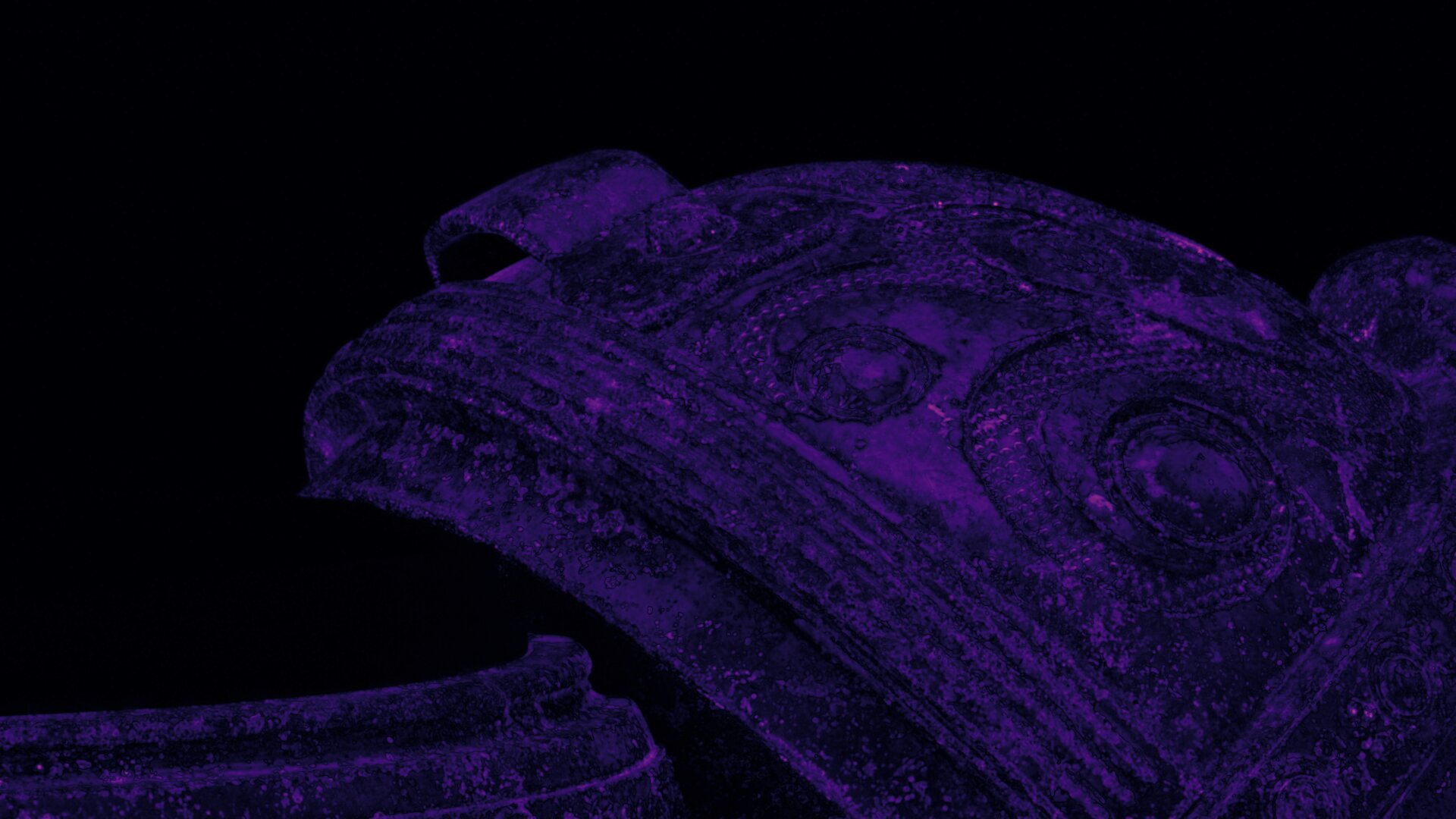}}
\put(-6.0,0){\rotatebox{90}{\hspace{9mm}View 2}}
\end{overpic}%
&\begin{overpic}[width=0.24575\textwidth]{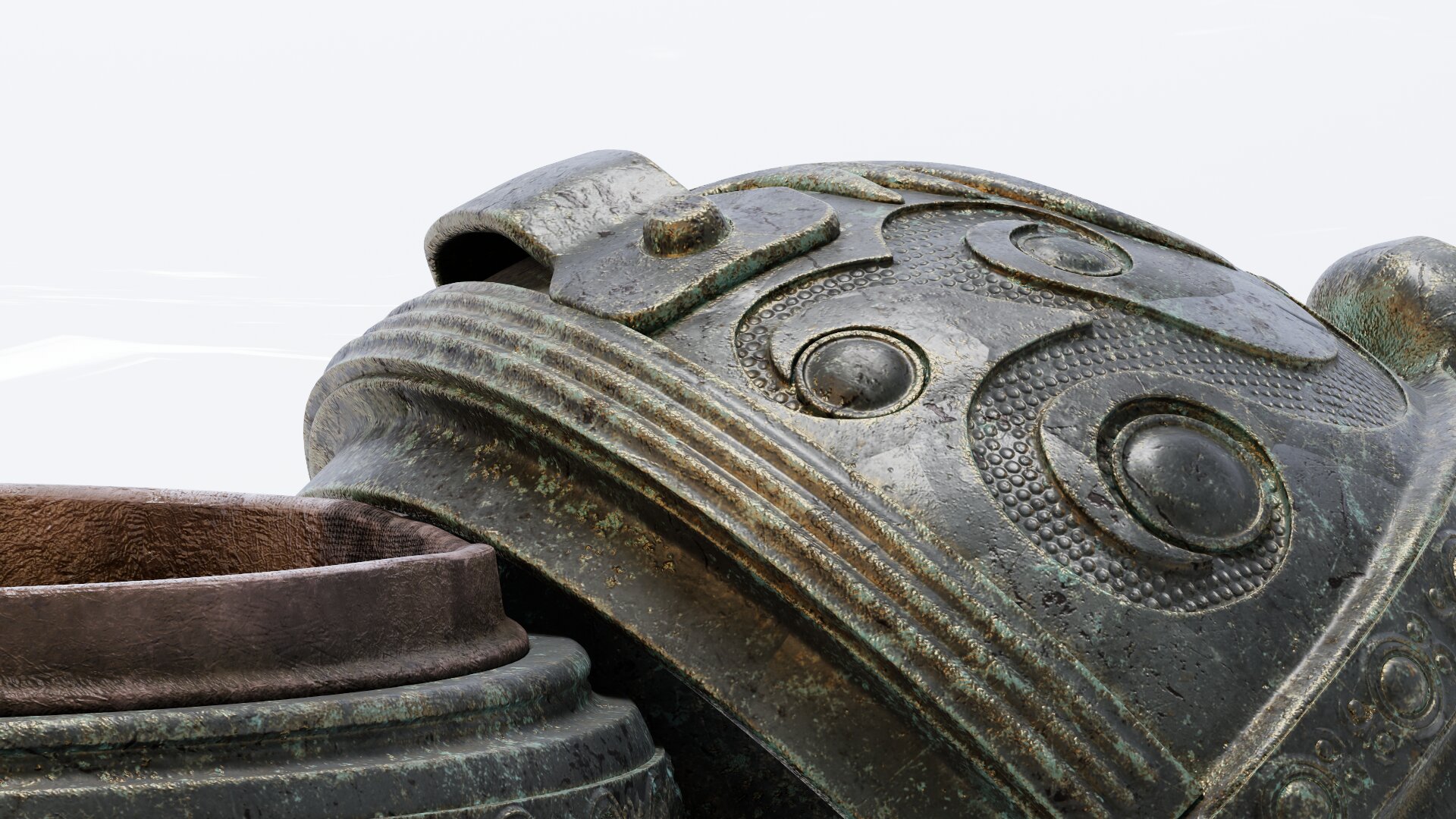}
\put(78,2){\textcolor{white}{$4.16$ ms}}
\put(0,0){\includegraphics[width=0.05\linewidth]{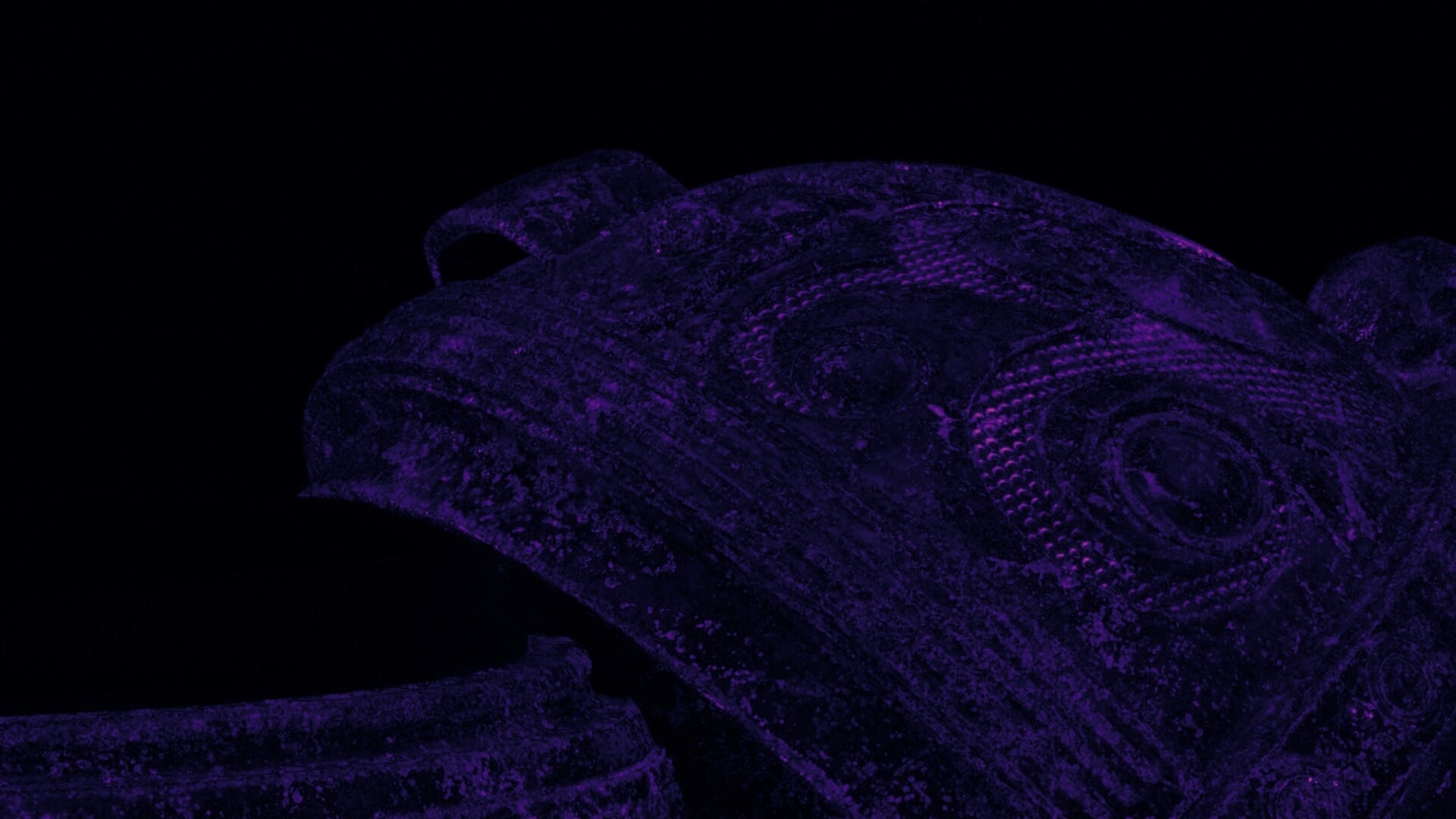}}
\end{overpic}%
&\begin{overpic}[width=0.24575\textwidth]{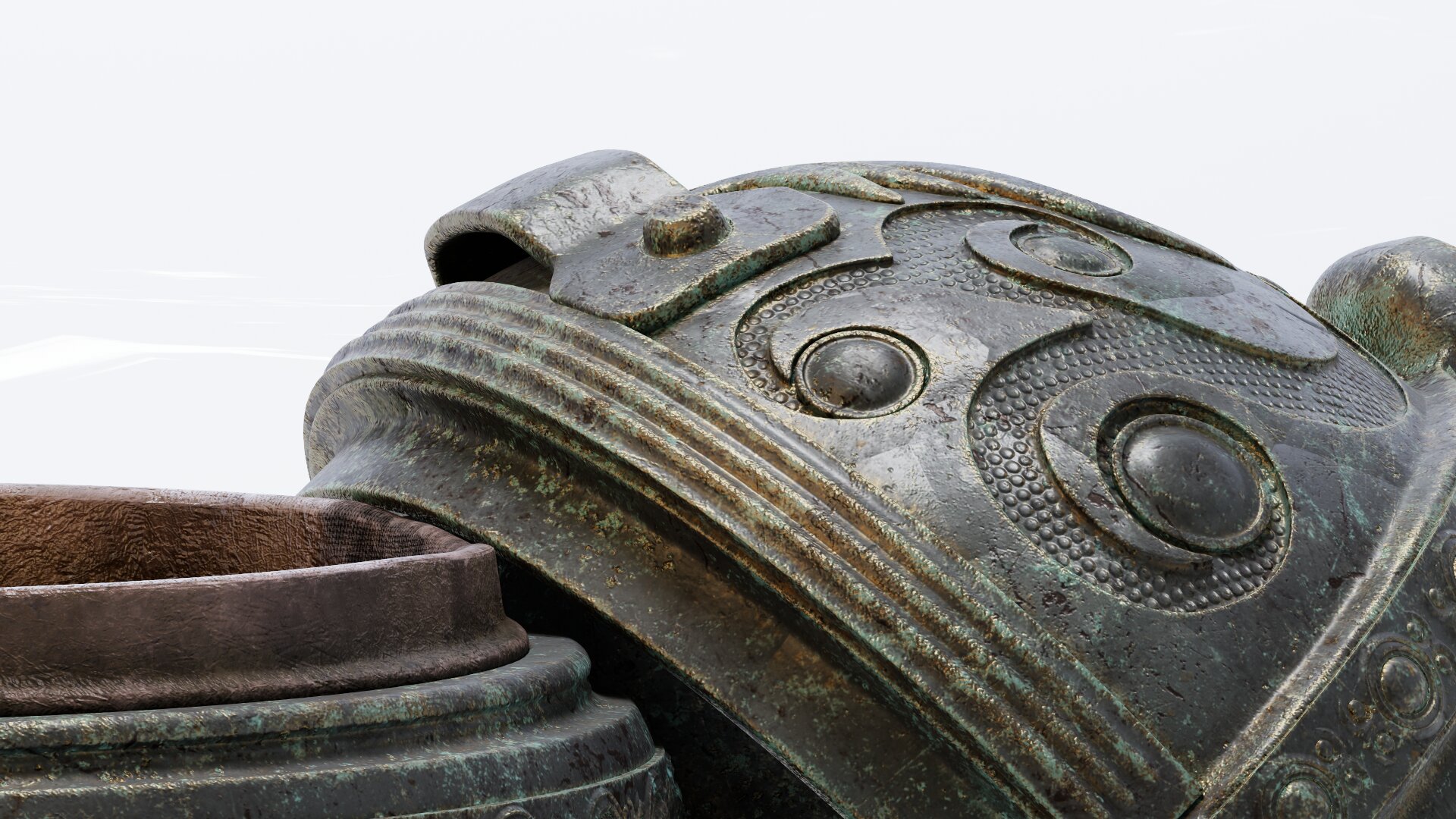}
\put(78,2){\textcolor{white}{$10.93$ ms}}
\put(0,0){\includegraphics[width=0.05\linewidth]{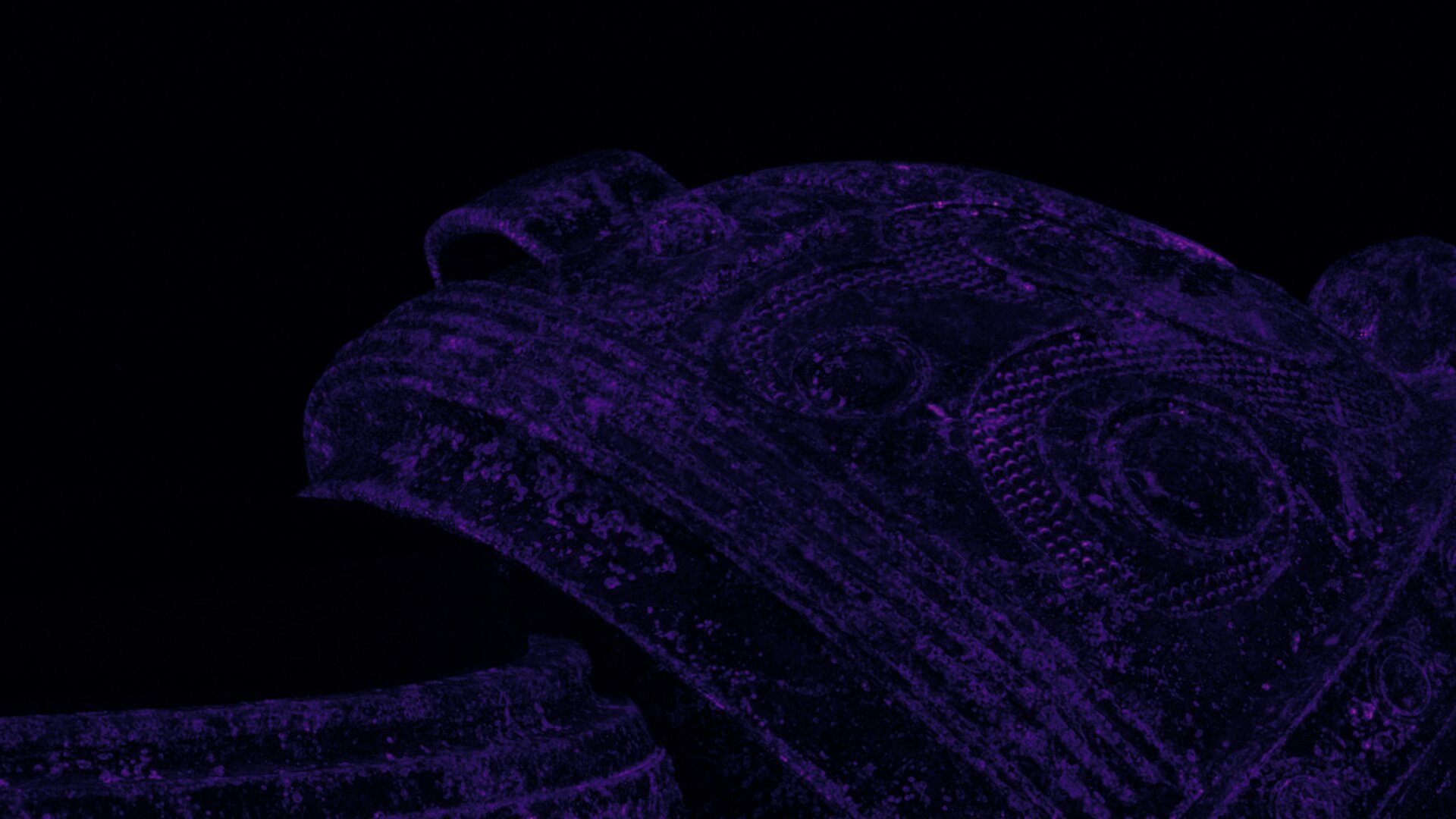}}
\end{overpic}%
&\begin{overpic}[width=0.24575\textwidth]{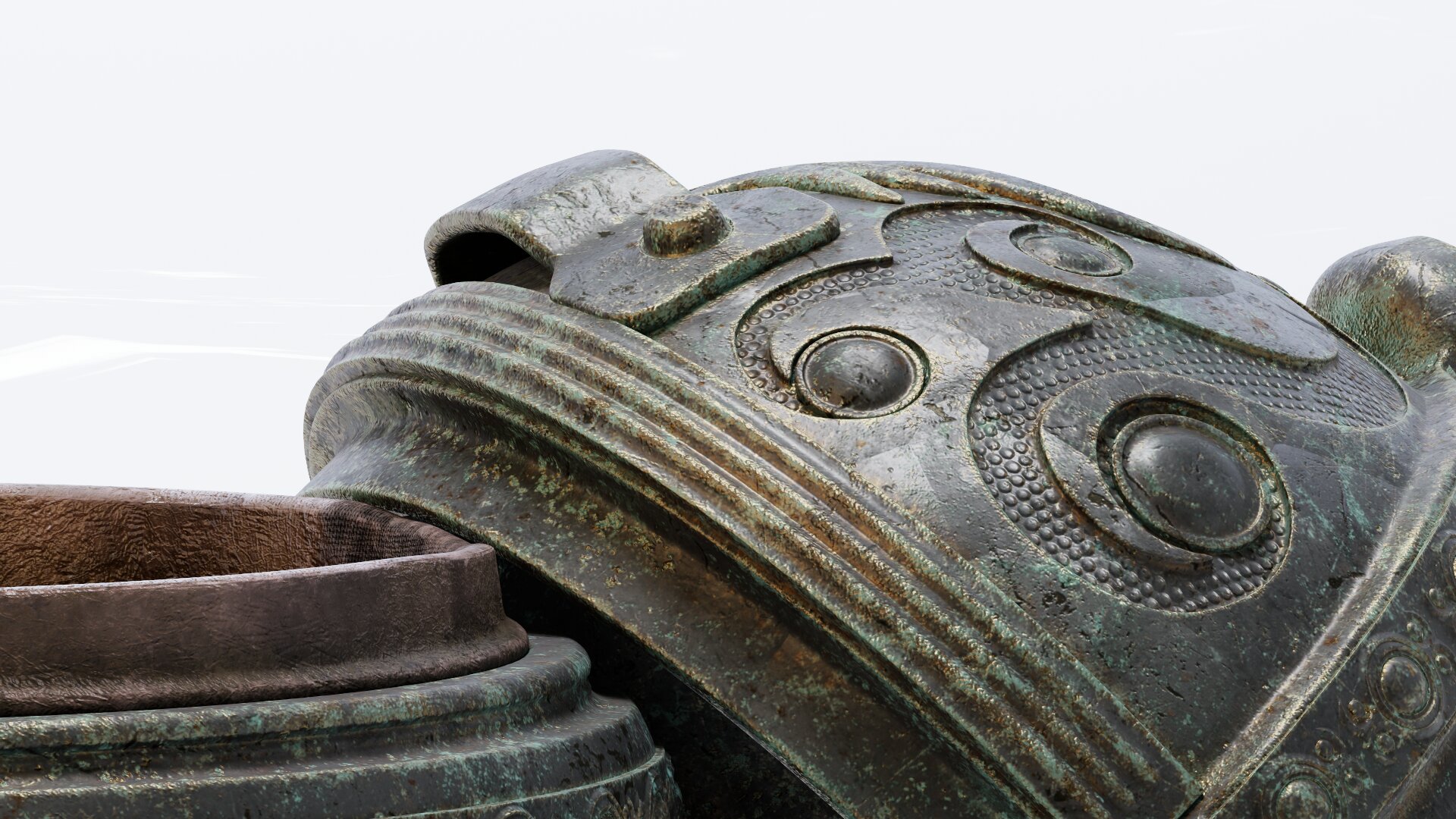}
\put(78,2){\textcolor{white}{$15.36$ ms}}
\end{overpic}%
\\
\end{tabular}

	\vspace{-2mm}
	\caption{
    The \inkwell{} scene where the metal uses the proposed neural BRDF. The remaining parts use analytical BRDFs. The first three columns show different sizes of the BRDF decoder, from fastest to the most accurate. In the corners we show a \FLIP error image and the rendering performance of an image with a \emph{single path sample per pixel} (1 SPP) at 1920~$\times$~1080 resolution using paths of up to length six. All images are rendered at 8192 SPP to suppress path tracing noise.
	}\label{fig:performance-comparison-inkwell}
\end{figure*}
\begin{figure*}[t!]
  	\input{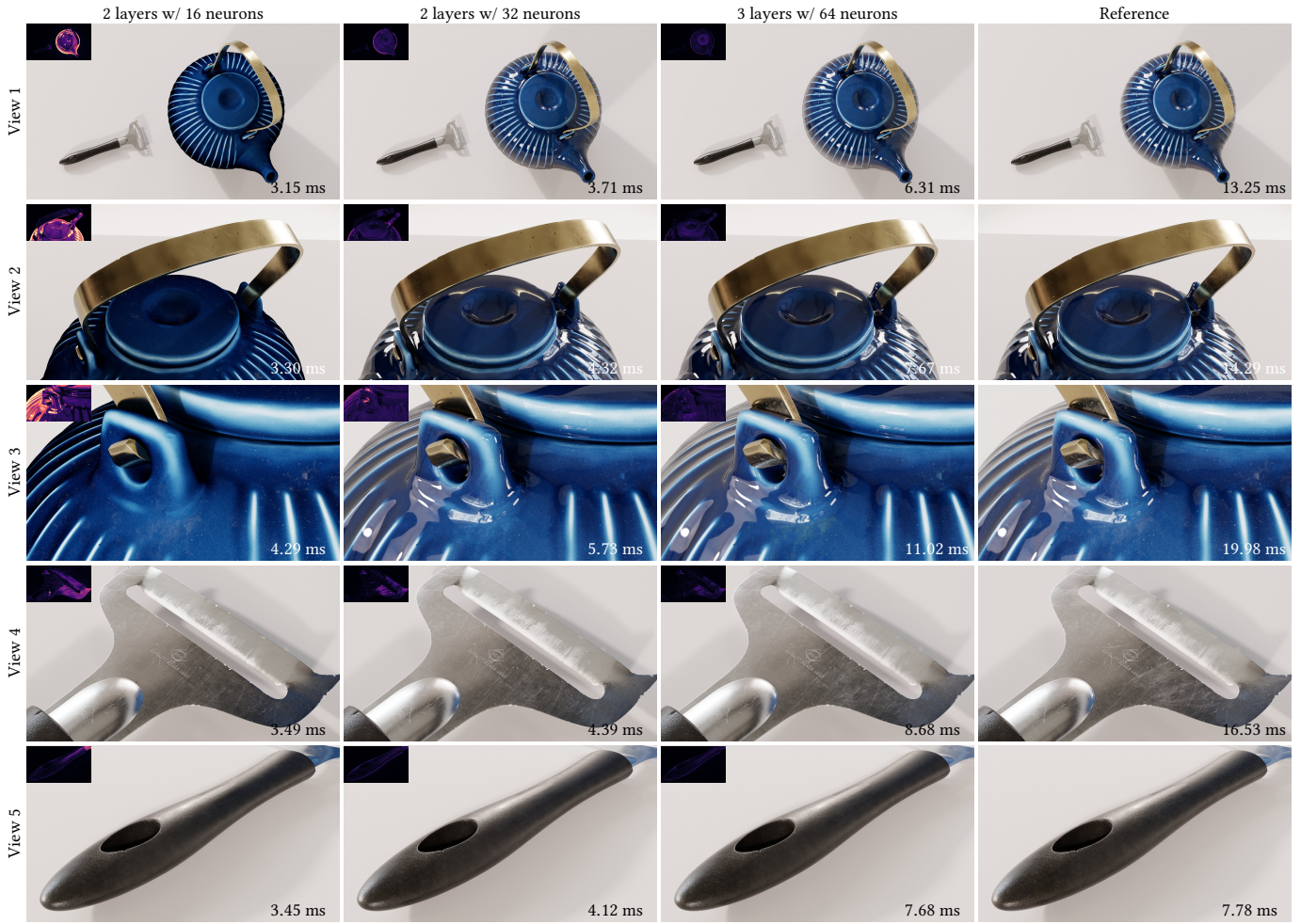}
  	\vspace{-2mm}
  	\caption{
    	The \stage{} scene with four materials that we approximate using the proposed neural BRDFs. We use a similar layout as in \autoref{fig:performance-comparison-inkwell}. \FLIP error images are in the corners, timings quantify the cost of rendering a 1 SPP image of the scene at 1920$\times$1080 resolution using paths of up to length six. All images are rendered at 8192 SPP to suppress path tracing noise. The rendering with neural BRDFs is \textbf{1.64}$\mathbf{\times}$ to \textbf{4.14}$\mathbf{\times}$ faster than the reference materials in full frame time (averaged over the views in \autoref{fig:performance-comparison-inkwell} and here). \revision{Please refer to the supplemental document for details on the scene and lighting setup.}
 	}\label{fig:performance-comparison-stage}
 	\vspace{-4mm}
\end{figure*}

\section{Runtime Analysis and Results}
\label{sec:results}

To study quality and performance, we implement our system for neural materials
in a real-time path tracer~\cite{Clarberg2022,ClarbergHPG2022} built on the
Falcor rendering framework~\cite{Kallweit2022}. %
The path tracer uses next-event estimation with MIS~\cite{Veach1995}, and each path calls the \emph{eval}, \emph{sample},
and \emph{evalPdf} material interface multiple times.

Our system is running on Direct3D~12 using hardware-accelerated ray tracing through DirectX Raytracing (DXR).
All results are generated on an NVIDIA GeForce RTX 4090 GPU at resolution 1920~$\times$~1080, unless otherwise noted.
We focus on evaluating quality and performance for path tracing with neural materials,
and therefore disable denoising and other features that can bias the results.

Performance is reported as total time in milliseconds (ms) for rendering  a 1920~$\times$~1080 image with \emph{one} path sample per pixel (SPP).
The timing in ms/SPP is representative for real-time path tracing, and can be scaled linearly 
to predict rendering time at higher SPP for applications such as high-quality preview rendering.
Path length is capped at six path vertices (camera and light included) and Russian roulette is turned off
for the purpose of these measurement.

\paragraph{Reference materials.}
In order to study rich materials, we added support for physically-based, layered material graphs expressed in the open standard MaterialX~\cite{Smythe2021}, a common interchange format for high-fidelity materials in VFX and movie production.
This allows authoring complex layered materials (c.f., \autoref{fig:material-overview}) in Houdini and other tools. %
All materials consist of multiple BRDFs combined through mixing or coating operations. Nearly all parameters are textured, with resolutions of 4k-8k per texture. Some materials stitch multiple (up to 14) 4k texture tiles for even higher resolution.
We programmatically converted the reference materials into an optimized Slang code that implements the shading graph as a weighted ($\wi$-dependent) combination of standard BRDF models.
Each material comprises multiple layers, where each layer is driven by a number of textures; the statistics are provided in \autoref{tab:material-stats}.

\subsection{Visual accuracy}

\revision{In \autoref{fig:8d-approximation}, we compare our proposed neural material parameterized by an 8-channel latent texture to a simple analytical model that combines a diffuse component with an isotropic Trowbridge-Reitz (GGX) lobe, which are driven by textures with 8 channels in total.
We tested two variants for the analytical model: numerically optimized parameters obtained using our existing training pipeline (which was tuned for training neural materials), and parameters that were manually optimized by a specialist. Both variants fail to capture the complexity of the reference, multi-layered material.
In particular, the diffuse albedo of the simple analytical model can only capture a slice of the view-dependent color of the ceramic glazing and is therefore accurate only for the specific view directions that match the chosen albedo.
The neural material offers a more faithful reproduction, overall striking a balance between the speed and quality of the high-quality but slow reference, and the lower-quality but fast analytical approximation.}

In Figures~\ref{fig:performance-comparison-inkwell} and \ref{fig:performance-comparison-stage}, we compare the visual quality and rendering performance of three configurations of the neural BRDF decoder
(the importance sampler always comprises 3 hidden layers with 32 neurons each). As expected, quality varies with the size of the decoder. The largest configuration, with 3 hidden layers and 64 neurons, reproduces the reference material well, with most details and colors captured accurately.
The errors appear mostly at grazing angles of near-specular materials, e.g., the ceramic \teapot{} body near to the silhouette.
We tested a number of hyper-parameter configurations, and while some successfully reduced the grazing angle artifacts (e.g., using $L_2$ loss), the quality elsewhere degraded, sometimes significantly.
In order to escape this ``zero-sum'' game, we posit that another graphics prior is needed for handling Fresnel effects; we leave this to future work.

We include \FLIP~\cite{Andersson2020} false-color error images in corners to illustrate the perceived difference when toggling between the neural and reference BRDFs renders; all images are also provided as part of the supplemental material to facilitate such inspection.
\autoref{tab:metrics_table_averaged} lists average errors using a variety of standard image error metrics. The supplemental also includes polar plots for the learned materials with different decoder sizes.

\begin{table}[t]
	\centering
	\small
	\caption{
		Image error metrics averaged \revision{over the converged renderings shown in Figures \ref{fig:performance-comparison-inkwell} and \ref{fig:performance-comparison-stage}, each of which was produced using 8192 SPP}. View-specific statistics are included in the supplemental material.
	}\label{tab:metrics_table_averaged}
	\begin{tabular}{lrrr}
\toprule
{} & 2 $\times$ 16 & 2 $\times$ 32 & 3 $\times$ 64 \\
\midrule
Mean \FLIP & 0.1087 & 0.0551 & 0.0444 \\
Mean abs. error & 0.0439 & 0.0145 & 0.0121 \\
Mean sqr. error & 1.3855 & 0.0107 & 0.0101 \\
Mean rel. abs. error & 0.1042 & 0.0429 & 0.0347 \\
Mean rel. sqr. error & 0.0353 & 0.0056 & 0.0035 \\
SMAPE & 0.1449 & 0.0468 & 0.0363 \\
\bottomrule
\end{tabular}

\end{table}

\begin{table}[t]
  	\centering
  	\small
  	\caption{
    	Full frame performance in ms/SPP with three different BRDF decoder architectures (importance sampler is always 3~$\times$~32). Column labels denote the number and width of hidden layers. Numbers in parenthesis show speed up over the reference material, reported in the last column.
  	}\label{tab:performance}
  	\begin{tabular}{lrrrr}
\toprule
& 2 $\times$ 16 & 2 $\times$ 32 & 3 $\times$ 64 & Ref. \\
\midrule
\inkwell{}, View 1 &
$3.64$ ($4.01\times$) &
$4.36$ ($3.34\times$) &
$9.94$ ($1.47\times$) &
$14.58$\\
\inkwell{}, View 2 &
$3.26$ ($4.71\times$) &
$4.16$ ($3.69\times$) &
$10.93$ ($1.41\times$) &
$15.36$\\
\stage{}, View 1 &
$3.15$ ($4.21\times$) &
$3.71$ ($3.57\times$) &
$6.31$ ($2.10\times$) &
$13.25$\\
\stage{}, View 2 &
$3.30$ ($4.33\times$) &
$4.32$ ($3.31\times$) &
$7.67$ ($1.86\times$) &
$14.29$\\
\stage{}, View 3 &
$4.29$ ($4.66\times$) &
$5.73$ ($3.49\times$) &
$11.02$ ($1.81\times$) &
$19.98$\\
\stage{}, View 4 &
$3.49$ ($4.74\times$) &
$4.39$ ($3.77\times$) &
$8.68$ ($1.90\times$) &
$16.53$\\
\stage{}, View 5 &
$3.45$ ($2.26\times$) &
$4.12$ ($1.89\times$) &
$7.68$ ($1.01\times$) &
$7.78$\\
\midrule
Average &
$3.51$ ($4.14\times$) &
$4.40$ ($3.31\times$) &
$8.89$ ($1.64\times$) &
$14.54$\\
\bottomrule
\end{tabular}

  	\vspace{-1mm}
\end{table}

\begin{table}[t]
	\centering
	\small
	\caption{
		Material shading performance in ms/SPP with two different BRDF decoder architectures (importance sampler is always 3~$\times$~32). Column labels denote the number and width of hidden layers. Numbers in parenthesis show speed up over the reference material, reported in the last column.
	}\label{tab:perf_shading}
	\begin{tabular}{lrrr}
\toprule
& 2 $\times$ 32 & 3 $\times$ 64 & Ref. \\
\midrule
\stage, View 3 &
$1.59$ ($10.19\times$) &
$6.02$ ($2.69\times$) &
$16.21$\\
\stage, View 4 &
$1.23$ ($12.82\times$) &
$5.06$ ($3.12\times$) &
$15.77$\\
\inkwell, View 1 &
$1.59$ ($6.99\times$) &
$6.01$ ($1.85\times$) &
$11.11$\\
\inkwell, View 2 &
$1.74$ ($7.25\times$) &
$7.15$ ($1.76\times$) &
$12.61$\\
\midrule
Average &
$1.54$ ($9.06\times$) &
$6.06$ ($2.30\times$) &
$13.93$\\
\bottomrule
\end{tabular}

	\vspace{-0.5mm}
\end{table}

\begin{figure}[t]
	\setlength{\tabcolsep}{0.0\textwidth}%
\renewcommand{\arraystretch}{1}
\begin{center}
    \footnotesize
    \begin{tabular}{cc}
        \stage{} timings (ms) & \inkwell{} timings (ms) \\
        \includegraphics[width=0.49\columnwidth]{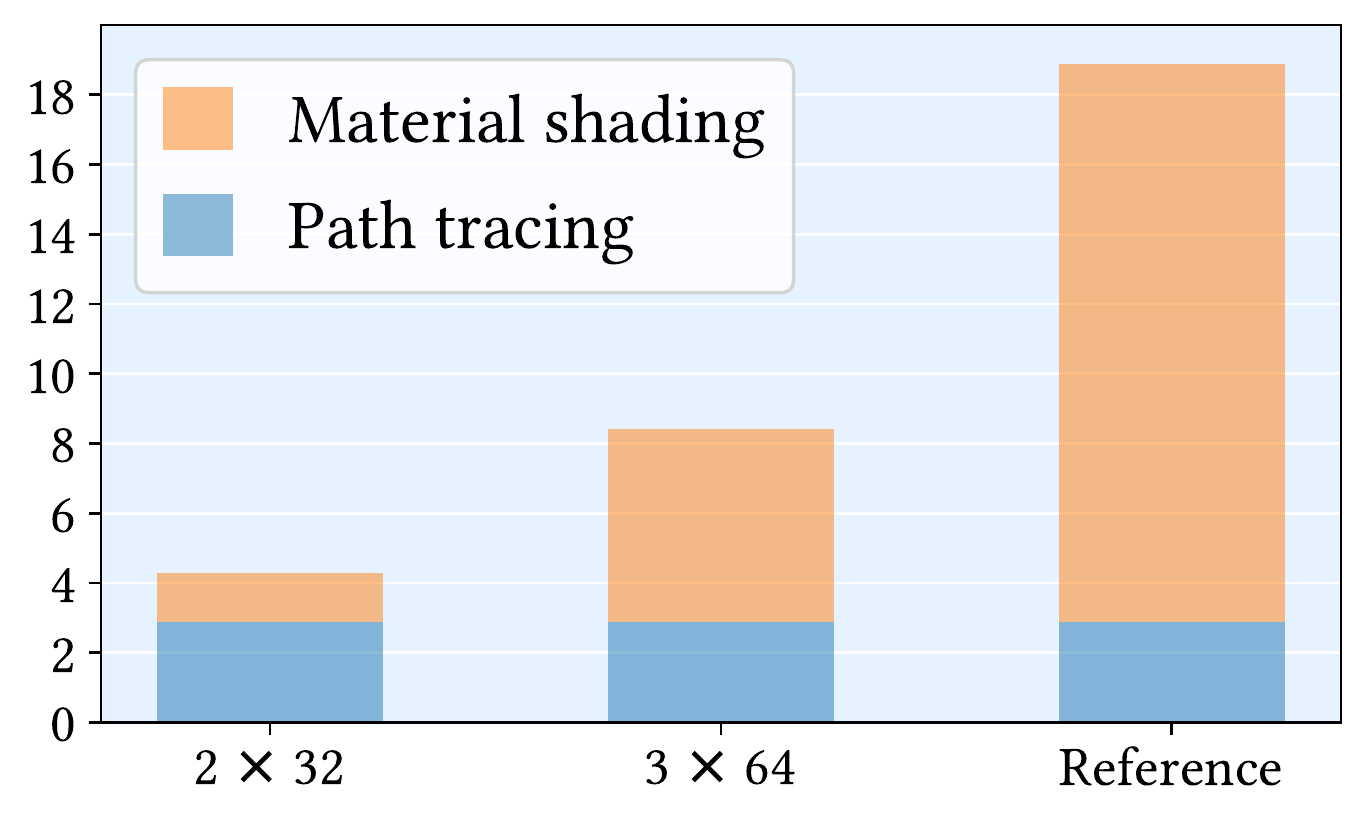}&
        \includegraphics[width=0.49\columnwidth]{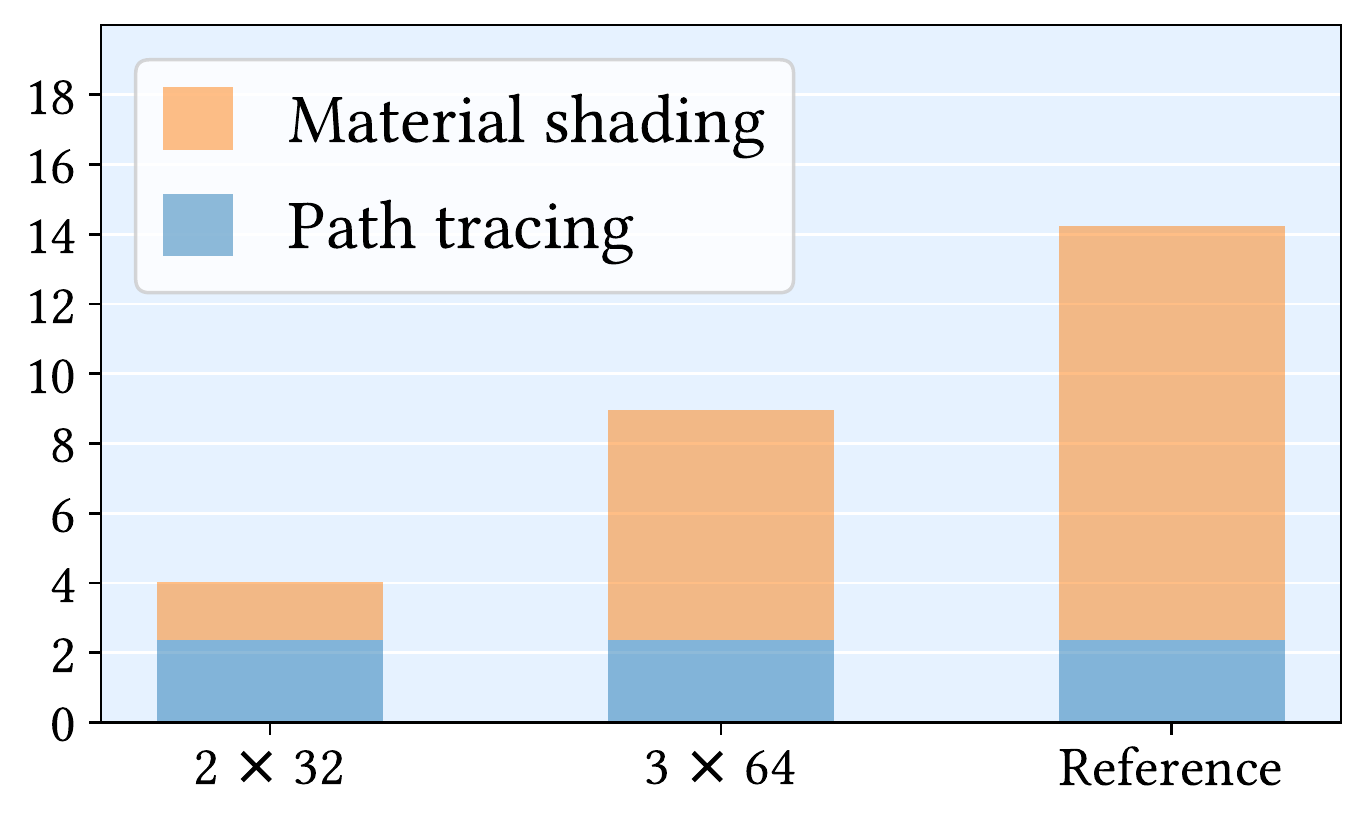}
    \end{tabular}
\end{center}

	\vspace{-4mm}
	\caption{
		Average path tracing and material shading time in ms, respectively, for rendering a 1~SPP image of the scene at
		1920$\times$1080 pixels resolution using paths up to six path vertices in length.
		Two different BRDF decoder architectures are profiled, and compared to the cost of shading using the reference materials.
	}\label{fig:perf_shading}
	\vspace{-2mm}
\end{figure}

\subsection{Runtime performance}

The smallest network yields the best rendering performance, albeit at reduced reconstruction accuracy. 
\autoref{tab:performance} lists the absolute performance in ms/SPP and the relative speed improvement over rendering a GPU-optimized implementation of the reference material (all running on NVIDIA GeForce RTX 4090 GPU).
The full frame rendering times with the neural BRDFs are $\mathbf{1.64\times}$ (3~$\times$~64) to $\mathbf{4.14\times}$ (2~$\times$~16) faster than the reference material on average.

The frame time includes both general path tracing operations (light sampling, ray tracing, 
and control logic) as well as material sampling and evaluation. To estimate how much time is
spent in material shading, and thus the relative speedups of our neural materials over 
the reference materials, we setup a dedicated benchmark. Since all neural material shaders 
in our system are running inline in the renderer, not as separate kernels, this has to be done with care;
we lock the path distribution to a simple cosine-weighted distribution, while ensuring that the compiler 
does not eliminate any of the material code.
As a baseline, we measure the pure path tracing cost using a material with constant color.

\autoref{tab:perf_shading} and \autoref{fig:perf_shading} summarize our findings for two representative
views of the \inkwell\ scene (\autoref{fig:performance-comparison-inkwell}, view 1 \& 2) and
\stage\ scene (\autoref{fig:performance-comparison-stage}, view 3 \& 4).
The shading times with the neural BRDFs are $\mathbf{2.30\times}$ (3~$\times$~64) to $\mathbf{9.06\times}$ (2~$\times$~32) faster than the reference materials on average, with over an order of magnitude speedup for several views and the mid-sized BRDF decoder (2~$\times$~32).

Overall, the performance and visual fidelity scale in a predictable manner as neural BRDFs accommodate trading quality for performance. Next, we analyze the scaling behavior in more detail.

\subsection{Scalability}

\begin{figure}[t]
	\setlength{\tabcolsep}{0.0\textwidth}%
\renewcommand{\arraystretch}{1}
\begin{center}
    \footnotesize
    \begin{tabular}{c}
        Rendering time (ms) for increasing number of neural materials\\
        \includegraphics[width=0.8\columnwidth]{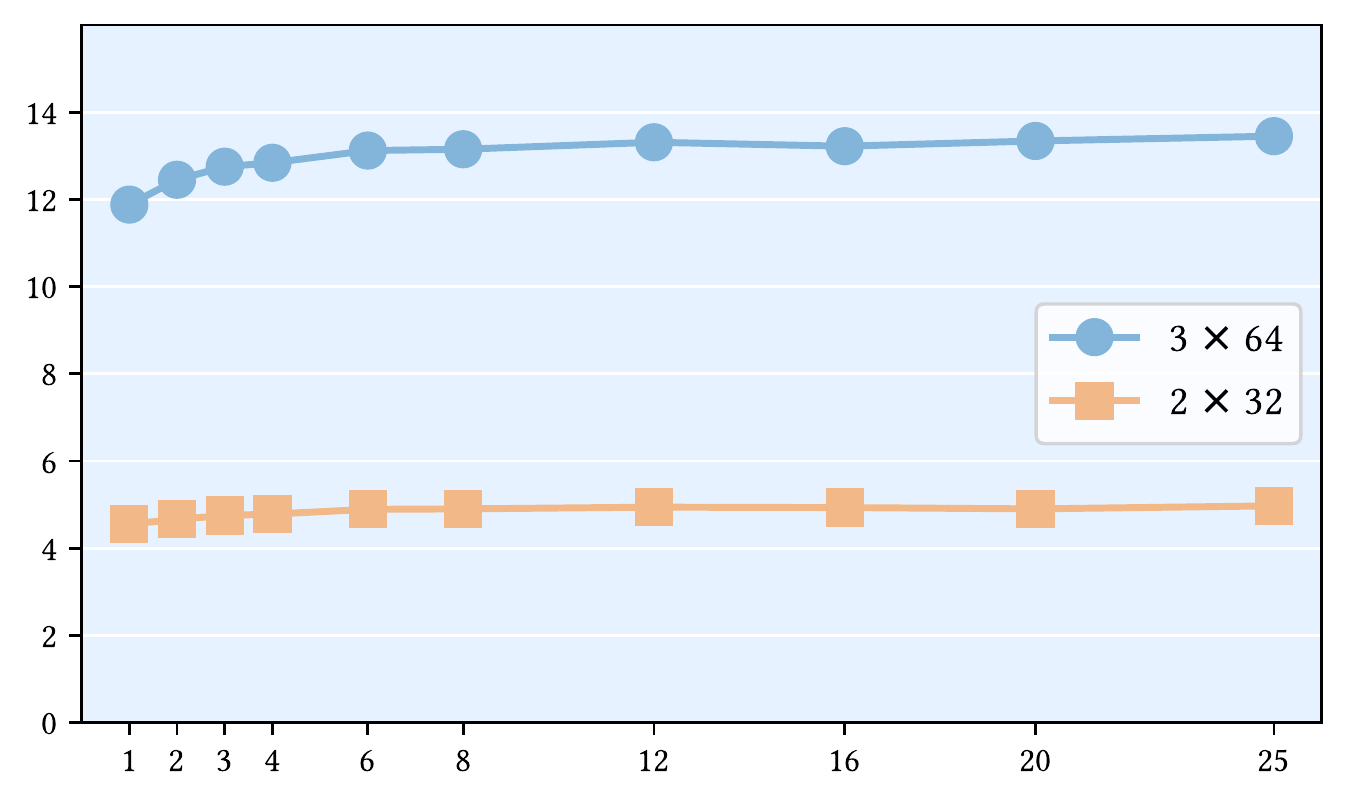}
    \end{tabular}
\end{center}

	\vspace{-4mm}
	\caption{
		Rendering times for path tracing a 1~SPP image of the \cakebox\ scene %
		with varying numbers of neural materials.
		The measurements show that our method is insensitive to the divergence introduced by path tracing
		scenes with many neural materials; rendering times stay near constant as material count increases.
		Two different BRDF decoder architectures are studied.
		The path distribution is kept fixed to isolate the effects on performance from scaling the number of materials.
	}\label{fig:cakebox_perf_scaling}
	\vspace{-3mm}
\end{figure}

\autoref{fig:cakebox_perf_scaling} shows that performance scales favorably
when increasing the number of neural materials. For this test we render the
\cakebox\ scene (\autoref{fig:system_merl_scene}) and vary the number of (different) neural materials, 
while keeping geometry and path distribution identical. Paths up to ten vertices in length
are traced and the scene also contains a small number of traditional materials,
in order to introduce significant execution and data divergence.

For very small numbers of neural materials, the network parameters fit in caches close to the shader cores,
whereas with more materials the parameters are increasingly streamed in from L2 or global memory.
Our approach based on a megakernel path tracer with local work reordering manages to extract enough 
coherency to amortize the cost of memory loads well.

\paragraph{Memory usage.} \revision{The memory footprint is dominated by the 8-channel, half-precision latent texture, requiring 256MB per 4k texture tile.
The network weights are comparably small, requiring 37kB for the 3x64 network configuration and 9.3kB for the 2x16 configuration.}

\paragraph{Discussion.}

It is difficult to do a direct comparison to previous work as our focus is different; we show that neural materials 
can run efficiently in real-time shaders even in divergent workloads such as path tracing.
There are few examples of inferencing in traditional shaders. One exception is \emph{deep shading}~\cite{Nalbach2017}
that runs a forward pass in GLSL for traditional deferred shading.
Research on neural appearance models have generally used CUDA kernels, either directly or via machine learning frameworks.

\citet{Fan2022} record all intersections to global memory and shade in a deferred manner,
precluding adaptiveness and paying the cost of memory transfers.
The authors report a single BRDF evaluation per pixel with resolution $1920\times 1080$ costing 5~ms on an NVIDIA RTX 2080Ti.
NeuMIP~\cite{Kuznetsov2021} implement an interactive CUDA/OptiX-based path tracer
and report similar performance of 5~ms per evaluation at the same resolution/GPU.
The paper is scarce on details; in personal communication it was stated that the reported 60 frames per second 
path tracing applies to relatively short paths in a simple scene with a single material. 
Scaling to multiple materials is not explored.

We believe the scalability, handling of divergent shaders, and integration in real-time shading languages
are important contributions of our work for ease of adoption of neural materials more widely.

\section{Limitations \& Future Work}

\paragraph{Energy conservation and reciprocity.} Because the neural material is only an approximate fit of the input material, it is not guaranteed to be energy conserving.
Although we have not observed this to be a problem in our tests, this could become an issue for high albedo materials with high orders of bounces (e.g.\ white fur). Enforcing energy conservation would require the network to output in a form that is analytically integrable, or integrates to a known value. The latter can be achieved with normalizing flows (as in \cite{Mueller2020}) at an increased evaluation cost.
Our BRDF model is currently not reciprocal, but reciprocity could be enforced with the modified Rusinkiewicz encoding of directions~\cite{Zheng2021}. We opted for the Cartesian parameterization of directions that was more numerically stable in our experiments and yielded better visuals.
    
\paragraph{Displacement.} We do not currently support effects that affect surface geometry, such as displacement mapping. We implemented the neural displacement approach of \citet{Kuznetsov2021}, and tested several variations that include geometric priors, but we found that this approach is always outperformed by fixed-function ray marching, both in terms of bandwidth and runtime.
None of these approaches were sufficiently fast to reach our performance goals, but we expect additional research to make them viable alternatives.

\paragraph{Filtering.} Although neural prefiltering is effective at preventing aliasing, we report that, while the finest level is very accurate, the coarser levels of the latent pyramid tend to produce softer appearance than the supersampled reference BRDF. This is likely because the inputs to the encoder correlate strongly with the appearance \emph{only} at the finest level. In case of coarser levels, the encoder consumes prefiltered material parameters, where the correlation is weaker and the auto-encoder thus performs worse. Finetuning improves the quality somewhat, but cannot escape the initial local minimum.

\paragraph{Alternative geometric priors.} We tested a number of alternative implementations of the rotation prior (\autoref{sec:rotation-op}), ranging from unconstrained, high-dimensional affine transforms inspired by the generality of self-attention layers~\cite{Vaswani2017} to rotation-only matrices.
Our final solution uses normalized (but not orthogonal) normal $\normal$ and tangent $\tang$ from the network output, with bitangent $\bitang = \normal \times \tang / \| \normal \times \tang \|$.
Additionally, we tested explicitly supervising the extracted TBN frames against frames of the reference material, with an optional asymmetric loss~\cite{Vogels2018}. This occasionally improved the results (e.g., for glints),
but the training requires extensive hyperparameter tuning; hence we excluded it from results.

\paragraph{Training stability and time} We occasionally found training to converge to local minima with large visual differences based on small perturbations of hyperparameters or weight initialization. For instance, the smallest network configuration could not reliably preserve the highly specular glazing of the \teapot{} so we chose to include a version without it in our results (\autoref{fig:performance-comparison-stage}). We want to investigate robustness more closely, also while scaling to a larger target material diversity. At the same time, we would like to significantly reduce training times (ideally from \emph{hours} to \emph{minutes}) to improve iteration times when developing further enhancements and to make the current iteration of the system more practical.

\paragraph{Refraction.} We evaluate our method only on purely reflective materials. Extending our model to transmissive materials poses the following challenge: physically based renderers require knowing the index of refraction of the material to maintain reciprocity after refracting. While the network could be trained to produce the index as an additional output, it is difficult to guarantee that this trained value matches the actual behavior of the BRDF; this topic deserves special attention in the future.

\section{Conclusion}

We present a complete real-time neural materials system.
The model jointly addresses evaluation, sampling, and filtering of highly complex and detailed materials.
We achieve this by combining ideas from prior works with new graphics priors and training strategies to achieve higher quality and faster training.
A key contribution of our work is that such comprehensive solutions can be implemented efficiently on modern graphics hardware; we propose to deploy the neural network to the innermost rendering loop to reduce bandwidth requirements.
In our tests, the neural BRDFs achieve state-of-the-art rendering performance, outperform optimized GPU implementations of reference multi-layered classical materials, and scale to multiple materials in a scene.
We believe the presented neural BRDFs can serve as ``baked'' versions of complex materials; as well as increased performance and lower memory consumption, this enables easy interchange of arbitrarily complex materials between different workflows and tools, simply by exchanging a fixed set of latent textures and a small table of MLP weights. Lastly, we hope this article will stimulate adoption of small neural networks in real-time rendering.
\vspace{-2mm}

\begin{acks}
	We want to thank Toni Bratincevic, Davide Di Giannantonio Potente, and Kevin Margo for their help creating the reference objects, Yong He for evolving the Slang language to support this project, Craig Kolb for his help with the 3D asset importer, Justin Holewinski and Patrick Neill for low-level compiler and GPU driver support, and Karthik Vaidyanathan for providing the TensorCore support in Slang. We also thank Eugene d'Eon, Steve Marschner, Thomas M{\"u}ller, Marco Salvi, and Bart Wronski for their valuable input. The material test blob in \autoref{fig:system_merl_scene} was created by Robin Marin and released under CC (https://creativecommons.org/licenses/by/3.0/).
\end{acks}

\bibliographystyle{ACM-Reference-Format}
\bibliography{paper}

\appendix

\section{Importance sampling details}
\label{app:sampling}

The following outlines the implementation details of our analytic proxy model used for importance sampling.

\paragraph{Probability density}
Like prior work~\cite{Sztrajman2021,Fan2022} our sampling density is aS linear blend between a diffuse and specular term
\begin{equation}
    \vspace{-0.5mm}
    \pdf(\wo) = w_\text{d} \cdot \pdf_\text{d}(\wo) + w_\text{s} \cdot \pdf_\text{s}(\wo),
    \quad\text{with $w_\text{d} + w_\text{s} = 1$.}
    \label{eq:pdf_full}
\end{equation}

The diffuse PDF $\pdf_\text{d}$ is a cosine-weighted distribution but tilted by a normal vector computed from a predicted 2D surface slope $(\mu_\text{d,x}, \mu_\text{d,y})$ as
\begin{equation}
    \mathbf{n}_\text{d} = \text{Normalize}([-\mu_\text{d,x}, -\mu_\text{d,y}, 1]).
\end{equation}

The specular PDF $\pdf_\text{s}$ is a standard microfacet density using a Trowbridge-Reitz (GGX) NDF~\cite{Trowbridge1975,Walter2007} with elliptical anisotropy and non-centered mean surface slopes~\cite{Dupuy2015}:
\begin{equation}
    \pdf_\text{s}(\wo)
    =
    D_\text{std}\left(\frac{\mathbf{M}^{-1} \wh}{|| \mathbf{M}^{-1} \wh ||} \right)
    \frac{\text{det}\left( \mathbf{M}^{-1} \right)}{|| \mathbf{M}^{-1} \wh ||^3}
    \frac{1}{4\left| \wo \cdot \wh \right|},
    \label{eq:pdf_spec}
\end{equation}
where $\wh = \text{Normalize}(\wi + \wo)$ is the half vector and $D_\text{std}$ is the isotropic NDF with unit roughness ($\alpha=1$), transformed based on
\begin{equation}
    \mathbf{M} = \begin{bmatrix}
        \alpha_\text{x} & 0 & -\mu_\text{s,x} \\
        \alpha_\text{y} \, \rho & \alpha_\text{y} \sqrt{1 - \rho^2} & -\mu_\text{s,y} \\
        0 & 0 & 1
    \end{bmatrix}.
\end{equation}
Here, the elliptical anisotropy is described by two orthogonal roughness values $\alpha_\text{x}$, $\alpha_\text{y}$ with correlation parameter $\rho$ and the mean of the NDF is offset by a 2D surface slope $(\mu_\text{s,x}, \mu_\text{s,y})$.

The last two terms in \autoref{eq:pdf_spec} are the Jacobian determinants accounting for the transformation (and subsequent normalization) of $\wh$, as well as the change of variables between $\wh$ and $\wo$.

\paragraph{Sampling}
The sample transform $W$ first selects one of the two PDF terms (\autoref{eq:pdf_full}) based on the relative weights $w_\text{d}$ and $w_\text{s}$.
If the diffuse component is chosen we simply generate a cosine-weighted outgoing direction $\wo$ and tilt it based on $\mathbf{n}_\text{d}$.
Otherwise, we perform specular reflection along a sampled half-vector
\begin{equation}
    \wh = \text{Normalize}(\mathbf{M} \cdot \warp_\text{std}(\mathbf{u}))
\end{equation}
where $\warp_\text{std}$ is the usual isotropic NDF sampling technique ($\alpha=1$).

\paragraph{Network prediction} We dropped the explicit dependence of $\pdf$ and $\warp$ on $\wi$ and $\x$ above for brevity, but our full set of 9 proxy parameters $\{ w_\text{d}, \mu_\text{d,x}, \mu_\text{d,y}, w_\text{s}, \alpha_\text{x}, \alpha_\text{y}, \rho, \mu_\text{s,x}, \mu_\text{s,y} \}$ are the result of an MLP evaluation that takes these as input. To ensure that all inferred parameters lie in their respective valid ranges ($\alpha \in [0, 1], \rho \in [-1, 1], \mu \in [-\infty, +\infty]$) we append an appropriate final activation to each network output based on quadratic approximations of $\tanh(x)$ and $\sinh(x)$. Lastly, $w_\text{d}$ and $w_\text{s}$ are processed by the $\text{softmax}$ function to form valid mixing weights that add up to one.

\end{document}